\documentclass[12pt]{article}
\usepackage{a4wide,epsfig,psfrag,amsmath,amssymb,cite,scalefnt}
\usepackage{color}

\newcommand{\gsim}{\;\rlap{\lower 3.5 pt \hbox{$\mathchar \sim$}} \raise 1pt
 \hbox {$>$}\;}
\newcommand{\lsim}{\;\rlap{\lower 3.5 pt \hbox{$\mathchar \sim$}} \raise 1pt
 \hbox {$<$}\;}

\newcommand{\ep}{\epsilon}

\newcommand{\mufs}{\mu_F}

\newcommand{\HPL}[2]{{\rm H}_{#1}(#2)}
\newcommand{\HS}[2]{{\rm S}_{#1}(#2)}

\begin{document}

\title{\vskip-3cm{\baselineskip14pt
    \begin{flushleft}
      \normalsize SFB/CPP-11-40\\
      \normalsize TTP11-21
  \end{flushleft}}
  \vskip1.5cm
  Production of scalar and pseudo-scalar Higgs bosons to 
  next-to-next-to-leading order at hadron colliders
}

\author{
  Alexey Pak,
  Mikhail Rogal,
  Matthias Steinhauser
  \\[1em]
  {\small\it Institut f{\"u}r Theoretische Teilchenphysik}\\
  {\small\it Karlsruhe Institute of Technology (KIT)}\\
  {\small\it 76128 Karlsruhe, Germany}
}

\date{}

\maketitle

\thispagestyle{empty}

\begin{abstract}
We consider the production of intermediate-mass CP-even and CP-odd Higgs bosons
in proton-proton and proton-anti-proton collisions. We extend the recently
published results for the complete next-to-next-to-leading order calculation 
for a scalar Higgs boson to the pseudo-scalar case and
present details of the calculation that might be useful for similar
future investigations. The result is based on an expansion in the limit of
a heavy top quark mass and a subsequent matching to the expression obtained
in the limit of infinite energy.
For a Higgs boson mass of 120~GeV the deviation from the infinite-top quark
mass result is small. For 300~GeV, however, the next-to-next-to-leading order
corrections for a scalar Higgs boson exceed the effective-theory result by
about 9\% which increases to 22\% 
in the pseudo-scalar case. Thus in this mass
range the effect on the total cross section amounts to about 2\% and
6\%, respectively, which may be relevant in future precision studies.

\medskip


\end{abstract}

\thispagestyle{empty}


\newpage


\section{Introduction}

One of the most urgent problems in the modern particle physics is
unveiling the origin of masses of elementary particles, which according
to the Standard Model (SM) is closely related to the Higgs boson. A lower
limit on the Higgs boson mass of about 114~GeV was set more than
ten years ago by the experiments on the Large Electron-Positron
Collider (LEP)~\cite{Barate:2003sz} and more recently mass values around 160~GeV
were excluded by the {\tt Tevatron}~\cite{Aaltonen:2011gs}.\footnote{We refer
  to~\cite{Baglio:2011wn} for critical comments on the {\tt Tevatron} analysis.}
Due to rarity of the process, discovering the Higgs boson production
requires subtle experimental methods and precise theoretical predictions.

The Standard Model contains only one physical CP-even Higgs boson. However,
many extensions of the SM, such as two-Higgs-doublet models or supersymmetric
models predict also charged and CP-odd Higgs bosons.
In this paper we consider the production of a pseudo-scalar Higgs boson in
the form of an external current with a generic Yukawa coupling. We
require that the coupling be proportional to the heavy quark mass,
while the coefficient can be specified within any desired model.

The dominance of the gluon-fusion process in the production of a scalar or
a pseudo-scalar Higgs boson was established in the end of the
1970's~\cite{Wilczek:1977zn,Ellis:1979jy,Georgi:1977gs,Rizzo:1979mf}.  Later,
in the beginning of the 1990's, several groups obtained the
next-to-leading order (NLO) QCD
corrections~\cite{Dawson:1990zj,Djouadi:1991tka,Spira:1995rr}. The latter
appeared to be large, modifying the LO prediction by as much as 100\%.
Thus, the accurate prediction necessitated the next-to-next-to-leading
order (NNLO) calculation.

The NLO results include the complete dependence on the partonic
center-of-mass energy ($\hat{s}$), the Higgs boson mass ($M_\Phi$) and
the top quark mass ($M_t$)~\cite{Spira:1995rr,Harlander:2005rq}.
Assuming an infinitely heavy top quark, one considerably simplifies the
calculation while introducing only a few per cent error.
The NNLO corrections were first computed in this limit,
for the scalar Higgs boson in
Refs.~\cite{Harlander:2000mg,Harlander:2002wh,Anastasiou:2002yz,Ravindran:2003um}
and for the pseudo-scalar in
Refs.~\cite{Harlander:2002vv,Anastasiou:2002wq,Ravindran:2003um}.
More recently, the missing top mass-suppressed corrections to the
scalar Higgs production were estimated and found
small~\cite{Harlander:2009mq,Pak:2009dg,Harlander:2009my}
compared to other uncertainties. In these papers the mass dependence
was recovered by interpolating between the expansion of the cross section
near the heavy top limit and the leading asymptotics in the $\hat{s}\to\infty$
limit, given in Ref.~\cite{Marzani:2008az,Harlander:2009my}.

In this paper we apply the technique used in Ref.~\cite{Pak:2009dg}
and provide similar top mass-suppressed corrections to the
pseudo-scalar Higgs boson production. We compute five expansion terms
of the cross section in $1/M_t^2$, and match the partonic
cross sections to the values of Ref.~\cite{Caola:2011wq}, derived in the
$\hat{s}\to\infty$ limit.

Similarly to the conclusions of
Refs.~\cite{Harlander:2009mq,Pak:2009dg,Harlander:2009my},
we find that the infinite-top quark mass approximation with
factorized exact LO mass dependence receives relatively small corrections.
We are not aware of any trivial explanations of this behaviour. Note that 
assuming no factorization of the exact LO quark mass dependence the
$1/M_t^0$ result augumented with $\hat{s}\to\infty$ behaviour deviates
far from the infinite-top mass result, and only after including at least the
$1/M_t^6$ corrections the agreement of hadronic cross sections reaches
the level of a few percent.

For completeness let us also mention several results that improve upon
the fixed-order QCD. Those include the soft-gluon resummation to
next-to-next-to-leading~\cite{Catani:2003zt} and
next-to-next-to-next-to-leading~\cite{Moch:2005ky,Ravindran:2005vv,Ravindran:2006cg}
logarithmic orders and the identification (and resummation) of certain
$\pi^2$ terms~\cite{Ahrens:2008nc} that significantly improves the
convergence of the perturbative series.
Recent numerical predictions of Higgs boson
production in gluon fusion both at the Tevatron and the LHC are summarized
in Ref.~\cite{Anastasiou:2008tj,deFlorian:2009hc,Anastasiou:2011pi}. For reviews,
see Refs.~\cite{Djouadi:2005gi,Dittmaier:2011ti}.

The remainder of the paper is organized as follows: in the next
Section we introduce our notation and the basic formalism.  
After that, we describe the treatment of
$\gamma_5$ for the case of the pseudo-scalar Higgs boson.  In
Section~\ref{sec::nlo} we concentrate on the NLO prediction and
compare our approximations to the exact result both at the partonic
and the hadronic levels.  Section~\ref{sec::part} is devoted to the
NNLO partonic corrections and Section~\ref{sec::hadr} discusses the
hadronic cross section. Results are presented for the {\tt LHC}
running at $14$~TeV center-of-mass
energies. Conclusions are presented in Section~\ref{sec::concl}.
In Appendix~\ref{app::2lMIs} we discuss the phase space master integrals
expanded to the $\epsilon$ order sufficient for a N$^3$LO calculation.
Technical details of the convolutions of various functions are given
in Appendix~\ref{app::conv}.


\section{\label{sec::prel}Preliminaries}


\subsection{Notation and the LO result}

In the full theory, the scalar and the pseudo-scalar Higgs bosons couple
to fermions via the following terms in the Yukawa Lagrange density:
\begin{eqnarray}
  {\cal L}_Y &=& 
  - g_q^{Y,H} m_q^0 \frac{H^0}{v^0} \bar{q}^0 q^0 
  - g_q^{Y,A} m_q^0 \frac{A^0}{v^0} \bar{q}^0 i\gamma^5 q^0
  \,,
  \label{eq::leff}
\end{eqnarray}
where the dimensionless couplings $g_q^{Y,H}$ and  $g_q^{Y,A}$
specify the coupling strength of the Higgs bosons to the heavy quark $q$.
The cross section of the pseudo-scalar Higgs boson production
is proportional to $(g_q^{Y,A})^2$; in the following discussion of
this cross section we drop this constant for convenience.

In the Standard Model $g_q^{Y,H} = 1$, $g_q^{Y,A} = 0$, but
in the MSSM, e.g., one has 
$g_t^{Y,H} \sim 1/\sin\beta$,
$g_b^{Y,H} \sim 1/\cos\beta$,
$g_t^{Y,A} \sim 1/\tan\beta$,
$g_b^{Y,A} \sim \tan\beta$
where $\tan\beta$ is the
ratio of the Higgs field vacuum expectation values.
Thus, the Higgs coupling to the top quark mass is suppressed
for large $\tan\beta$ and our
analysis is only valid for small values; for larger
values also the contribution from bottom quarks has to be considered
(see, e.g., the recent publications~\cite{Harlander:2010wr,Degrassi:2011vq}).

The superscript ``0'' in Eq.~(\ref{eq::leff}) indicates bare quantities.
Since we only consider QCD corrections, there are no counterterms
for $v^0$, $H^0$ and $A^0$, and the Higgs field vacuum expectation
value is $v=2^{-1/4}G_F^{-1/2}$. We also introduce the variables
\begin{eqnarray}
  \rho &=& \frac{M_\Phi^2}{M_t^2}\,,~~
  x = \frac{M_\Phi^2}{\hat{s}}\,,
  \label{eq::xrho}
\end{eqnarray}
where $\hat{s}$ is the partonic center-of-mass energy, $\Phi$ denotes either
scalar ($H$) or pseudo-scalar ($A$) Higgs boson and $M_t$ is the
top quark pole mass. The divergences in the loop
integrals are regularized using dimensional regularization with
$d = 4 - 2\epsilon$ space-time dimensions.

The partonic cross section is commonly factorized as
\begin{eqnarray}
  \hat{\sigma}_{ij\to \Phi + X}(x) &=& \hat{A}_{\rm LO}^\Phi \left(
    \delta(1-x)
    + \left(\frac{\alpha_s}{\pi}\right) \Delta_{ij}^{\Phi,(1)}  
    + \left(\frac{\alpha_s}{\pi}\right)^2 \Delta_{ij}^{\Phi,(2)} + \ldots
  \right)
  \,,
  \label{eq::hatsigma}
\end{eqnarray}
where the factor $\hat{A}_{\rm LO}^\Phi$ contains the constants and
the complete non-trivial LO $\rho$-dependence.
Such factorization provides an excellent agreement between the exact
and the approximated results for the hadronic cross section at the
NLO and the NNLO~\cite{Harlander:2009mq,Pak:2009dg,Harlander:2009my}.
The well-known LO result is
\begin{eqnarray}
  \hat{A}_{\rm LO}^\Phi &=& \frac{G_F~\alpha_s^2}{288\sqrt{2}\pi} 
  f_0^\Phi(\rho)
  \,,
\end{eqnarray}
with
\begin{eqnarray}
  f_0^H(\rho) &=& 
  \left\{
    \begin{array}{ll}
      \frac{36}{\rho^2}\left[1 + \left(1 -
          \frac{4}{\rho}\right)\arcsin^2
        \left(\frac{\sqrt{\rho}}{2}\right)\right]^2, 
      &
      \qquad(\rho\le4)\,,
      \\
      \frac{9}{4\rho^2}\left|4-(1-\frac{4}{\rho})
      \left[\ln \frac{1+\sqrt{1-4/\rho}}{1-\sqrt{1-4/\rho}}-i\pi\right]^2\right|^2, 
      &
      \qquad(\rho>4)\,,
    \end{array}
  \right.
  \\
  f_0^A(\rho) &=& 
  \left\{
    \begin{array}{ll}
      \frac{36}{\rho^2}\arcsin^4
        \left(\frac{\sqrt{\rho}}{2}\right), 
      &
      \qquad(\rho\le4)\,,
      \\
      \frac{9}{4\rho^2}\left|\ln \frac{1+\sqrt{1-4/\rho}}{1-\sqrt{1-4/\rho}}-i\pi\right|^4, 
      &
      \qquad(\rho>4)\,.
    \end{array}
  \right.
\end{eqnarray}

In what follows, whenever we refer to the infinite-top quark mass result we
assume the factorization of the exact LO mass dependence as given in 
Eq.~(\ref{eq::hatsigma}).


\subsection{\label{sub::opt}Optical theorem and asymptotic expansion}

\begin{figure}[t]
  \centering
  \includegraphics[width=0.3\linewidth]{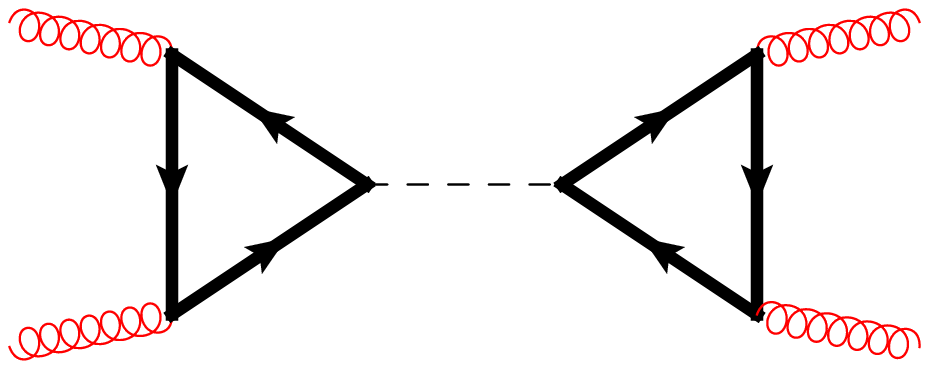}\hfill
  \includegraphics[width=0.3\linewidth]{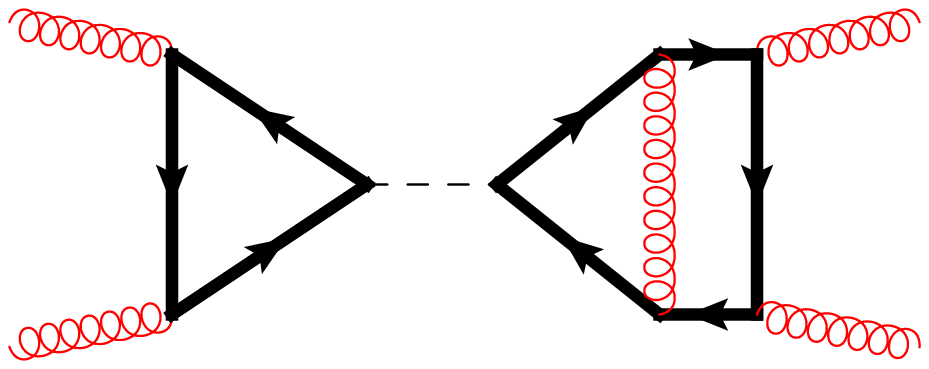}\hfill
  \includegraphics[width=0.3\linewidth]{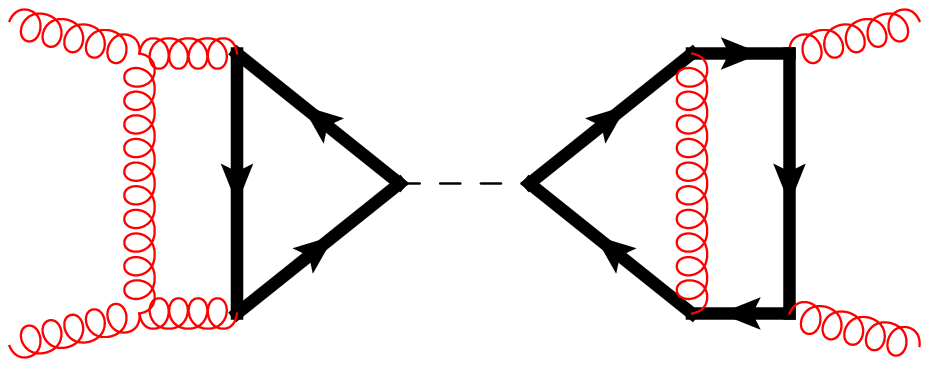}
  \\[1em]
  \includegraphics[width=0.3\linewidth]{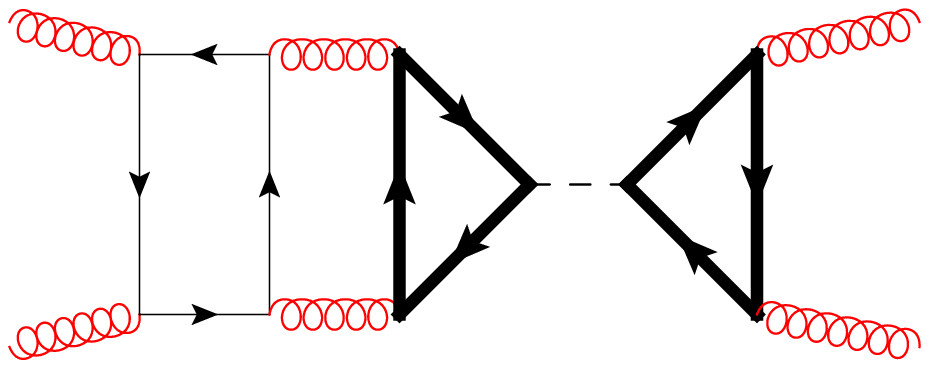}\hfill
  \includegraphics[width=0.3\linewidth]{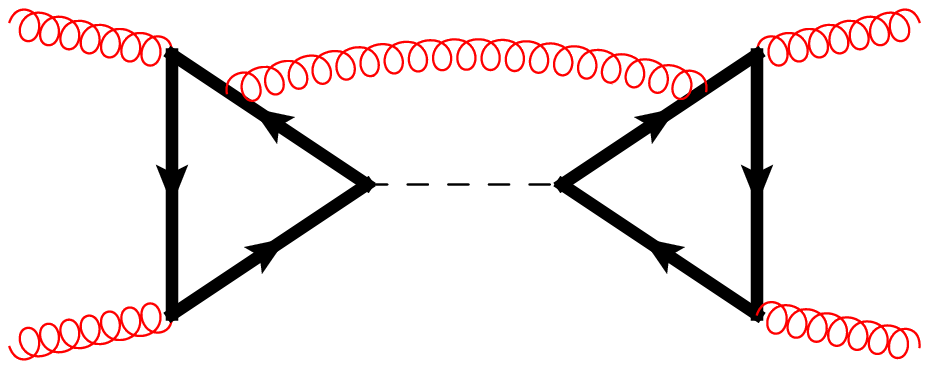}\hfill
  \includegraphics[width=0.3\linewidth]{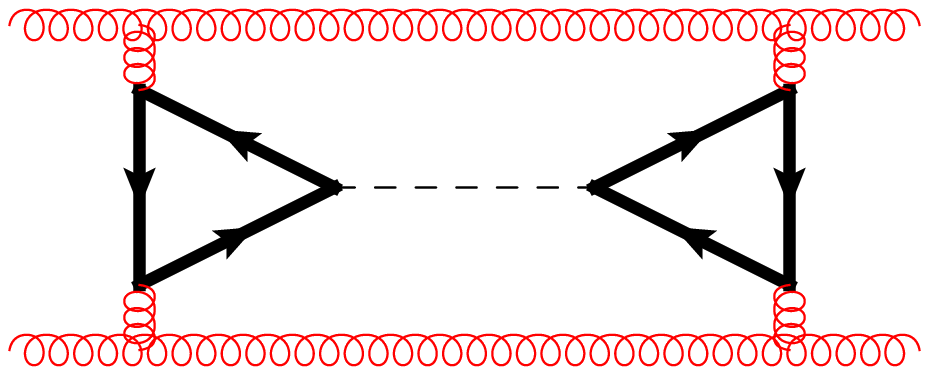}
  \\[1em]
  \includegraphics[width=0.3\linewidth]{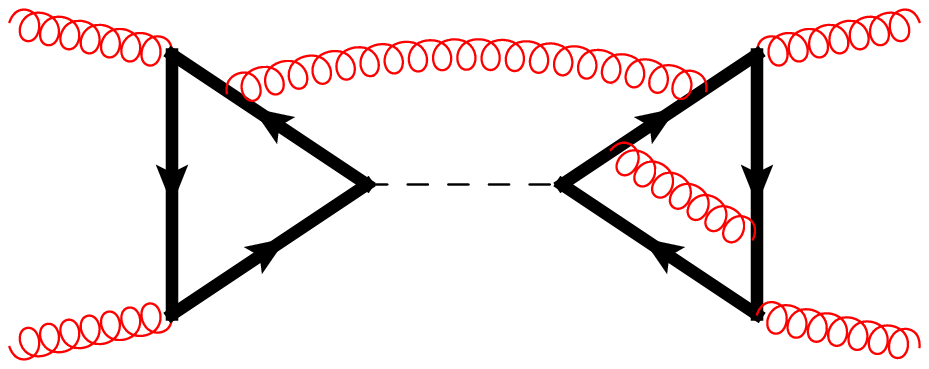}\hfill
  \includegraphics[width=0.3\linewidth]{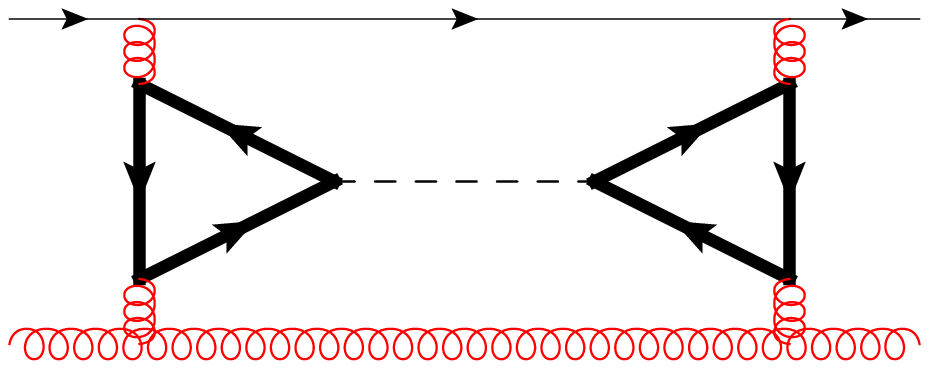}\hfill
  \includegraphics[width=0.3\linewidth]{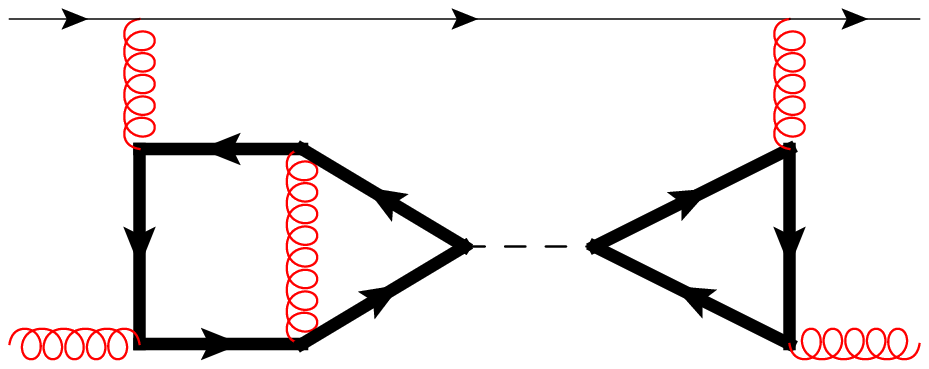}
  \\[1em]
  \includegraphics[width=0.3\linewidth]{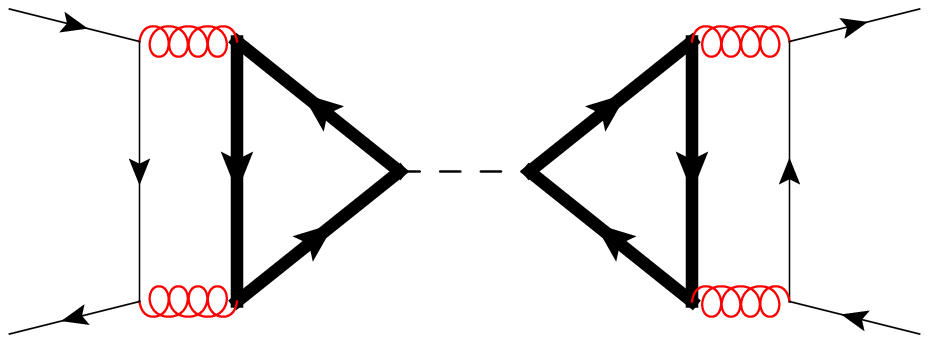}\hfill
  \includegraphics[width=0.3\linewidth]{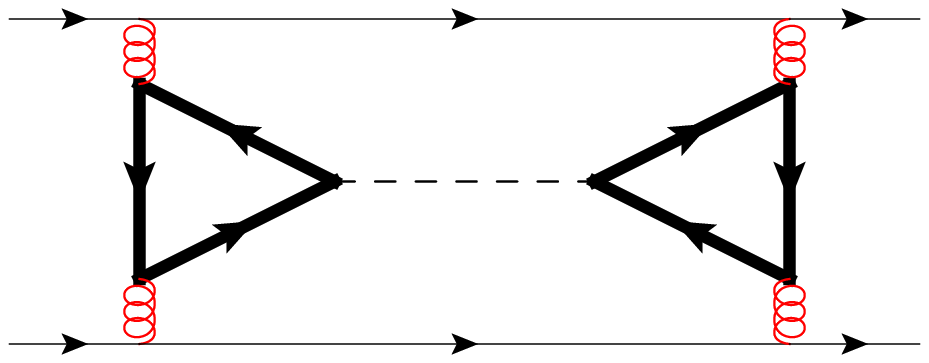}\hfill
  \includegraphics[width=0.3\linewidth]{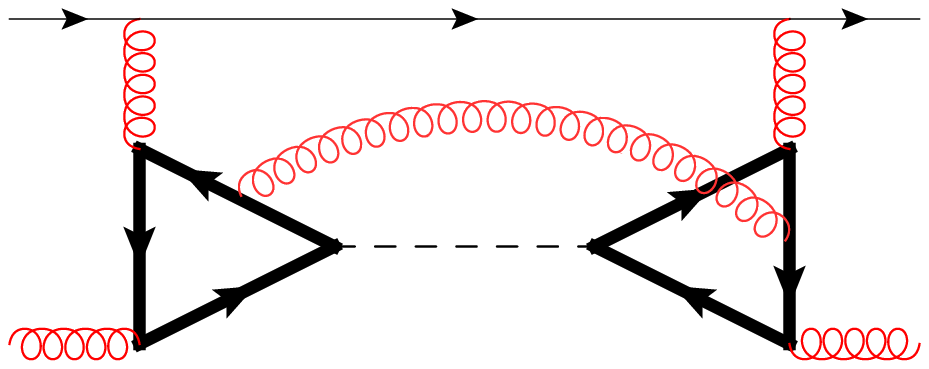}
  \caption[]{\label{fig::diag}
    Sample forward scattering diagrams
    whose cuts correspond to the LO, NLO and NNLO corrections to
    $gg\to \Phi +gg$, $qg\to \Phi +qg$ and $qq\to \Phi + qq$. 
    Dashed, curly and thick (thin) solid lines represent
    Higgs bosons, gluons and top (light) quarks, respectively.}
\end{figure}

Already at the NLO one has to consider real and virtual contributions
which individually contain quadratic poles in $\epsilon$.
In our approach we consider the forward scattering amplitude and use the
optical theorem in order to derive the inclusive total cross section for the
production of Higgs bosons (only the cuts that cross the Higgs boson line
should be considered). At the NLO, the possible initial and final states are
$gg\to \Phi + \mbox{(0 or 1 gluon)}$, $qg\to \Phi + q$, and $q\bar{q}\to \Phi + g$.
At the NNLO, we in addition have reactions $gg\to \Phi + \mbox{($gg$ or $q\bar{q}$)}$,
$qg\to \Phi + qg$, $q\bar{q}\to \Phi + \mbox{($gg$ or $q\bar{q}$)}$, $qq\to \Phi + qq$,
and $qq^\prime \to \Phi + qq^\prime$. Here $q$ and $q^\prime$ stand for
different massless quark flavours.\footnote{It is understood that the ghosts
  are always considered together with gluons.} 
Sample diagrams of the corresponding forward-scattering amplitudes
are shown in Fig.~\ref{fig::diag}.

The expressions are simplified by the forward-scattering kinematics
implied by the optical theorem. The proper projectors applied to the
external massless particles reduce each amplitude
to a scalar expression depending only on $\rho$ and $x$.
Yet, the imaginary part of the double-scale four-loop integral with
the exact dependence on both variables is still out of reach with the
present methods, and we apply the asymptotic expansion~\cite{Smirnov:2002pj}
in $\rho\to 0$ which corresponds to the limit $M_t^2\gg M_\Phi^2,\hat{s}$.
After the expansion, every four-loop integral factorizes into one-,
two- or three-loop vacuum bubble with the single mass scale $M_t$,
and the tree-level (for the virtual corrections), one- or two-loop
box graph depending on $x$.

The integrals with various denominator exponents that appear
during the asymptotic expansion are reduced to a few master integrals
using the integration-by-parts (IBP) relations, in which we treat cut
lines as normal propagators. The two- and three- particle cuts are
again re-introduced in the master integrals and evaluated separately.

\begin{figure}[t]
  \centering
  \includegraphics[width=0.3\linewidth]{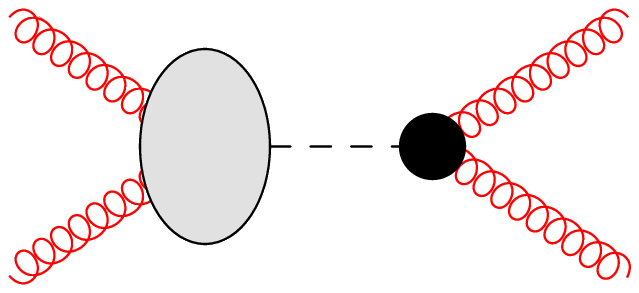}\hfill
  \includegraphics[width=0.3\linewidth]{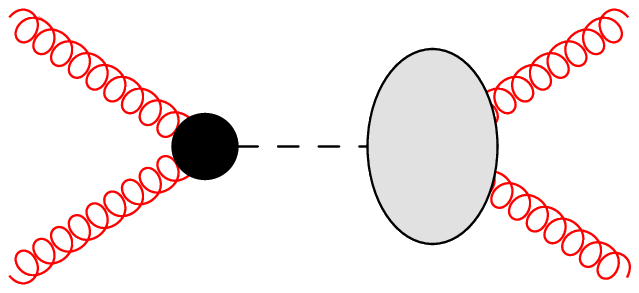}\hfill
  \includegraphics[width=0.3\linewidth]{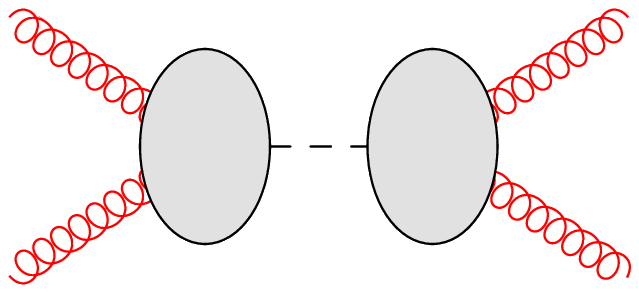}
  \\
  \hspace{2cm} (a)\hfill (b)\hfill (c) \hspace{2cm}
  \caption[]{\label{fig::opt_virt}
    Possible Feynman diagrams for the virtual corrections to the
    $gg$-initiated partonic cross section. The shaded blobs represent massless
    one- or two-loop diagrams, the black dots mark the effective vertices
    that have no imaginary part.}
\end{figure}

In the case of the virtual corrections the imaginary part of the
Feynman diagrams arises solely from the factor
\begin{eqnarray}
  \left(-1+i0\right)^{a\epsilon} &=&
    1 - i \left(\pi a \epsilon\right)
      - \frac{\left(\pi a \epsilon\right)^2}{2}
  + {\cal O}(\epsilon^3)\,,
  \label{eq::imag}
\end{eqnarray}
where in our case $a=1$ or $a=2$. Schematically, the occurring diagrams
can be divided into three cases sketched for the $gg$-initiated diagrams
in Fig.~\ref{fig::opt_virt}.
In the cases (a) and (b) the calculation is straightforward, the
loop integrals are evaluated and the factor of Eq.~(\ref{eq::imag})
expanded in $\epsilon$. If the grey blob develops an imaginary part,
we discard it since it corresponds to a cut outside the Higgs line.
The case (c) has massless one-loop integrals on the both
sides of the Higgs boson propagator. Here in order to
reproduce the corresponding product of a one-loop amplitude and
a complex conjugate amplitude, one has to replace the
factor $(-1+i0)^{2 \epsilon}$ with $(-1+i0)^{\epsilon} (-1-i0)^{\epsilon} = 1$.


\subsection{Treatment of $\gamma_5$}

The optical theorem simplifies the treatment of $\gamma^5$ which appears
in the coupling of the pseudo-scalar Higgs boson to quarks. We follow the
prescription of Ref.~\cite{Larin:1993tq} for the pseudo-scalar current
renormalization, replacing
\begin{eqnarray}
  \gamma_5 \to \frac{i}{24}
    \epsilon_{\mu\nu\rho\sigma} \gamma^{[\mu}\gamma^{\nu}\gamma^{\rho}\gamma^{\sigma]}.
  \label{eq::gamma5}
\end{eqnarray}
The square brackets denote total anti-symmetrization. The 24
terms on the right-hand side of Eq.~(\ref{eq::gamma5}) can be simplified via
\begin{eqnarray}
  \frac{1}{24}
  \gamma^{[\mu}\gamma^{\nu}\gamma^{\rho}\gamma^{\sigma]}
  &=&
  \frac{1}{4}
  \left(
    \gamma^{\mu}\gamma^{\nu}\gamma^{\rho}\gamma^{\sigma}
  + \gamma^{\sigma}\gamma^{\rho}\gamma^{\nu}\gamma^{\mu}
  - \gamma^{\nu}\gamma^{\rho}\gamma^{\sigma}\gamma^{\mu}
  - \gamma^{\mu}\gamma^{\sigma}\gamma^{\rho}\gamma^{\nu}
  \right).
  \label{eq::gamma5_2}
\end{eqnarray}
Next, one should factor out the $\epsilon$ tensor and compute the remaining
amplitude in $d$ dimensions. At the very end, when the expressions
are finite in the limit $\epsilon\to 0$, one multiplies the result
with the $\epsilon$ tensor and applies the finite renormalization
constant of Ref.~\cite{Larin:1993tq}.

Each forward-scattering amplitude contains two factors of $\gamma_5$.
After the replacements of Eq.~(\ref{eq::gamma5}) we have the product
of two $\epsilon$ tensors that can be immediately re-written as the
product of four metric tensors with antisymmetrized indices:
\begin{eqnarray}
  \epsilon^{\alpha\beta\gamma\delta} \epsilon_{\mu\nu\rho\sigma}
  &= - g^{[\alpha}_{\mu} g^{\beta}_{\nu} g^{\gamma}_{\rho}
  g^{\delta]}_{\sigma}
 =& - \begin{vmatrix}
 g^{\alpha}_{\mu} & g^{\beta}_{\mu} & g^{\gamma}_{\mu} & g^{\delta}_{\mu}\\
 g^{\alpha}_{\nu} & g^{\beta}_{\nu} & g^{\gamma}_{\nu} & g^{\delta}_{\nu}\\
 g^{\alpha}_{\rho} & g^{\beta}_{\rho} & g^{\gamma}_{\rho} & g^{\delta}_{\rho}
 \\
 g^{\alpha}_{\sigma} & g^{\beta}_{\sigma} & g^{\gamma}_{\sigma} &
 g^{\delta}_{\sigma}
  \end{vmatrix}
  \,.
\end{eqnarray}
The right-hand side of this equation is defined in $d$ dimensions and can
be used during the calculation of the Feynman diagrams which eliminates
explicit projectors.


\subsection{Alternative approach to virtual corrections}

As a cross-check, we evaluated the virtual corrections using two
different approaches. First, we use the optical theorem as described above
with a simple implementation of $\gamma_5$ according to Ref.~\cite{Larin:1993tq},
treating the virtual and the real corrections on the same ground.
The finite result is then obtained before applying the finite $\gamma_5$
renormalization constant.

In the second method we consider the (pseudo-scalar) Higgs-gluon-gluon vertex
diagrams and expand in the (formal) limit $M_t\gg M_H$. This is similar
to the calculation of Ref.~\cite{Pak:2009bx} for the scalar Higgs boson.
The amplitudes are multiplied with projectors that couple to Lorentz
indices of the gluons and the four additional indices that remain after
the epsilon-tensor removal~\cite{Chetyrkin:1998mw}.

More explicitly, the pseudo-scalar Higgs-gluon-gluon amplitude has the form
\begin{eqnarray}
  A_{gg\to A}^{\alpha\beta,ab} &=&
  A^{ab}_{gg\to A} \,\,
  \epsilon^{\alpha\beta\mu\nu} q_{1,\mu}q_{2,\nu}\,,
  \label{eq::gga}
\end{eqnarray}
where $\alpha$ and $a$, $\beta$ and $b$ are the Lorentz and the colour
indices of the incoming gluons. After the replacement Eq.~(\ref{eq::gamma5})
$A_{gg\to A}^{\alpha\beta,ab}$ becomes
\begin{eqnarray}
  A^{ab}_{gg\to A,\alpha\beta} &=&
  \epsilon_{\mu\nu\rho\sigma} A_{gg\to A,\alpha\beta}^{\mu\nu\rho\sigma,ab}\,.
\end{eqnarray}
Together with Eq.~(\ref{eq::gga}) it gives
\begin{eqnarray}
  A_{gg\to A,\alpha\beta}^{\mu\nu\rho\sigma,ab} &=&
  \frac{1}{24}
  A^{ab}_{gg\to A} \,\,q_{1}^{[\mu} q_{2}^{\nu} g^{\rho}_\alpha
  g^{\sigma]}_{\beta}\,,
\end{eqnarray}
where $q_1$ and $q_2$ are the incoming momenta of the gluons.
Now the (Lorentz) scalar amplitude $A^{ab}_{gg\to A}$ can be obtained from
$A_{gg\to A,\alpha\beta}^{\mu\nu\rho\sigma,ab}$ via the projector
$P^{\alpha\beta}_{\mu\nu\rho\sigma}$~\cite{Chetyrkin:1998mw}:\footnote{Note
  that the formulae presented in Ref.~\cite{Chetyrkin:1998mw} 
  apply to the coefficient function of the effective theory whereas here
  we invesigate the virtual corrections in the full theory.}
\begin{eqnarray}
  A^{ab}_{gg\to A} &=&
  P^{\alpha\beta}_{\mu\nu\rho\sigma}
  A_{gg\to A,\alpha\beta}^{\mu\nu\rho\sigma,ab}\,,\nonumber\\
  \label{eq::viaproj}
  P^{\alpha\beta}_{\mu\nu\rho\sigma} &=&
   - \frac{q_{1,[\mu} q_{2,\nu} g^\alpha_\rho
  g^\beta_{\sigma]}}{(d-2) (d-3) (q_1\cdot q_2)^2}\,.
\end{eqnarray}
In the actual calculations we use Eq.~(\ref{eq::gamma5}) in the initial
diagram, then drop $\epsilon_{\mu\nu\rho\sigma}$ and
obtain $A_{gg\to A,\alpha\beta}^{\mu\nu\rho\sigma,ab}$.
Using Eq.~(\ref{eq::viaproj}) and summing the diagrams, we then
arrive at $A_{gg\to A}^{ab}$ which has no open Lorentz indices.
The virtual contribution to the total cross section is finally
obtained by squaring the amplitude Eq.~(\ref{eq::gga}) and
integrating over the phase space. Accounting for the averaging
factors, we find:
\begin{eqnarray}
  \hat{\sigma}_{\rm virt} &=& \frac{\pi}{256}
  \frac{(d-3)}{(d-2)}\left|A^{ab}_{gg\to A}\right|^2.
\end{eqnarray}


\subsection{Software}

To evaluate Feynman diagrams we use the well-tested chain of computer
algebra programs developed for the scalar Higgs case~\cite{Pak:2009dg}.
The original diagrams, the corresponding asymptotic expansion prescriptions,
and the master integrals are identical to the scalar Higgs
case, with only trivial changes. Thus, here we only briefly outline
the procedure.

First, the diagrams are generated with {\tt QGRAF}~\cite{Nogueira:1991ex}
supplemented by additional scripts that eliminate unnecessary graphs.  Each
diagram is then expanded in the limit $M_t^2\gg \hat{s},M_\Phi^2$ using two
independent programs: a combination of {\tt q2e} and {\tt
  exp}~\cite{Harlander:1997zb,Seidensticker:1999bb}, and a separate {\tt Perl}
program.  This turns the original four-loop three-scale integrals into
products of single-scale (one-, two-, and three-loop) vacuum bubbles and
double-scale forward-scattering integrals (of one and two loops).  The latter
are reduced to master integrals with our own implementation of the Laporta
algorithm~\cite{Laporta:1996mq,Laporta:2001dd}.  Master integrals have been
computed in Ref.~\cite{Anastasiou:2002yz}; in this calculation we extended
them by one order in $\epsilon$ which might be useful for the future
calculations.


\subsection{Initial state singularities}

The renormalized sum of the real and the virtual diagrams
is not finite in the limit $\ep\to 0$. The remaining poles
originate from the collinear divergences in the initial state:
the initial gluon may split into a quark-antiquark pair,
and the quark participate in the Higgs boson production.
The corresponding contribution is determined by
\begin{eqnarray}
  \hat{\sigma}_{ij\to \Phi + X}^{\rm renormalized}(x)
  &=& \mathcal{R}\left[\hat{\sigma}_{ik}^{\rm bare} \otimes P_{kj}
    + P_{ik} \otimes \hat{\sigma}_{kj}^{\rm bare}\right],
  \\ \nonumber
  P_{ij}(x) &=& \delta_{ij} \delta(1 - x)
    + \frac{\alpha_s^{(5),{\rm bare}}}{\pi} P_{ij}^{(1)}(x) + \ldots,
\end{eqnarray}
where $P_{ij}^{(k)}(x)$ is the splitting function that describes
the probability of parton $j$ to emit parton $i$
with the fraction $x$ of its initial energy. $\mathcal{R}$ is the
renormalization operation, and $\alpha_s^{(5),{\rm bare}}$ is defined
as prescribed by the definition of the splitting functions
and consistently with the definition of the parton distribution
functions (PDFs) which absorb the non-perturbative and non-singular
features of the initial state.

In our calculation, we use the common definition consistent with
the MSTW08 PDF set, where $\alpha_s^{(5),{\rm bare}}$ is the bare coupling
in the effective theory with decoupled top quark and $n_f = 5$
massless quarks. The convolution of functions that enter into
$P_{ij}^{(k)}(x)$ and $\hat{\sigma}_{kj}^{\rm bare}$ is
defined as
\begin{eqnarray}
  \left[f \otimes g \right](x)
    = \int_0^1 {\rm d}x_1 {\rm d}x_2 \delta(x - x_1 x_2) f(x_1) g(x_2).
\end{eqnarray}
Appendix~\ref{app::conv} contains the details of evaluation
of these integrals.
In this context, see also Refs.~\cite{Anastasiou:2002yz,Ravindran:2003um} and
references therein. 


\subsection{From partonic to hadronic cross sections}

In order to find the hadronic cross section one has to convolute the
partonic cross section with the PDFs which can be written in the form
\begin{eqnarray}
  \sigma_{pp\to \Phi+X}(s) &=&
  \sum_{ij} \int_{M_\Phi^2/s}^1 {\rm d}x\,
  \bigg[\frac{{\rm d}{\cal L}_{ij}}{{\rm d}x}\bigg](x,\mufs)~
  \hat{\sigma}_{ij\to \Phi+X}(x,\mufs,\mu_R)\,,
  \label{eq::sighad}
\end{eqnarray}
where the sum includes all distinct production channels,
$ij\in\{gg,~ qg,~ q\bar{q},~ qq,~ qq^\prime\}$. The
luminosity function ${\rm d}{\cal L}_{ij}/{\rm d}x$
contains the symmetry factors and the convolution of
PDFs. For example, the quark-gluon luminosity is given by\footnote{This
  definition applies to $pp$ collisions at the {\tt LHC}; its 
  modification for $p\bar{p}$ collisions at the Tevatron is obvious.}
\begin{eqnarray}
  \bigg[\frac{d{\cal L}_{qg}}{{\rm d}x}\bigg](x,\mufs)
  &=&
  2 \sum_{q\in\{u,...,b,\bar{u},...,\bar{b}\}}
  \int_0^1 {\rm d}x_1 \int_0^1 {\rm d}x_2\,
  f_{g/p}(x_1,\mufs) f_{q/p}(x_2,\mufs)\,
  \\ \nonumber && \times
  \delta\left({M^2_\Phi\over s x} - x_1 x_2\right) {M^2_\Phi \over s x^2}\,.
  \label{eq::lumi} 
\end{eqnarray}

In Ref.~\cite{Pak:2009dg} the luminosities ${\rm d}{\cal L}_{ij}/{\rm d}x$
have been discussed in some detail; for our purposes it is important to
remember that ${\rm d}{\cal L}_{gg}/{\rm d}x$ is practically
zero for $x \lesssim 10^{-3}$.

We split the hadronic cross section according to
\begin{eqnarray}
  \sigma_{pp^\prime\to \Phi+X}(s) &=&
  \sigma^{\rm LO}_\Phi + \delta\sigma^{\rm NLO}_\Phi 
  + \delta\sigma^{\rm NNLO}_\Phi
  \,,
  \label{eq::deltasigma}
\end{eqnarray}
and add subscript ``$\infty$'' or ``exact'' to the quantities in 
Eq.~(\ref{eq::deltasigma}) when referring to the infinite-top
quark mass result or the exact expression, respectively.
A subscript ``$n$'' indicates that corrections through order $1/M_t^n$
have been included.
We leave out the subscript $\Phi$ in case when the meaning is clear from the
context. 

We use LO, NLO and NNLO PDFs by the MSTW2008
collaboration~\cite{Martin:2009iq} in order to obtain the respective
predictions for $\sigma_{pp^\prime\to \Phi+X}$ in Eq.~(\ref{eq::deltasigma}).
The choice of the PDFs determines
the values of $\alpha_s(M_Z)\equiv\alpha_s^{(5)}(M_Z)$:
\begin{eqnarray}
  \alpha_s^{\rm LO}(M_Z) &=& 0.139384\,,~~
  \alpha_s^{\rm NLO}(M_Z) = 0.120176\,,~~
  \alpha_s^{\rm NNLO}(M_Z) = 0.117068\,.
\end{eqnarray}
The appropriate $\overline{\rm MS}$ beta function then
determines $\alpha_s(\mu_R)$ that enters the formulae.



\section{\label{sec::nlo}NLO cross section}

\subsection{\label{sub::nlo_part}Partonic cross section}

In this subsection we discuss the partonic NLO cross section as a
function of $x=M_\Phi^2/\hat{s}$.  As mentioned in the previous
section, we computed the partonic cross section in the limit
$M_t^2\gg \hat{s}, M_\Phi^2$. By construction these approximations
poorly converge for $\hat{s} > (2 M_t)^2$, i.e.
in the energy region where top quark pairs can be produced.
In terms of the variable $x$, the threshold is given by
$x_{\rm thr}=M_\Phi^2/(4M_t^2)$ which is $x_{\rm thr} = 0.12$
for $M_\Phi=120$~GeV and $x_{\rm thr} = 0.75$ for $M_\Phi=300$~GeV.
For $x < x_{\rm thr}$ we do not expect our expansion in $1/M_t^2$
to converge.

To cure this problem we follow the suggestions discussed in the
literature and match the heavy-top expansion to the leading
term in the high-energy expansion obtained in
Refs.~\cite{Marzani:2008az,Harlander:2009my} 
and~\cite{Caola:2011wq} for the scalar and pseudo-scalar Higgs boson,
respectively. The matching procedure has been successfully
applied to the scalar Higgs boson in
Refs.~\cite{Harlander:2009mq,Pak:2009dg,Harlander:2009my}. 
In the pseudo-scalar case the matching was done in
Ref.~\cite{Caola:2011wq} based on the leading order
approximation in $M_t^2\gg \hat{s}, M_A^2$. In the following we
discuss the effect of the power-suppressed terms and compare to the
exact result as implemented in {\tt HIGLU}~\cite{Spira:1995mt}.

Before we discuss our results in detail, let us describe our
matching procedure. The expansion near the heavy top limit of the quantities 
$\Delta_{ij}(x)$ in Eq.~(\ref{eq::hatsigma}) has the form\footnote{In what
  follows we omit the superscripts ``$\Phi$'' and ``$(1)$''.}
\begin{eqnarray}
  \Delta_{ij}^{\rm exp}(x) = \Delta_{ij,0}(x)
  + \rho \Delta_{ij,1}(x)
  + \rho^2 \Delta_{ij,2}(x) + \ldots
  \label{eq::sig^e}
\end{eqnarray}
and is expected to converge for $x > x_{\rm thr}$.

The $x\to 0$ limit of the NLO partonic cross section has the form
\begin{eqnarray}
  \Delta_{ij}(x) \stackrel{x\to 0}{=} C_{ij} + \mathcal{O}(x)\,,
\end{eqnarray}
with coefficients $C_{ij}$ computed in
Refs.~\cite{Marzani:2008az,Harlander:2009my,Caola:2011wq}.
In order to combine these results we find some matching value
$x_m \sim x_{\rm thr}$ and coefficient $D_{ij}$, so that
$\Delta_{ij}(x) = C_{ij} + D_{ij}x$, $x < x_m$, and
$\Delta_{ij}(x) = \Delta_{ij}^{\rm exp}(x)$, $x > x_m$.
As a matching condition, we require that
$\Delta_{ij}^{\rm exp}(x_m) = C_{ij} + D_{ij} x_m$ and
$\frac{d}{dx} \Delta_{ij}^{\rm exp}(x_m) = D_{ij}$, i.e.
we ensure that the transition between the $x\to 0$ and
$M_t\gg \hat{s}, M_\Phi^2$ approximations is smooth.
The matching prescription for the channels with
quarks in the initial state is less intelligent due to
slower convergence below threshold and the oscillating
functions. Here we set $x_m=0.9\,x_{\rm thr}$
and determine $D_{ij}$ from the requirement
$\Delta_{ij}^{\rm exp}(x_m) = C_{ij} + D_{ij} x_m$.

\begin{figure}[t]
  \centering
  \begin{tabular}{cc}
    \includegraphics[width=0.45\linewidth]{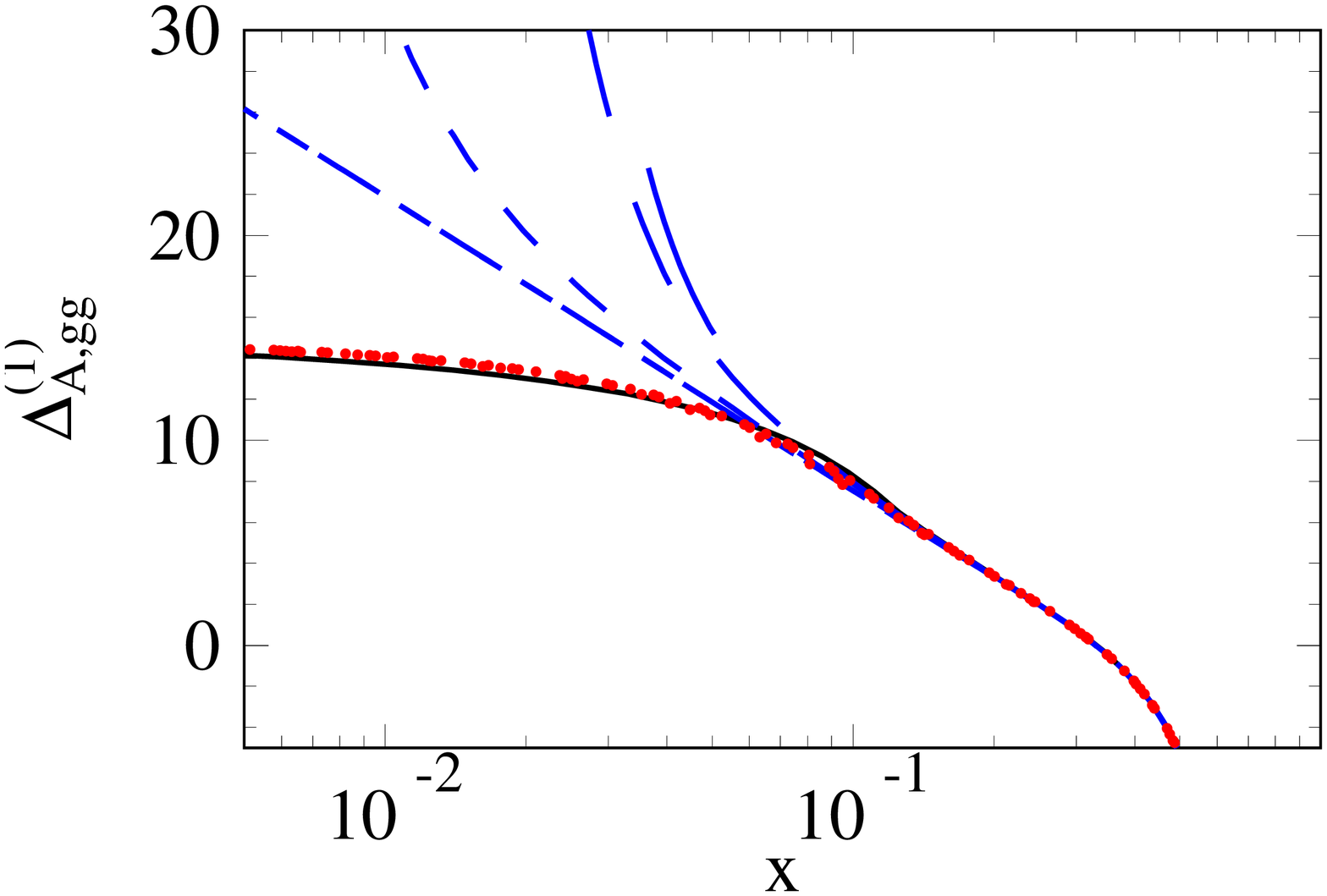}
    &
    \includegraphics[width=0.45\linewidth]{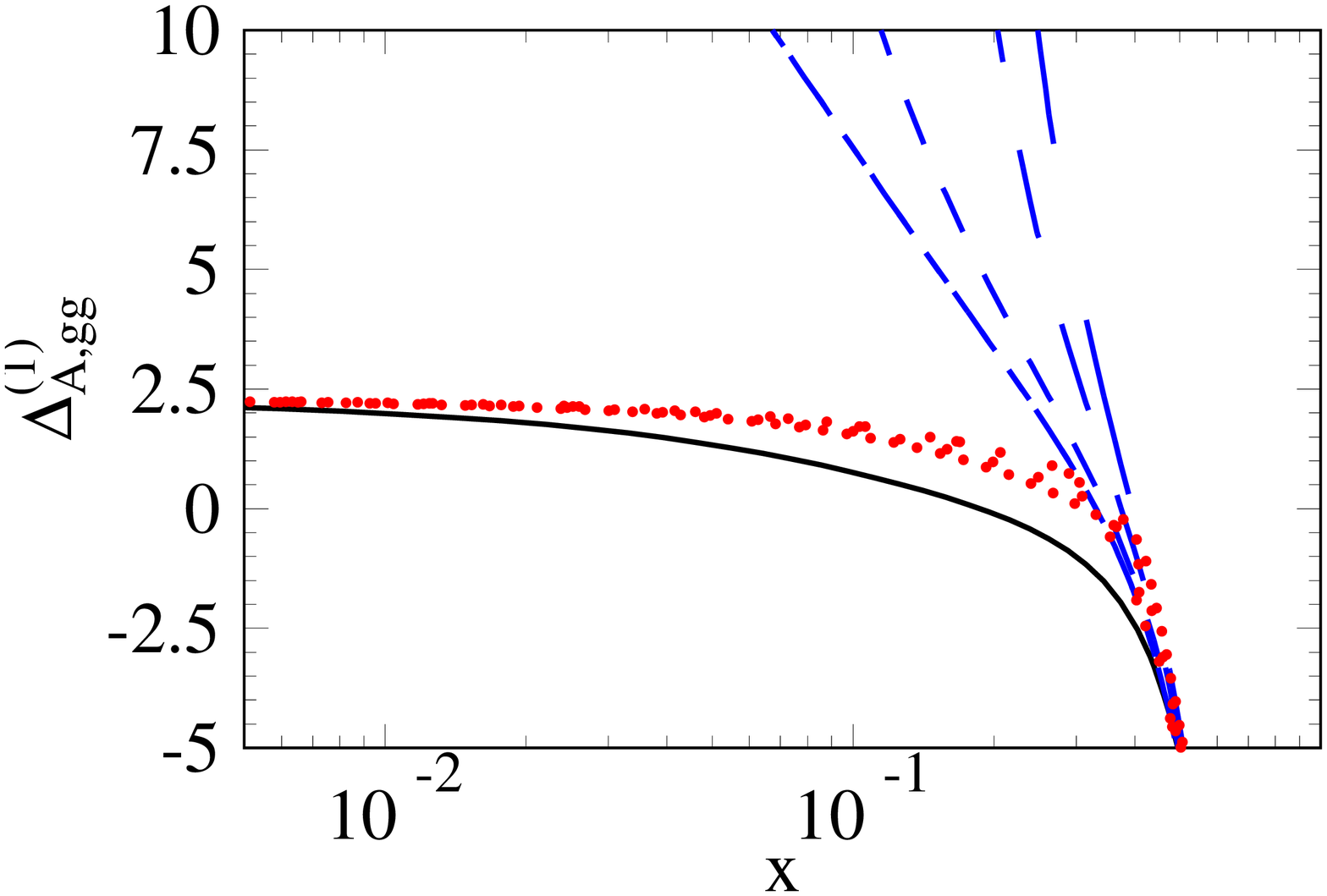}
    \\
    \includegraphics[width=0.45\linewidth]{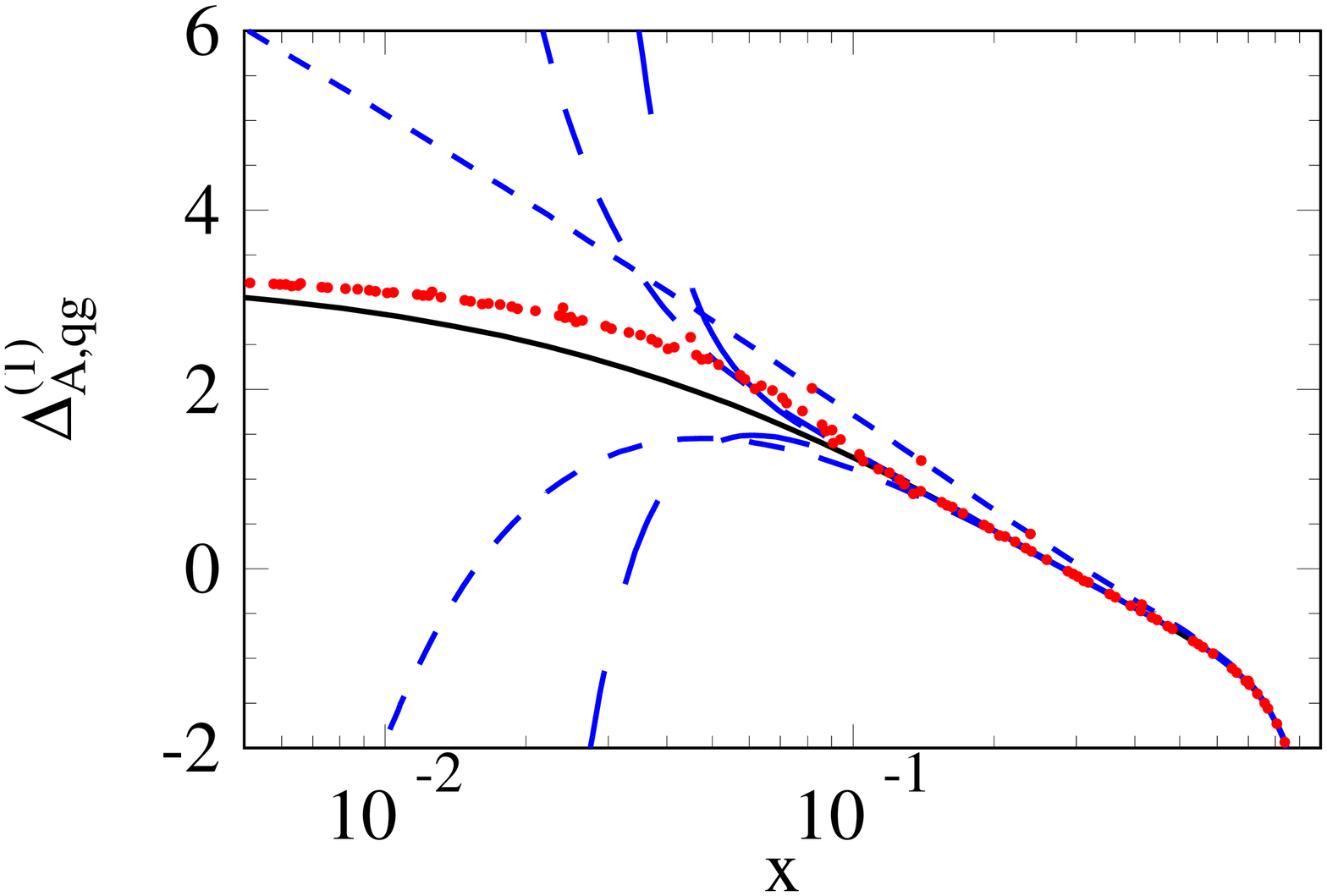}
    &
    \includegraphics[width=0.45\linewidth]{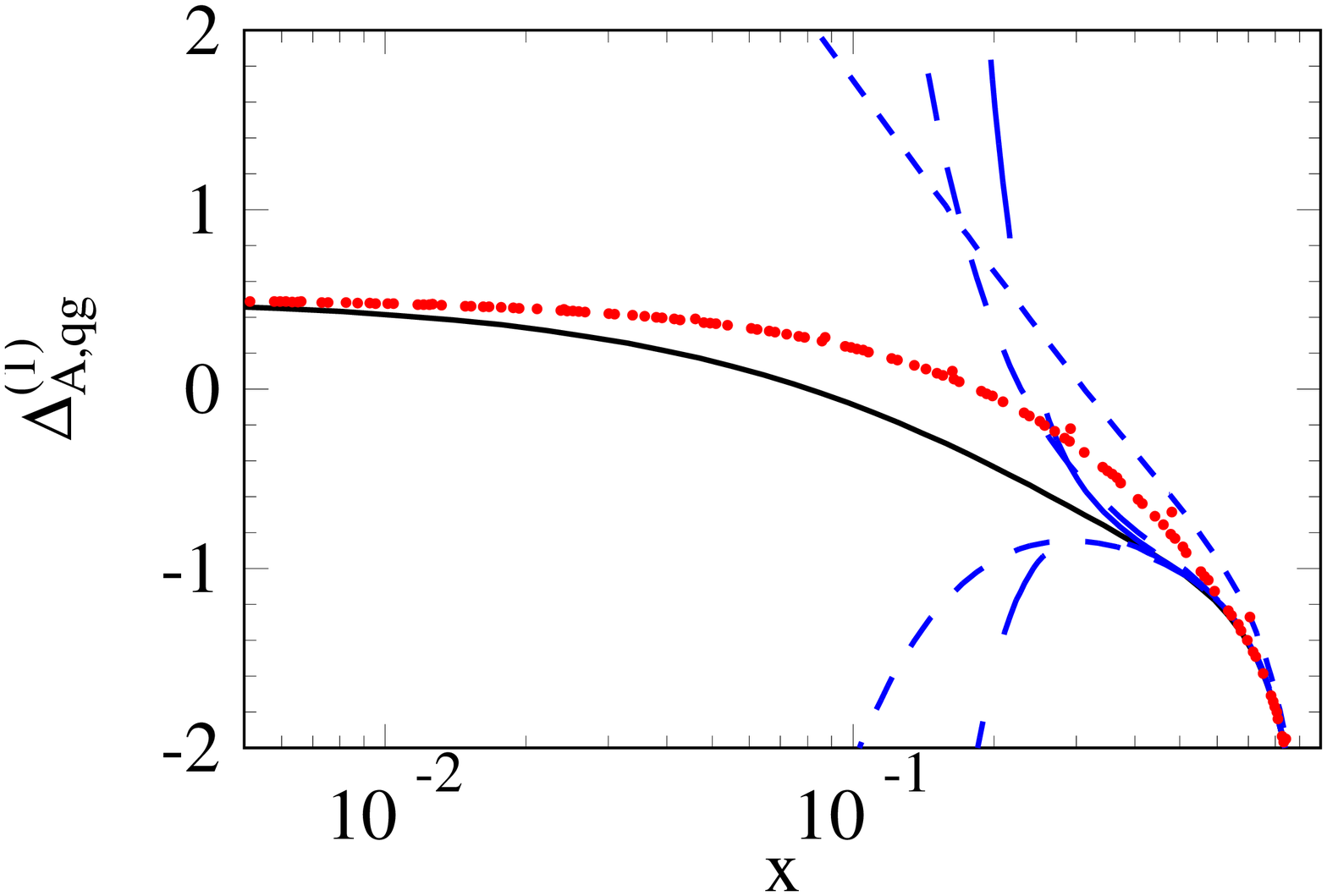}
    \\
    \includegraphics[width=0.45\linewidth]{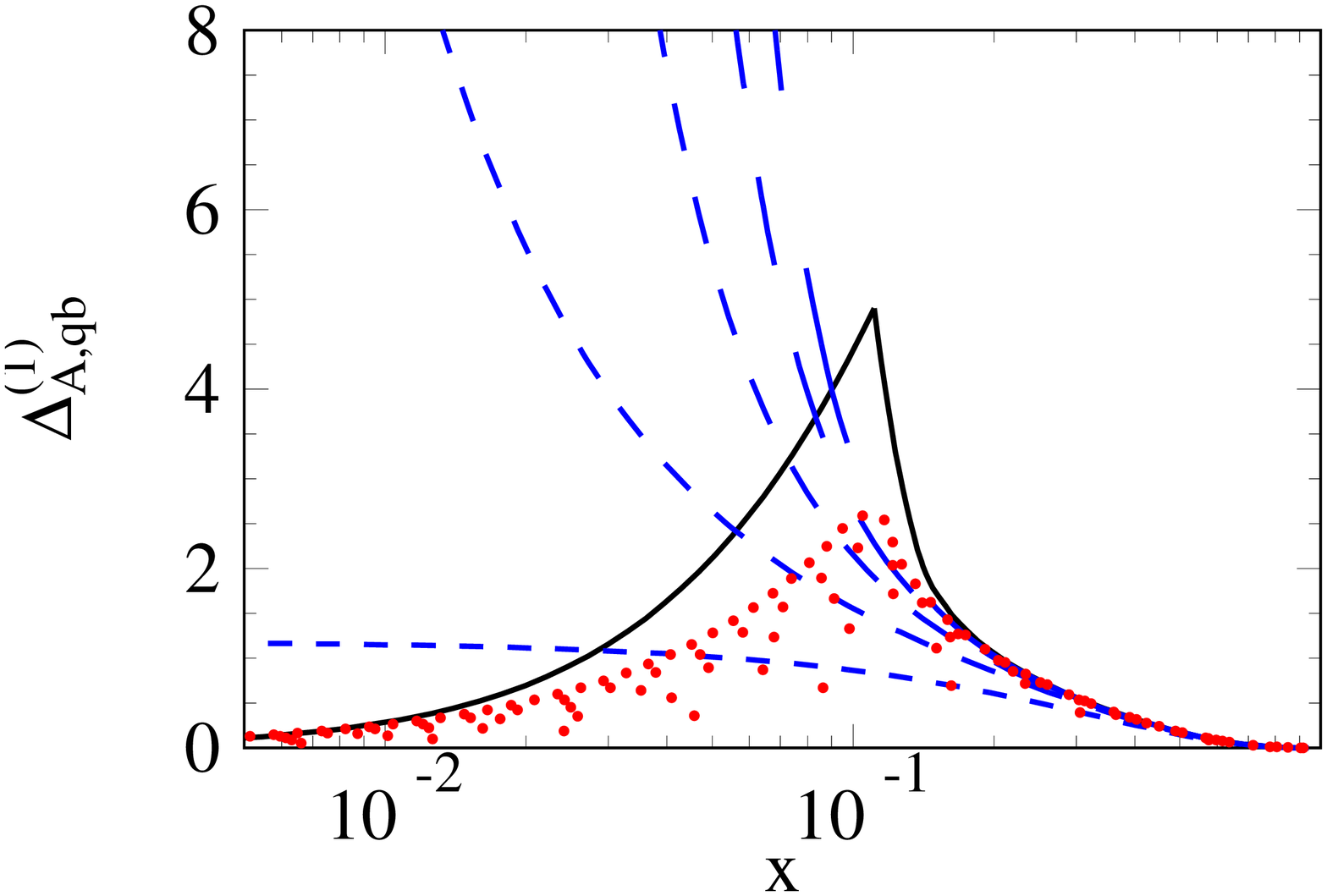}
    &
    \includegraphics[width=0.45\linewidth]{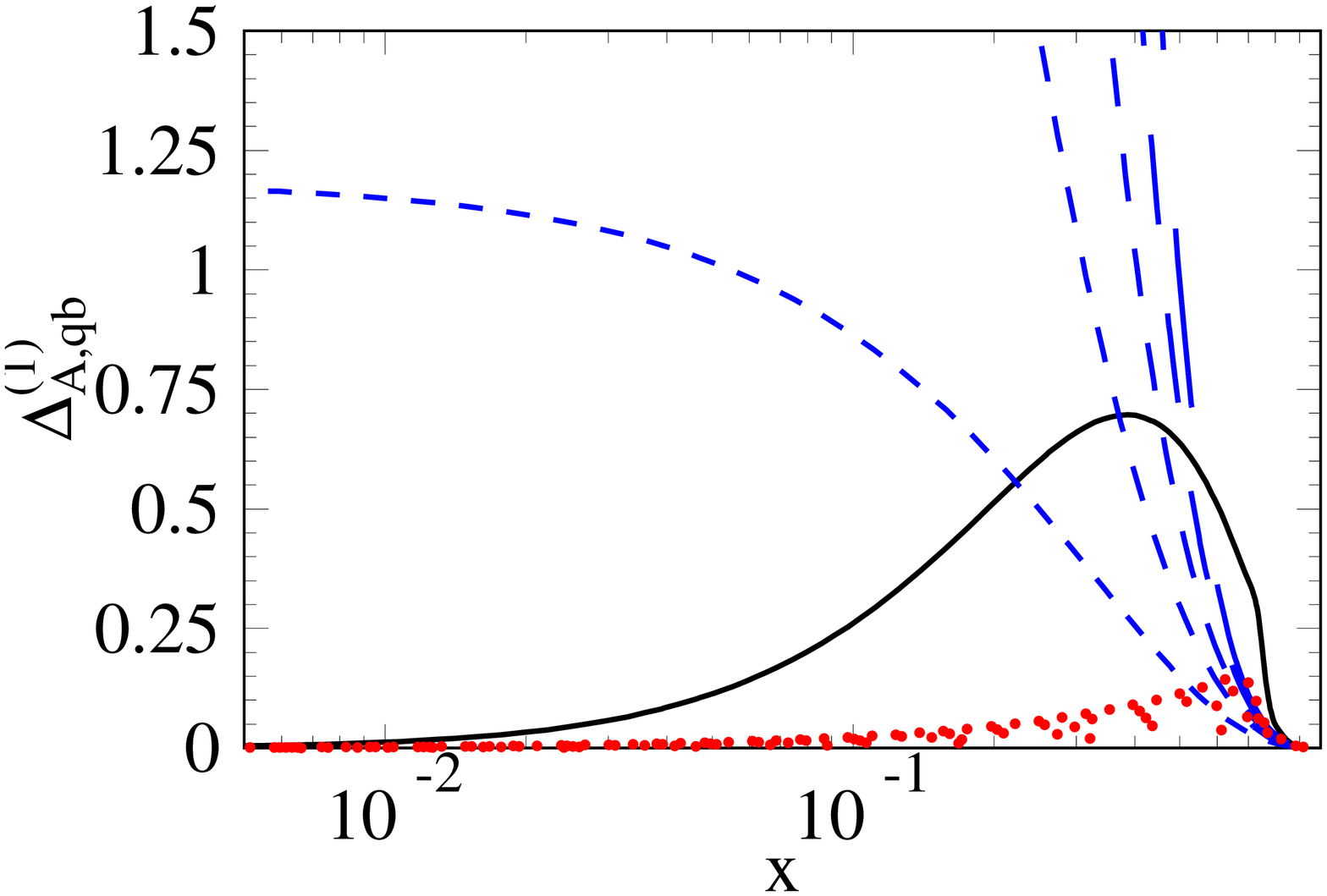}
  \end{tabular}
  \caption[]{\label{fig::NLOpart}Partonic NLO cross sections for the
    $gg$ (top), $qg$ (middle) and $q\bar{q}$ (bottom) 
    channel as functions of $x$ for
    $M_A=120$~GeV (left) and $M_A=300$~GeV (right column). 
    The expansion
    for $\rho\to0$ (dashed lines) is compared with the exact result (solid
    lines). Lines with longer dashes include higher order terms in $\rho$.
    The interpolation results (see text) are shown as dotted lines which
    (for $gg$ and $qg$) demonstrate an
    irregular structure since several lines are plotted on top of each
    other.
    The threshold values $x_{\rm thr}$ are 0.12 and 0.75, respectively.
    The subscript ``qb'' corresponds to the $q\bar{q}$ channel.
  }
\end{figure}

In Fig.~\ref{fig::NLOpart} we show the partonic cross sections for
the pseudo-scalar Higgs boson production for
$M_A=120$~GeV (left column) and $M_A=300$~GeV (right column).
The approximations obtained in the limit of large top quark mass
(dashed lines) show
good convergence below $x_{\rm thr}$ and diverge, as expected,
for small values of $x$.
For the $gg$ and $qg$ channel the matching procedure leads to good
approximations to the exact results for both values of $M_A$. 
The minor
differences in the $gg$ cannel only lead to  
small deviations at the hadronic level. On the other hand 
the deviation of the matched result
to the exact one in the $qg$ channel leads to a shift of about 15\%
in the 
hadronic contributions as discussed below. At the LHC with $\sqrt{s}=14$~TeV
the contribution of the
$qg$ channel to the NNLO part of the total cross section varies between about
$(-8)\%$ and $(-17)\%$ for Higgs masses between 110~GeV and 300~GeV.\footnote{These
  numbers are basically identical for a scalar and pseudo-scalar Higgs boson.}
Thus, an accuracy of 15\% at NNLO induces an uncertainty of at most
3\% in the NNLO piece and is hence negligible in the sum of the LO, NLO and NNLO
contribution. 

The situation is different for the $q\bar{q}$ channel.
The partonic cross section has a characteristic peak at $x\approx
x_{\rm thr}$ which is reproduced neither by the $x\to 0$ nor by
the $\rho\to 0$ approximation. 
Due to the small contribution
of this production channel (below 1\% at NLO) 
the accuracy of the infinite-top quark mass result is sufficient.

It is worth mentioning that the quantities $\Delta_{A,ij}^{(1)}$ shown in
Fig.~\ref{fig::NLOpart} for the production of a pseudo-scalar Higgs boson
are very similar to the corresponding scalar ones. This is both true for the
shape and the numerical size of the corrections.
For the $gg$ channel it would not be possible to distinguish the curves 
for $\Delta_{gg}^{A,(1)}(x)$ and $\Delta_{gg}^{H,(1)}(x)$ for $x>0.001$; a
visible difference only occurs for $x\to0$.

\begin{table}[t]
  \begin{center}
    \begin{tabular}{c|rr|rr}
      $\rho^n$ & $M_H=120$~GeV & $M_H=300$~GeV & $M_A=120$~GeV & $M_A=300$~GeV \\
      \hline
      $n=0$ &  15.3696  &   15.3696&  15.8696 &  15.8696\\
      $n=1$ & + 0.1208  &  + 0.7547& + 0.2464 & + 1.5400\\
      $n=2$ & + 0.0072  &  + 0.2814& + 0.0183 & + 0.7130\\
      $n=3$ & + 0.0005  &  + 0.1296& + 0.0016 & + 0.3896\\
      $n=4$ & + 0.0000  &  + 0.0669& + 0.0002 & + 0.2312\\
      \hline
      exact &  15.4981  &   16.6956&  16.1360 &  19.1723
    \end{tabular}
    \caption{\label{tab::delta}Contribution to the coefficient of $\delta(1-x)$
      in the normalization of Eq.~(\ref{eq::hatsigma}) from the various
      expansion terms $\rho^n = (M_\Phi^2/M_t^2)^n$. The renormalization scale
      has been set to $M_\Phi$.
      }
  \end{center}
\end{table}

It is interesting to separately look at the $\delta$-function
contribution which is not shown in Fig.~\ref{fig::NLOpart}. In
Tab.~\ref{tab::delta} we present the coefficient of $\delta(1-x)$ in
the normalization of Eq.~(\ref{eq::hatsigma}) both for the scalar and
pseudo-scalar case.
One observes that the bulk of the contribution
is given by the leading term whereas the power-suppressed terms add an
additional part of only 8\% and 17\% even for $M_\Phi=300$~GeV in
the scalar and 
the pseudo-scalar case, respectively.
We anticipate that this behaviour is different at
NNLO where the higher order terms provide more sizeable contributions
as discussed in Section~\ref{sec::part}.

It is interesting to note that for $M_\Phi=300$~GeV 
the sum of the first five expansion terms in
Tab.~\ref{tab::delta} reproduces the exact result to better than 1\%
in the scalar and 2\% in the pseudo-scalar case. For the latter
expansion, the error can be accounted for by doubling the contribution
of $n=4$ term.

\subsection{\label{sub::nlo_hadr}Hadronic cross section}

In this subsection we compare the hadronic cross section obtained from
our approximate expansions with the one obtained using the 
exact partonic results, implemented
in {\tt HIGLU}~\cite{Spira:1995mt} and integrated with the same PDFs.

\begin{figure}[t]
  \centering
  \begin{tabular}{cc}
    \includegraphics[width=0.48\linewidth]{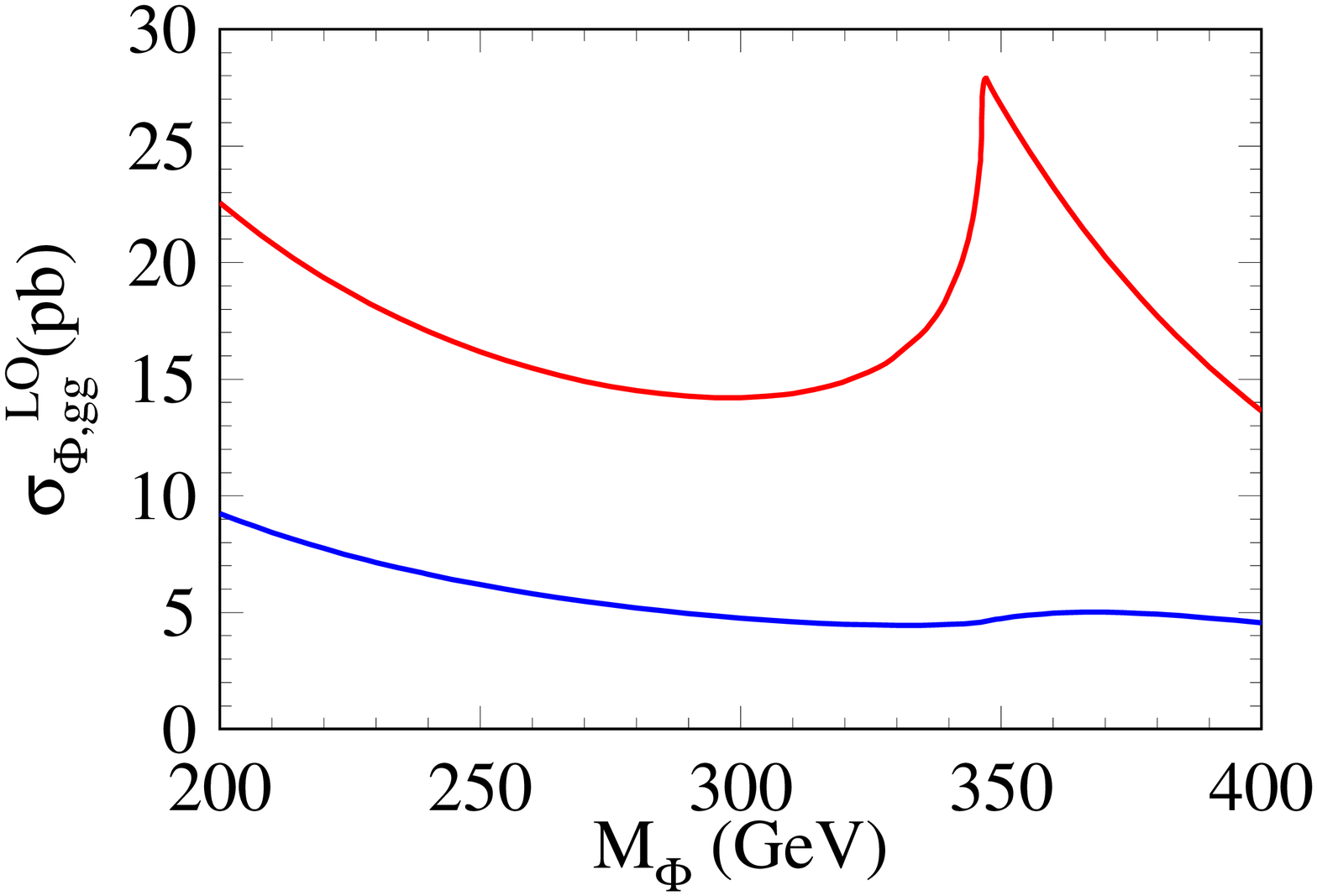}
    &
    \includegraphics[width=0.48\linewidth]{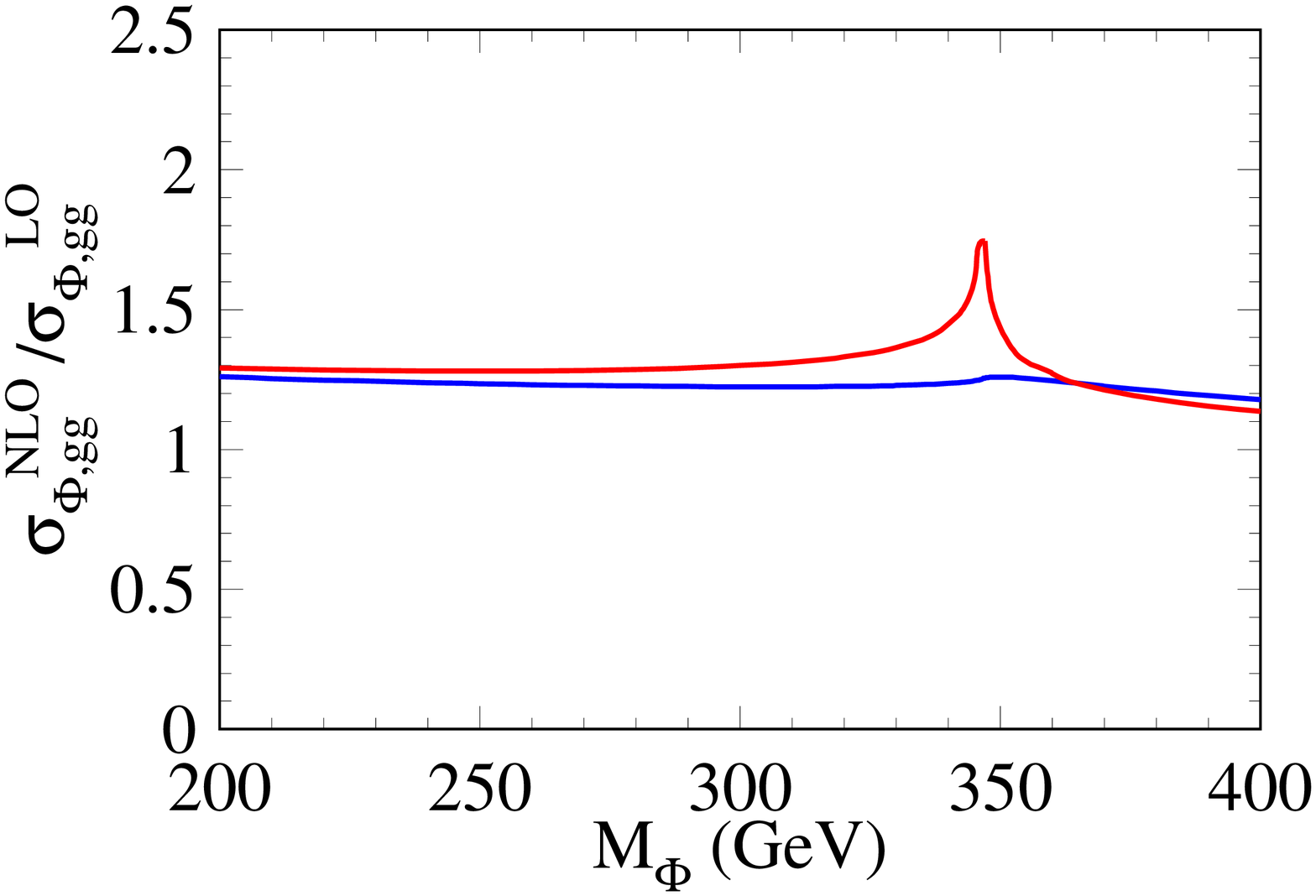}
    \\ (a) & (b)
  \end{tabular}
  \caption{\label{fig::lo}(a) LO cross section for the production of a
    pseudo-scalar (top) and scalar (bottom) Higgs boson. (b) ratio of NLO
    and LO prediction for the total cross section. For these plots the exact
    results from~\cite{Spira:1995mt} have been used.}
\end{figure}

In Fig.~\ref{fig::lo}(a) we plot the exact LO cross section as a
function of $M_\Phi$ for $200~\mbox{GeV}\le M_\Phi\le 400~\mbox{GeV}$ 
in order to demonstrate the
threshold behaviour. Whereas there is a smooth transition in the scalar
case, one observes a strong enhancement of the cross section for the
pseudo-scalar Higgs boson.  In Fig.~\ref{fig::lo}(b) the purely NLO part of
the cross section is shown where the LO contribution is divided out. This is
actually the quantity which is approximated by our procedure. No strong
enhancement is visible for $M_A\lsim 300$~GeV which is the upper limit
considered in this paper. Thus we expect that in the
pseudo-scalar case the matching procedure works as well as for the scalar
one.

When discussing the quality of the matching procedure it is convenient to
consider ratios of cross sections. We normalize either the exact or the
matched expressions to the (un-matched) infinite-top quark mass result since
these kind of ratios are also possible at NNLO.  In the following plots the
dashed lines represent the matched results including successively higher order
in $1/M_t$ when going from short to long dahes. The solid line corresponds to
the exact result.

\begin{figure}[t]
  \centering
  \begin{tabular}{cc}
    \includegraphics[width=0.45\linewidth]{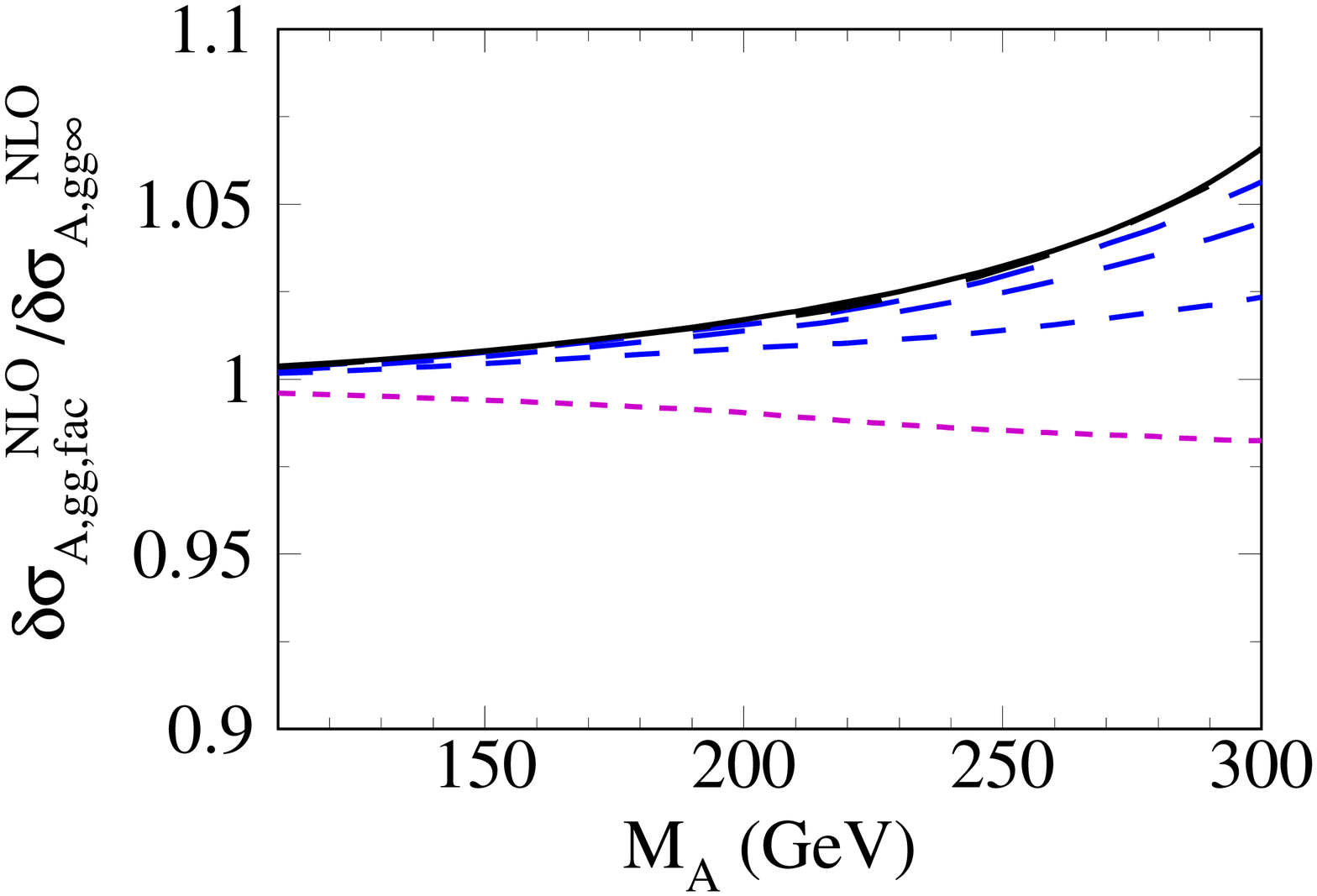}
    &
    \includegraphics[width=0.45\linewidth]{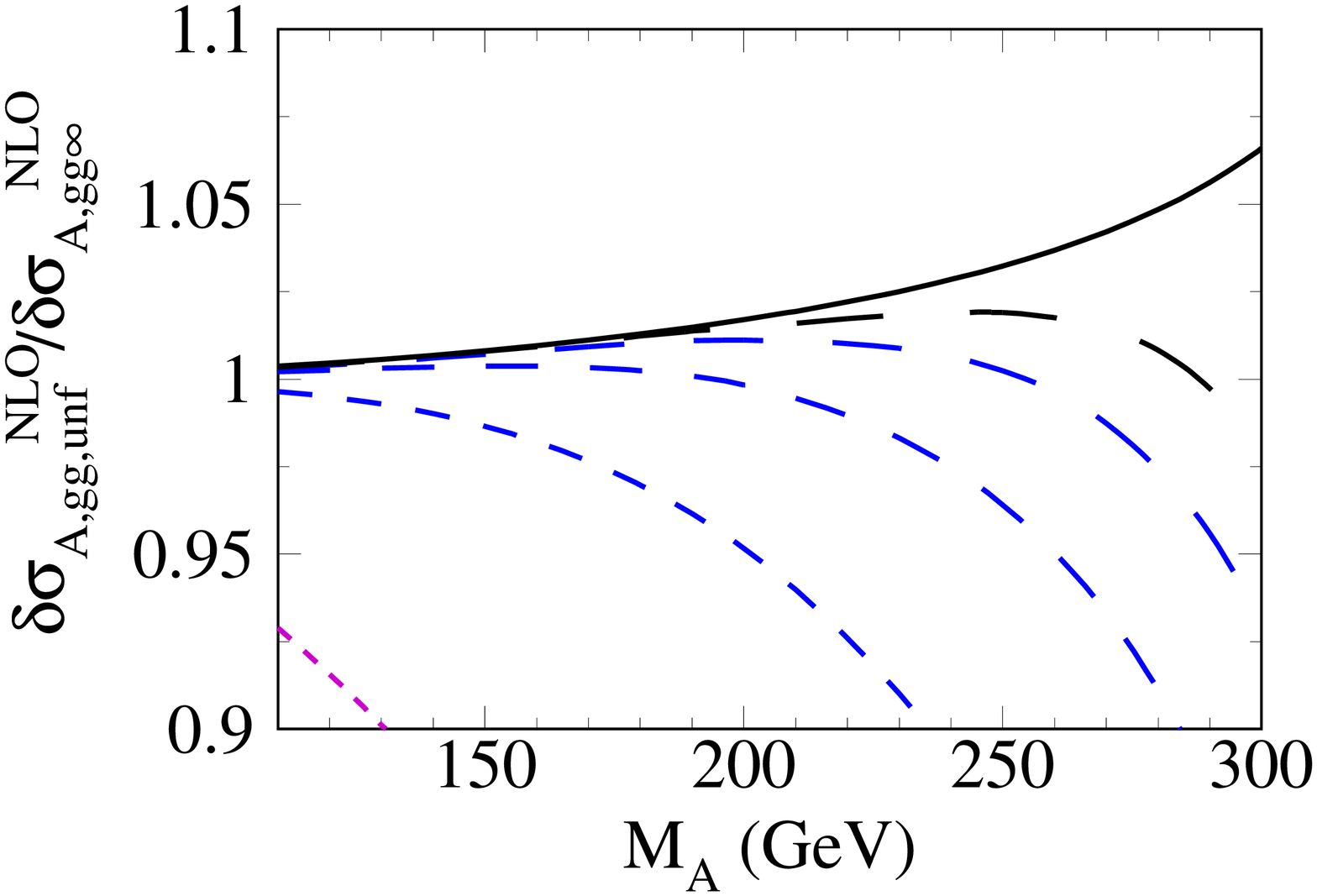}
    \\ (a) & (b) \\
    \includegraphics[width=0.45\linewidth]{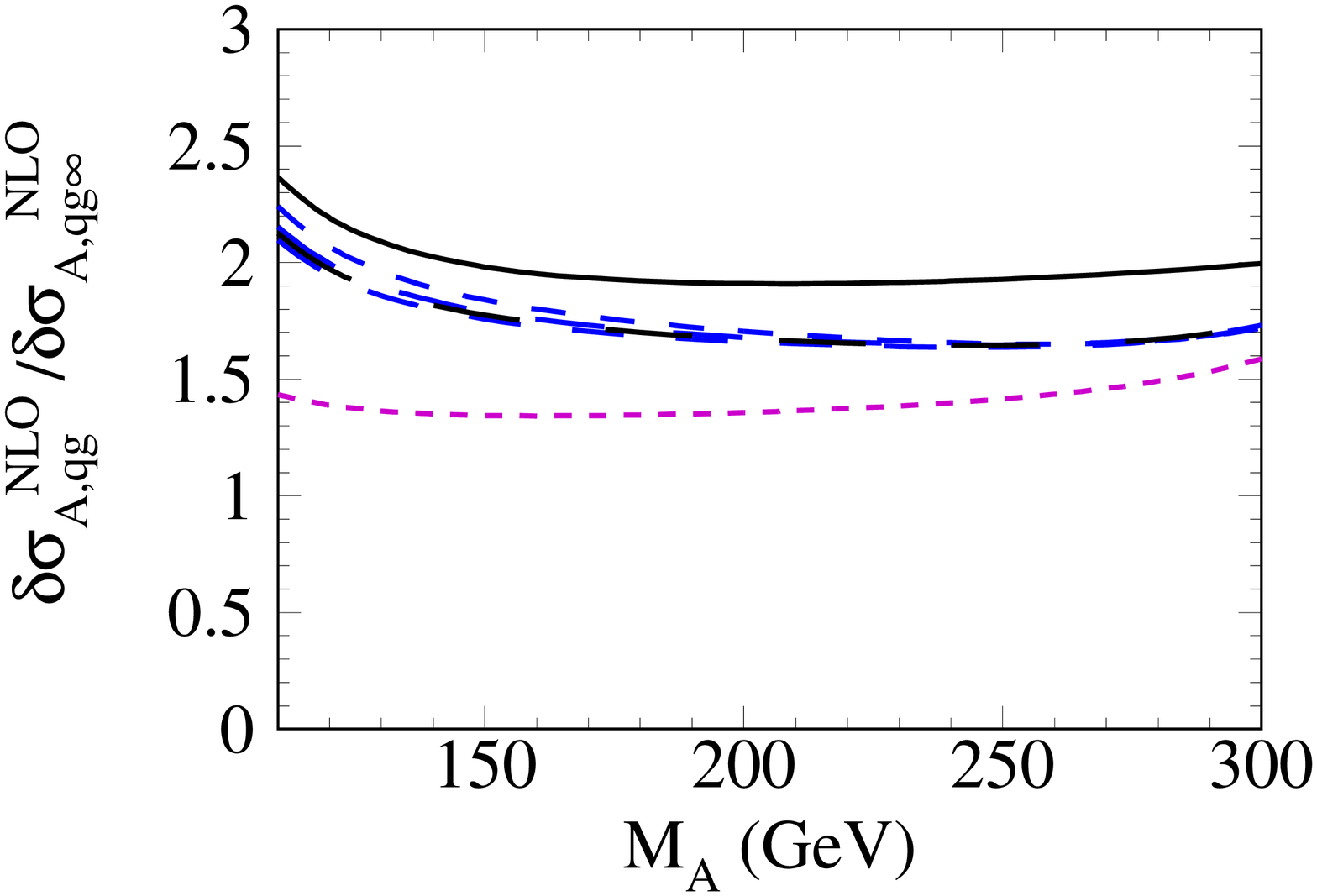}
    &
    \includegraphics[width=0.45\linewidth]{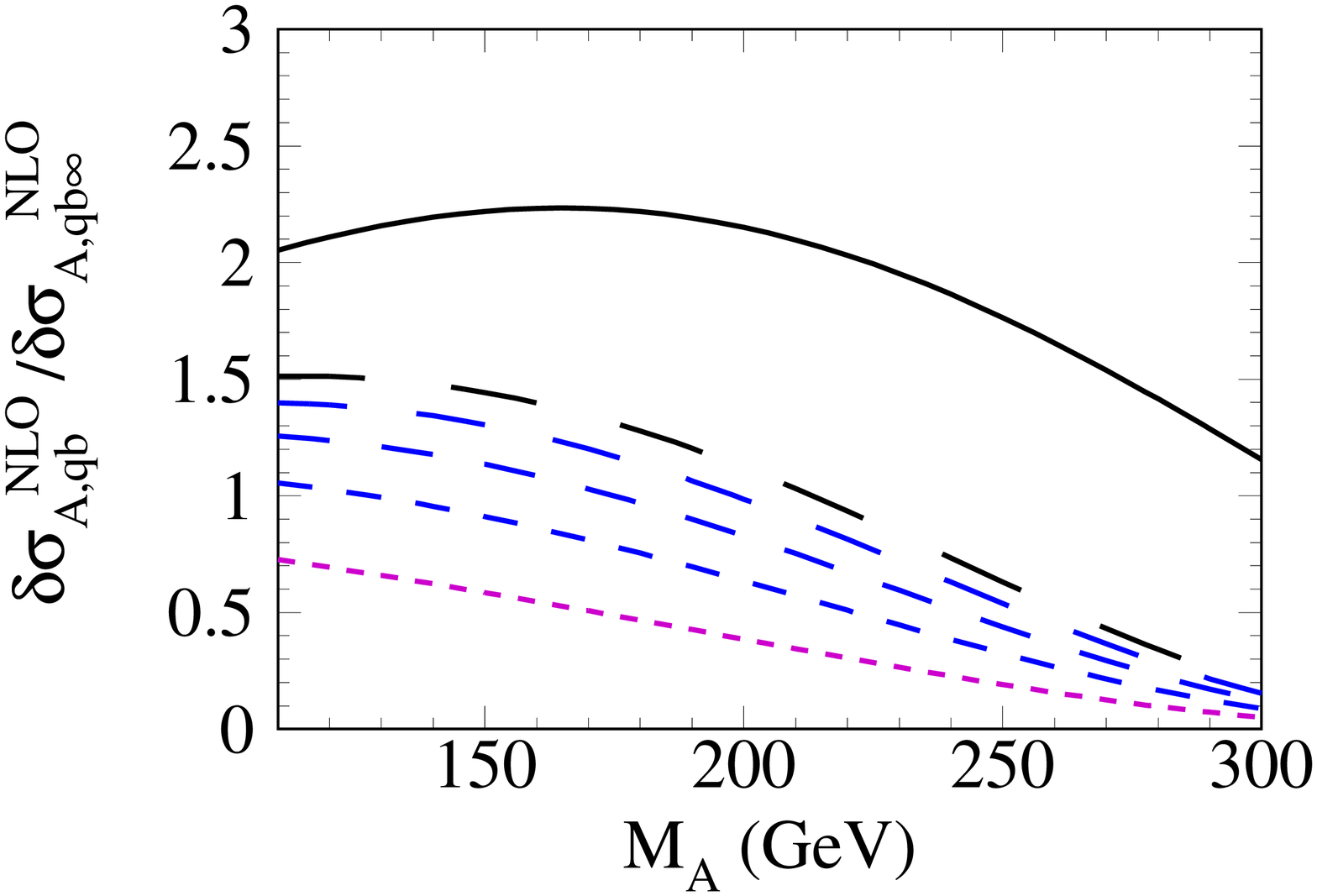}
    \\ (c) & (d)
  \end{tabular}
  \caption[]{\label{fig::nlo_hadr_1} Ratio of the NLO parts of the 
    hadronic cross sections for the $gg$ (Fig.~(a) and~(b)),
    $qg$ (Fig.~(c)) and $q\bar{q}$ (Fig.~(d)) channel.
    In (a) the exact LO result is factored out also in the matched
    result in the numerator whereas in (b) this option is abandoned.}
\end{figure}

In Fig.~\ref{fig::nlo_hadr_1}(a) the NLO $gg$ part is shown where the
factorized expressions are used in the numerator. For $M_A=120$~GeV the
deviation between the various curves is below per cent level. For
$M_A=300$~GeV, however, one observes a deviation between the exact
result and the infinite-top mass approximation of about 6\%. The exact
result and the matched result based on the leading order deviate
by about 8\% whereas there is perfect agreement with the 
matched result including $1/M_t^8$ terms.
(The corresponding dashed curve is below the solid line.)

An alternative comparison between the matched results on the one hand
and the infinite-top mass and exact result on the other hand
is shown in Fig.~\ref{fig::nlo_hadr_1}(b) for the $gg$ channel
where in the numerator of the considered ratio the
fully expanded result is used and successively more terms in $1/M_t$
are included. A large deviation 
of the matched result including only the leading $M_t$ dependence
(short dashes) from the exact result is observed. 
However, after including higher order mass corrections in
the matching procedure good agreement up to $M_A\approx 250$~GeV is
observed. For Higgs boson masses close to 300~GeV a significant deviation of
the matched and exact result is visible which can be traced back to the
expansion of the LO result. Thus at NNLO we adopt the approximation of
Fig.~\ref{fig::nlo_hadr_1}(a) and factor the exact LO result out
after including the $1/M_t$ mass corrections.

Figs.~\ref{fig::nlo_hadr_1}(c) shows the result for the $qg$ channel.
The matched result agrees with the exact one with an accuracy of about
15\% whereas the infinite-top quark mass result is about a factor two
smaller than the prediction based on {\tt HIGLU}. 
For $\sqrt{s}=14$~TeV the $qg$ part amounts between $(-2)\%$ ($M_A=110$~GeV) and
$(-7)\%$ ($M_A=300$~GeV) of the total NLO part. Thus, if at NNLO a similar
behaviour is observed it is important to incorporate the matched
$qg$ channels in precision predictions.

As expected from the above discussion of the partonic cross section it
is not possible to obtain a good approximation to the exact result for
the $q\bar{q}$ channel which is shown in
Fig.~\ref{fig::nlo_hadr_1}(d). For lower pseudo-scalar
Higgs boson masses there is
also a significant deviation between the exact and the infinite-top
mass result, for $M_A=300$~GeV, however, they (accidentally, as can be seen in
the lower right panel of Fig.~\ref{fig::NLOpart}) agree
perfectly well.


\section{\label{sec::part}Partonic NNLO cross section}

In this Section we discuss the partonic cross sections for the various
channels at NNLO. We follow the matching procedure outlined in
Section~\ref{sub::nlo_part} for the NLO calculation, with minor
modifications due to the different $x\to 0$ limit. At the NNLO\footnote{We
  again omit the superscripts ``$\Phi$'' and ``(2)''.} 
\begin{eqnarray}
  \Delta_{ij}(x) \stackrel{x\to 0}{=} E_{ij}\ln x + \mathcal{O}(1)\,,
\end{eqnarray}
and Ref.~\cite{Caola:2011wq} provides coefficients $E_{ij}$.
We again choose the matching value $x_m \sim x_{\rm thr}$ such that the
transition between the $x\to 0$ and $M_t^2\gg \hat{s}, M_\Phi^2$ approximations
is smooth: $\Delta_{ij}^{\rm exp}(x_m) = E_{ij}\ln x_m + F_{ij}$,
$\frac{d}{dx} \Delta_{ij}^{\rm exp}(x_m) = E_{ij}/x_m$, and
$\Delta_{ij}^{\rm exp}(x_m)$
is given in Eq.~(\ref{eq::sig^e}).

This prescription applies to the $gg$ channel. For the other
initial states we again use $x_m=0.9\,x_{\rm thr}$ and require that
$\Delta_{ij}^{\rm exp}(x_m)$ agree with the corresponding infinite-energy
result.

\begin{figure}[t]
  \centering
  \begin{tabular}{cc}
    \includegraphics[width=0.45\linewidth]{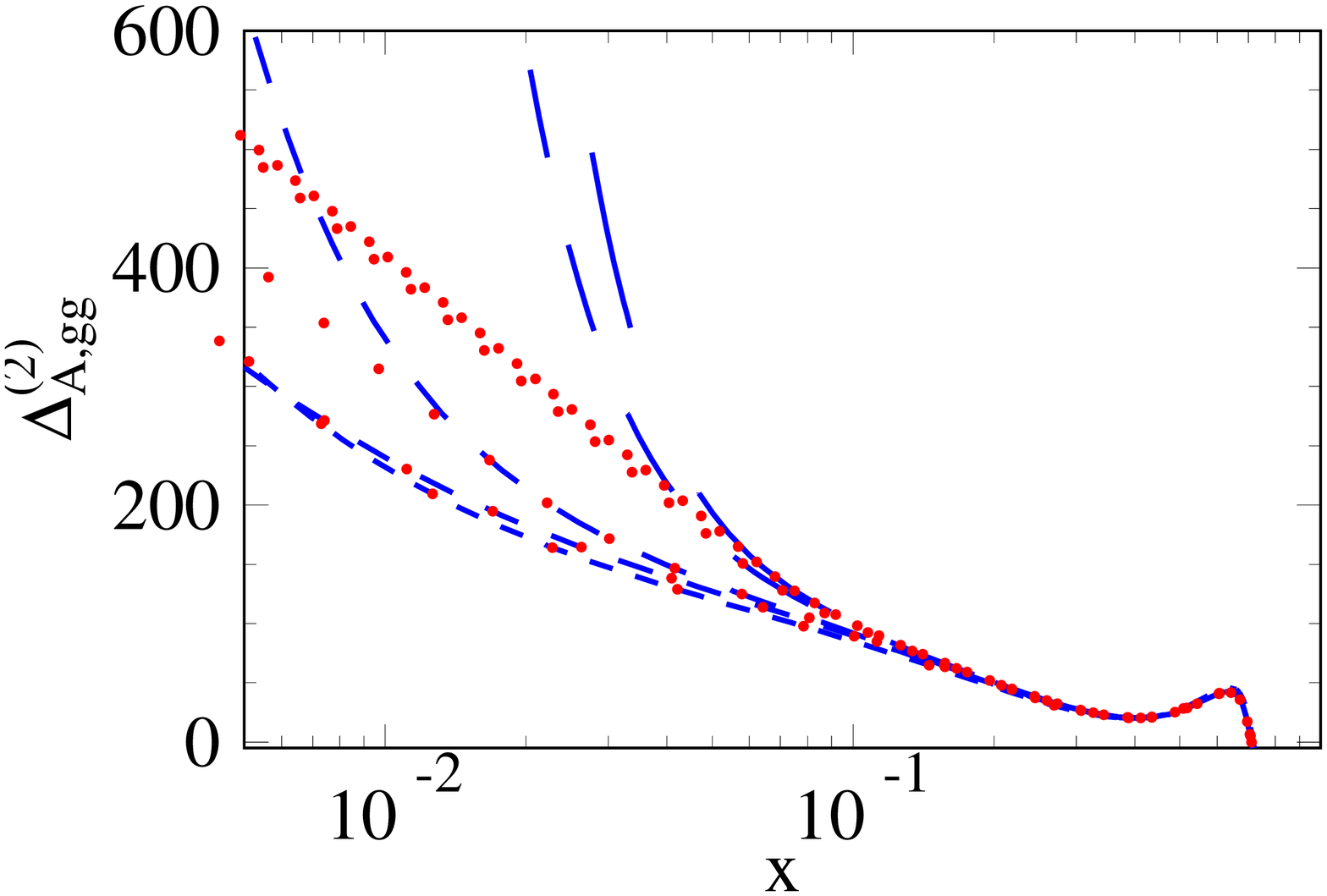}
    &
    \includegraphics[width=0.45\linewidth]{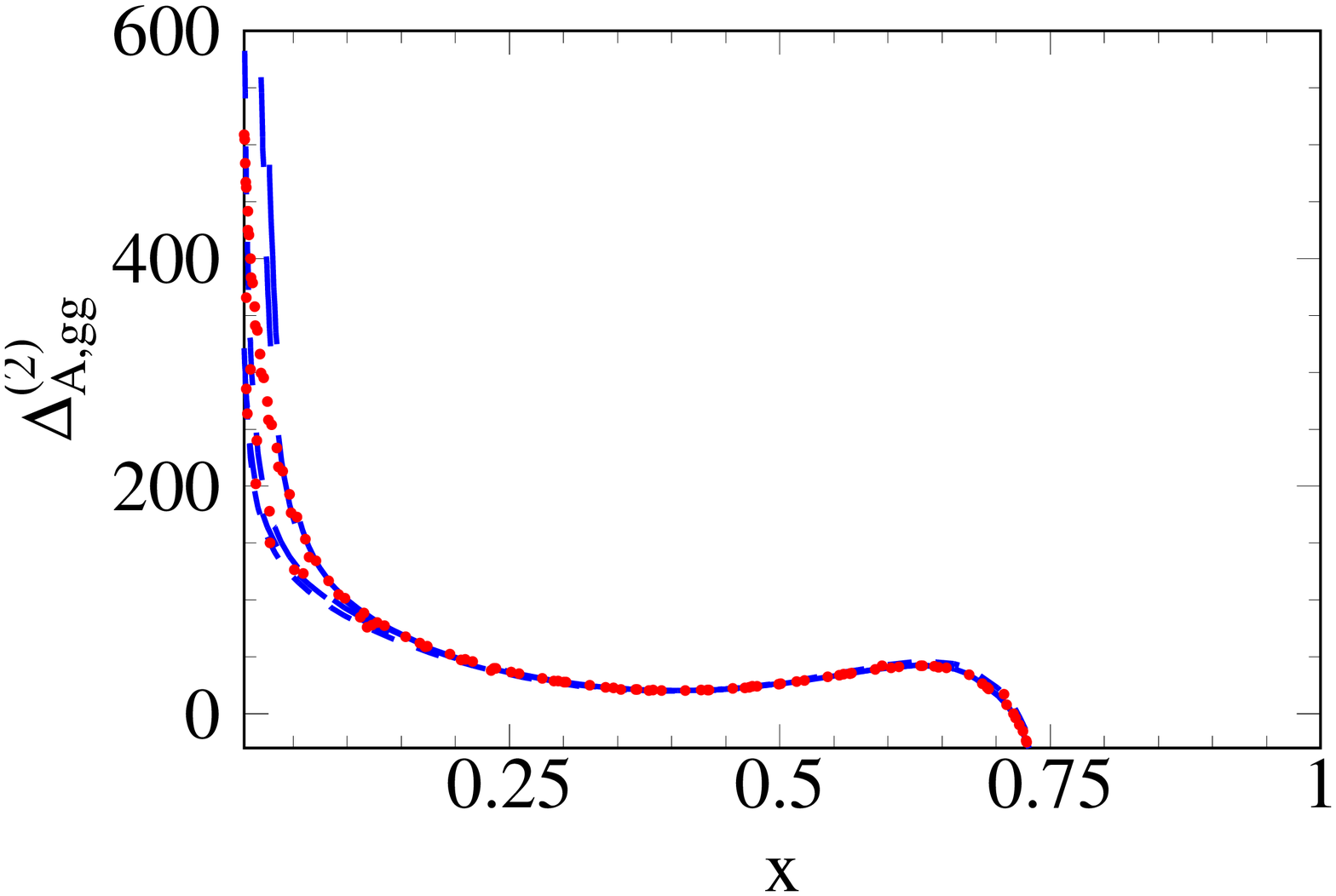}
    \\
    \includegraphics[width=0.45\linewidth]{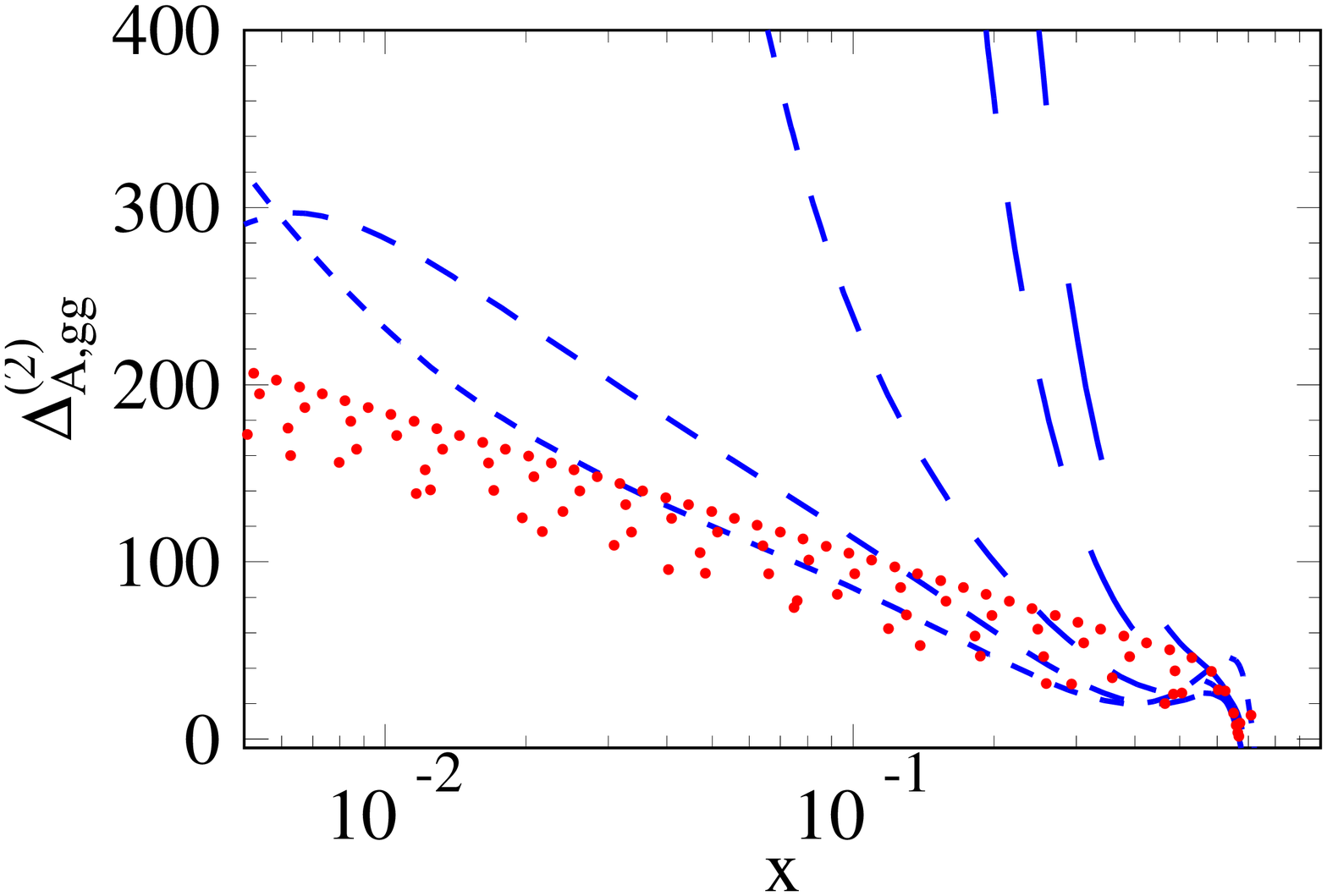}
    &
    \includegraphics[width=0.45\linewidth]{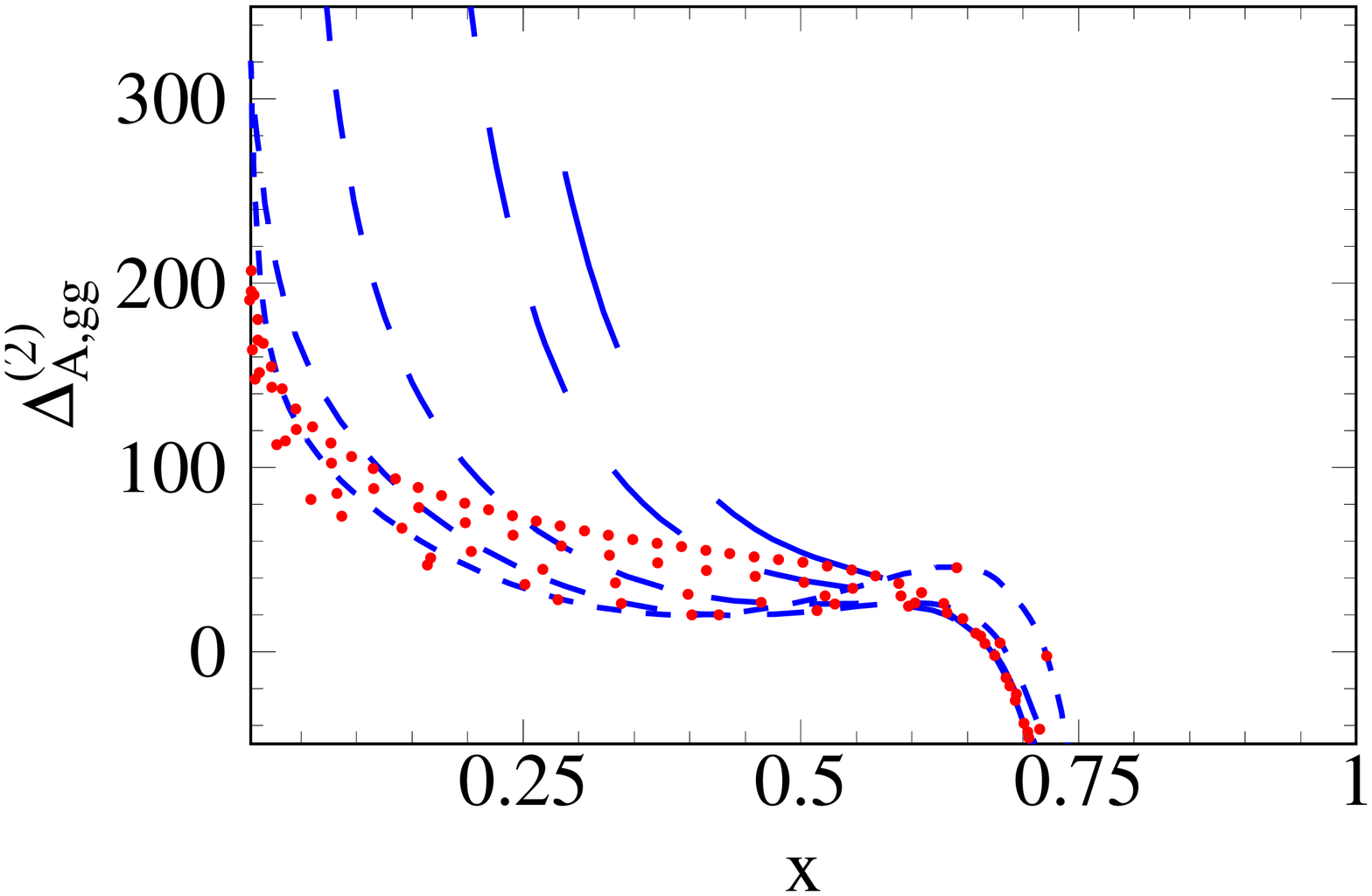}
  \end{tabular}
  \caption[]{\label{fig::nnlo_gg}
    Partonic cross section for the $gg$ channel for $M_A=120$~GeV (top)
    and $M_A=300$~GeV (bottom). The plots on the left employ a logarithmic
    and on the right a linear $x$-scale.
    The notation is adopted from Fig.~\ref{fig::NLOpart}.}
\end{figure}

In Fig.~\ref{fig::nnlo_gg} we show the partonic cross section 
for the numerically most important $gg$ channel at
$M_A=120$~GeV and $M_A=300$~GeV, with linear and logarithmic scales
of the $x$ axis.
The dashed lines represent the results for $\sigma_{ij}^{\rm exp}$
(cf. Eq.~(\ref{eq::sig^e})); longer dashes indicate higher order
$\rho$-expansion.
As expected, these results diverge for $x<x_{\rm thr}$.
The results interpolated as described above are shown with
dotted lines, shorter distances between the dots correspond to
higher-order $\rho$-expansion in the matching procedure.

For $M_A=120$~GeV the partonic cross section has a
maximum at $x\approx 0.6$ which is significantly higher than 
$x_{\rm thr}\approx 0.12$ and thus it is 
nicely reproduced from $\sigma_{ij}^{\rm exp}$. The approximations start diverging
at $x\approx x_{\rm thr}$. As far as the matched results are concerned one
observes stabilization starting from the one including the $\rho^2$ terms.

For $M_A=300$~GeV we have $x_{\rm  thr}\approx 0.75$ which is close to the
steep rise of the partonic cross section. The expansion results show good
convergence properties down to $x\approx0.6$ and start to diverge
around $x=0.45$, significantly below $x_{\rm  thr}$.
As far as the matched results are concerned a similar behaviour as for 
$M_A=120$~GeV is observed: The inclusion of
higher order terms in $\rho$ stabilizes the matching procedure, leading to
firm NNLO results for the partonic cross section.

\begin{figure}[t]
  \centering
  \begin{tabular}{cc}
    \includegraphics[width=0.45\linewidth]{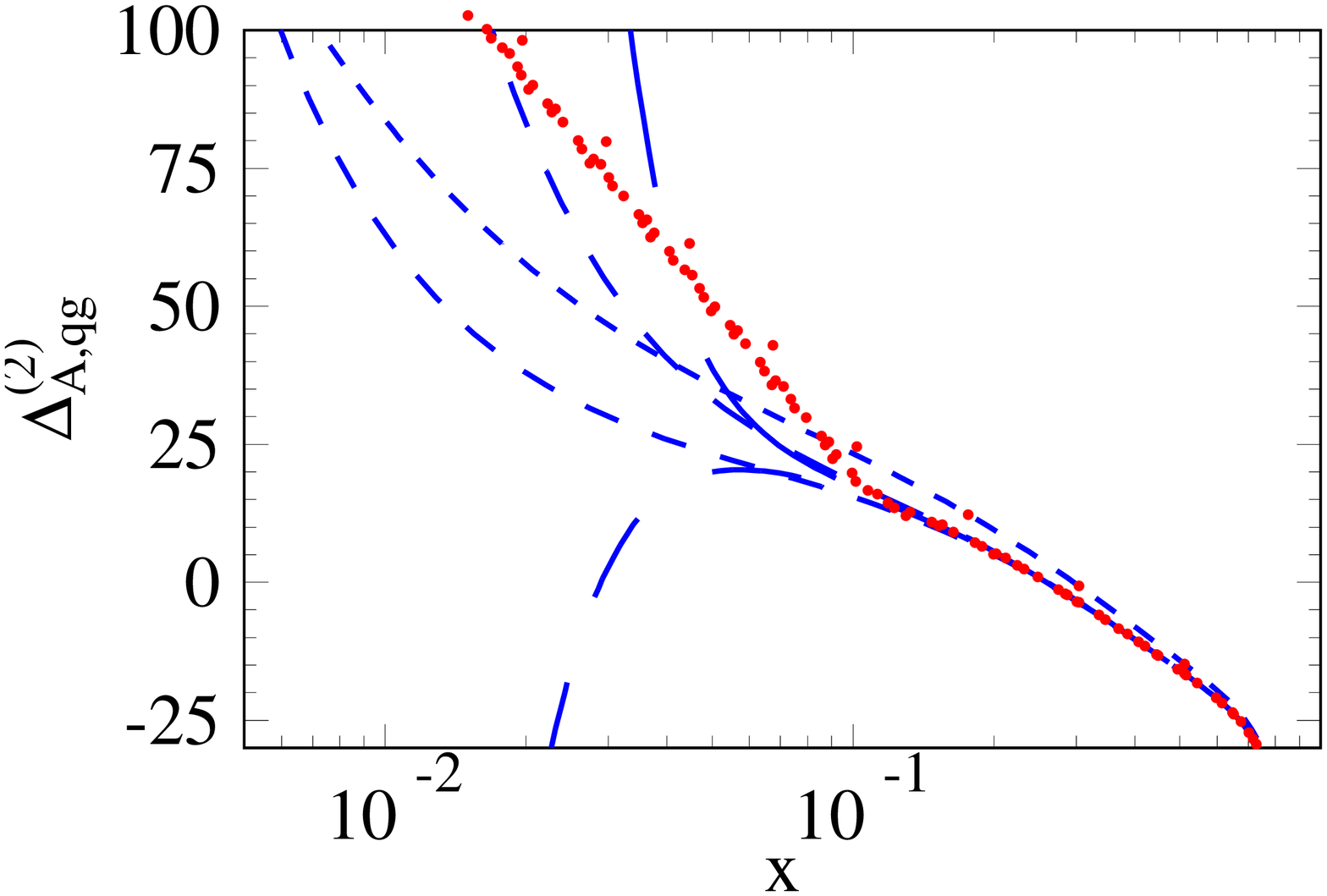}
    &
    \includegraphics[width=0.45\linewidth]{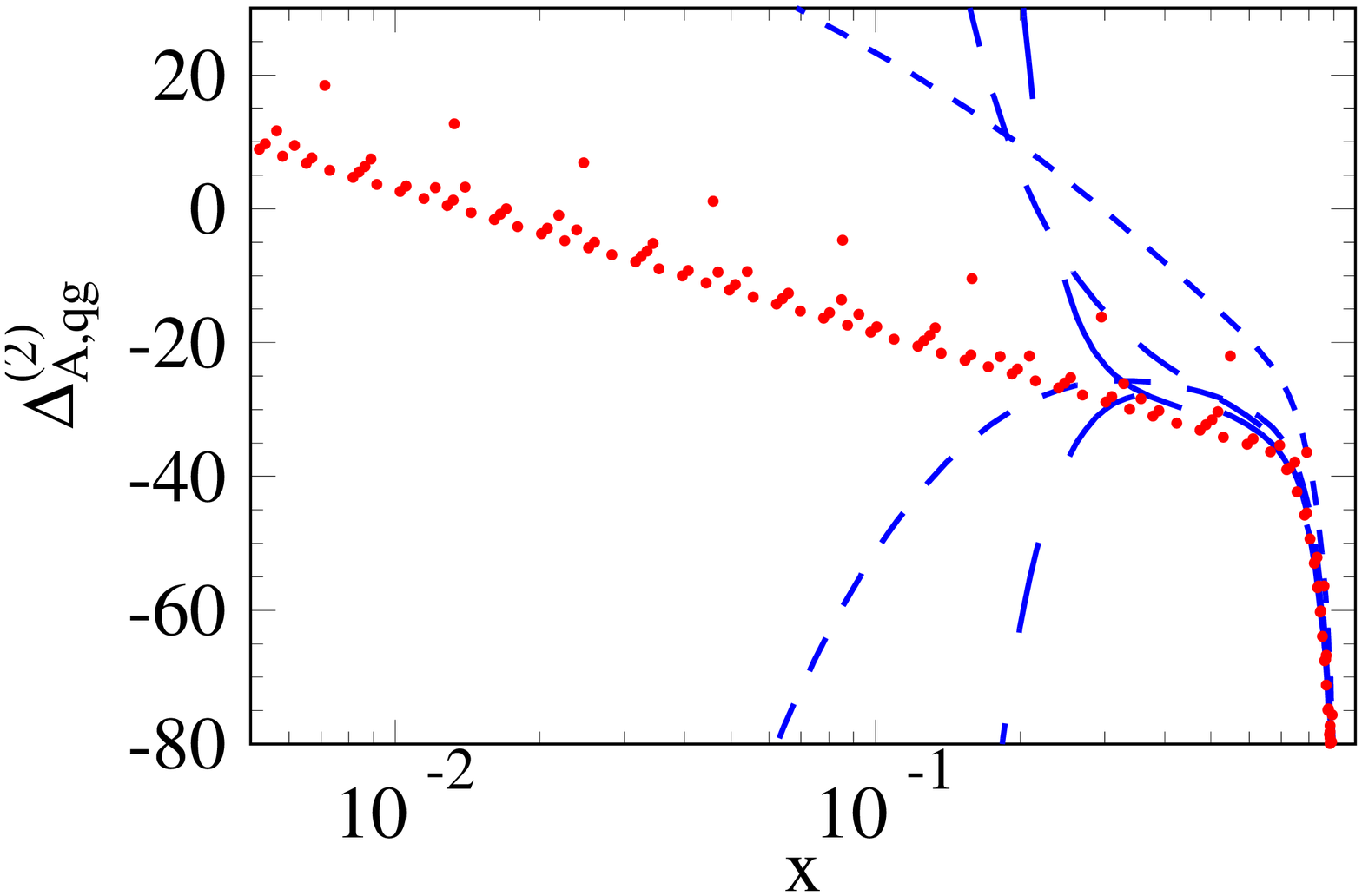}
  \end{tabular}
  \caption[]{\label{fig::nnlo_qg}
    Partonic cross section for the $qg$ channel for $M_A=120$~GeV (left)
    and $M_A=300$~GeV (right panel).
    The notation is adopted from Fig.~\ref{fig::NLOpart}.}
\end{figure}

Let us next discuss the $qg$-initiated partonic cross section shown
in Fig.~\ref{fig::nnlo_qg} for $M_A=120$~GeV and $M_A=300$~GeV.
Again, very nice convergence happens at $x>x_{\rm thr}$ and the
matching to $\hat{s}\to\infty$ results is rather stable. The approximation
is shown with dotted lines, the shorter distances between the dots
denote higher-order terms in $\rho$. One observes that the
latter are important for reliable results. Both for 
$M_A=120$~GeV and $M_A=300$~GeV
there is a visible difference between the curves including only $\rho^0$ terms
and the ${\cal O}(\rho)$ result. Further corrections
are relatively small.
Judging from the behaviour at NLO it can be expected that the results for the
partonic $qg$ channel approximate the (unknown) exact result quite well with
an uncertainty below 15\%.

\begin{figure}[t]
  \centering
  \begin{tabular}{cc}
    \includegraphics[width=0.45\linewidth]{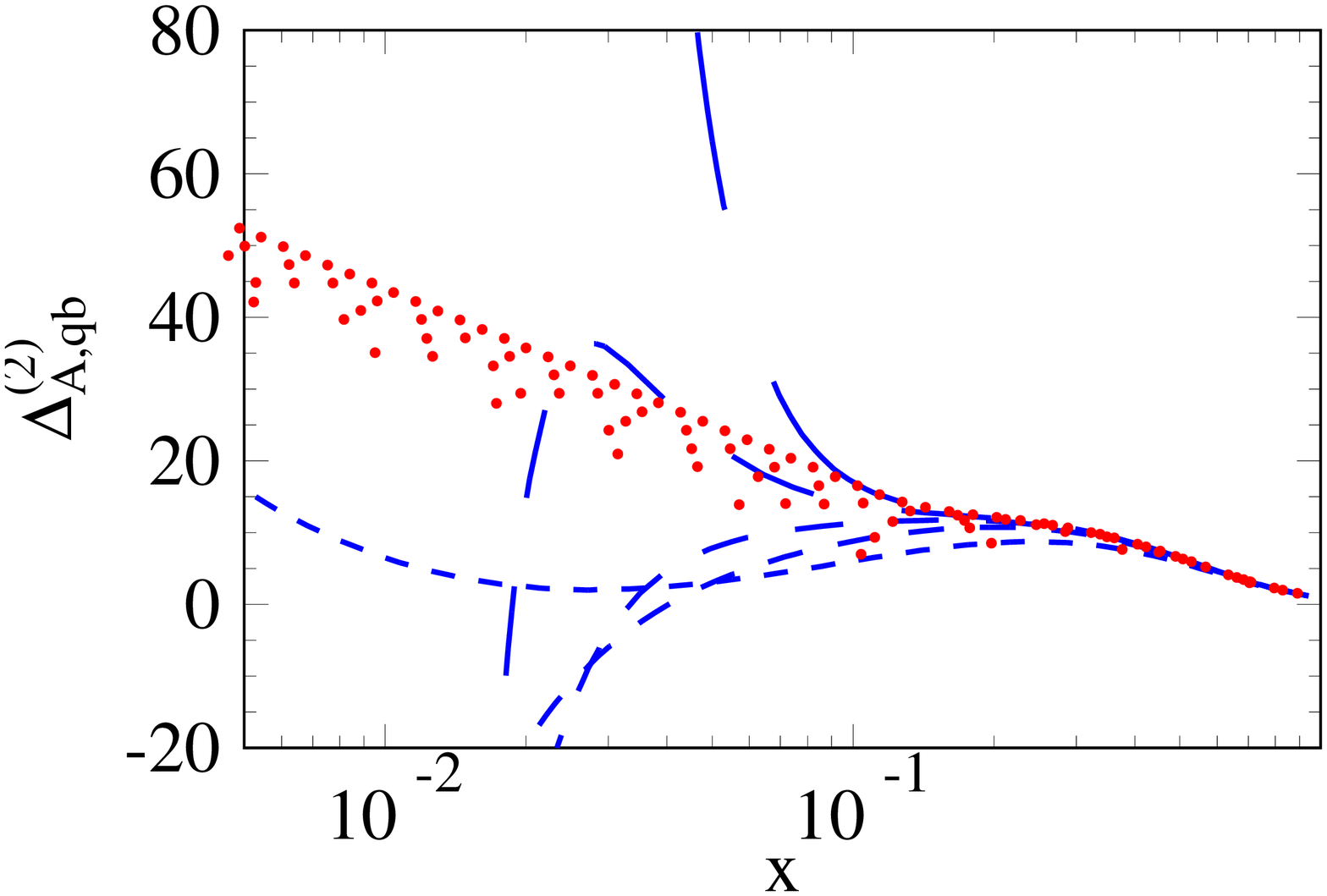}
    &
    \includegraphics[width=0.45\linewidth]{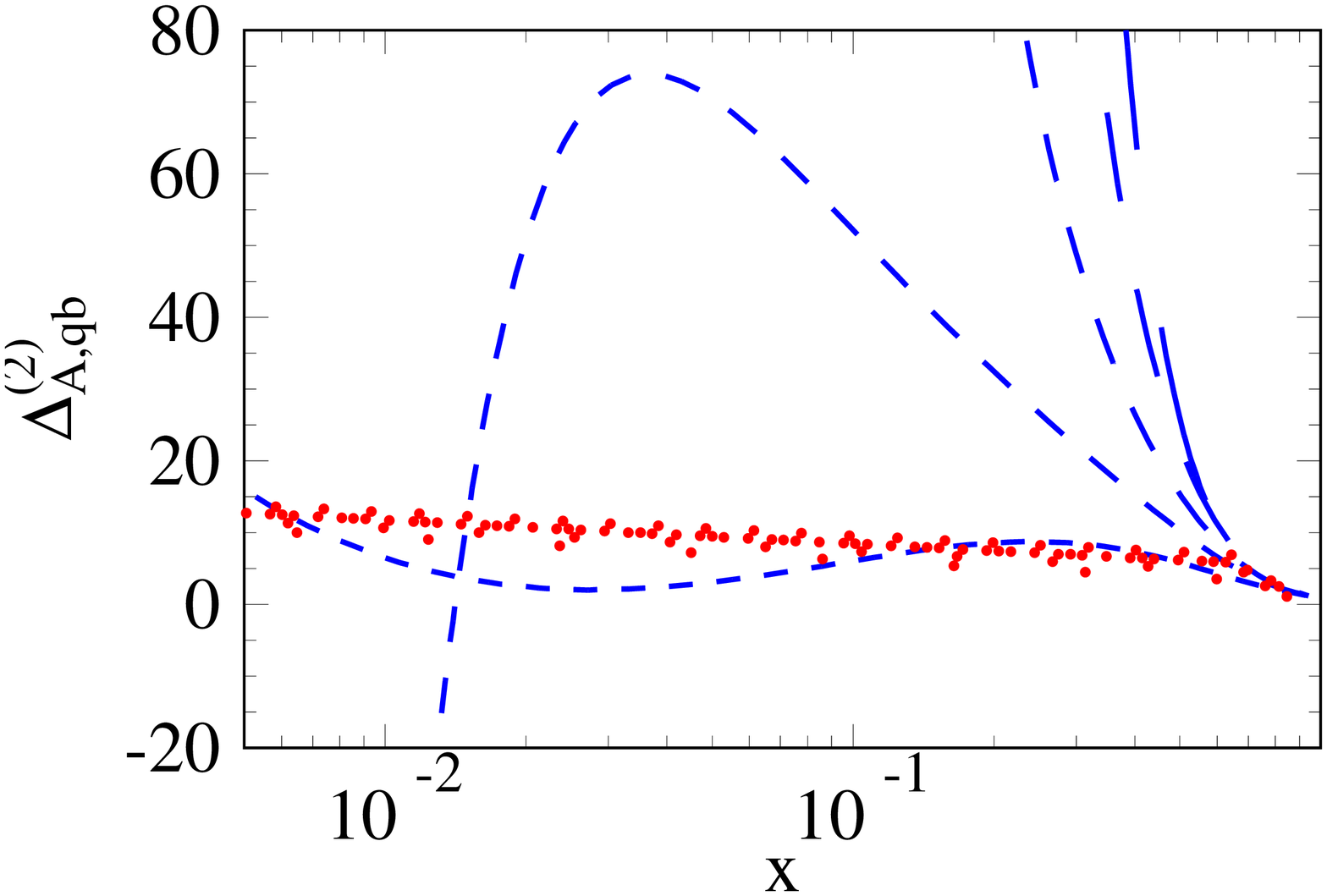}
    \\
    \includegraphics[width=0.45\linewidth]{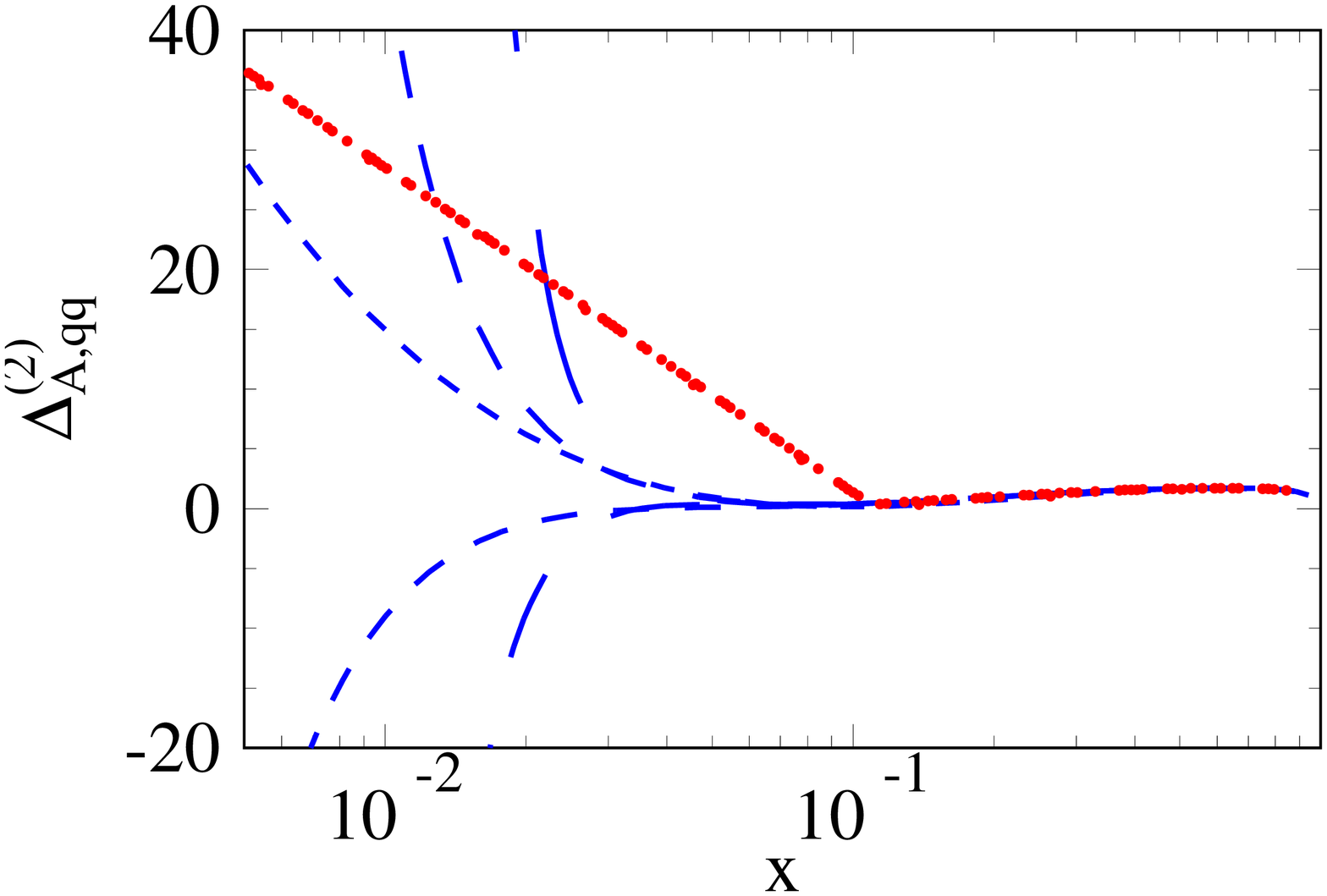}
    &
    \includegraphics[width=0.45\linewidth]{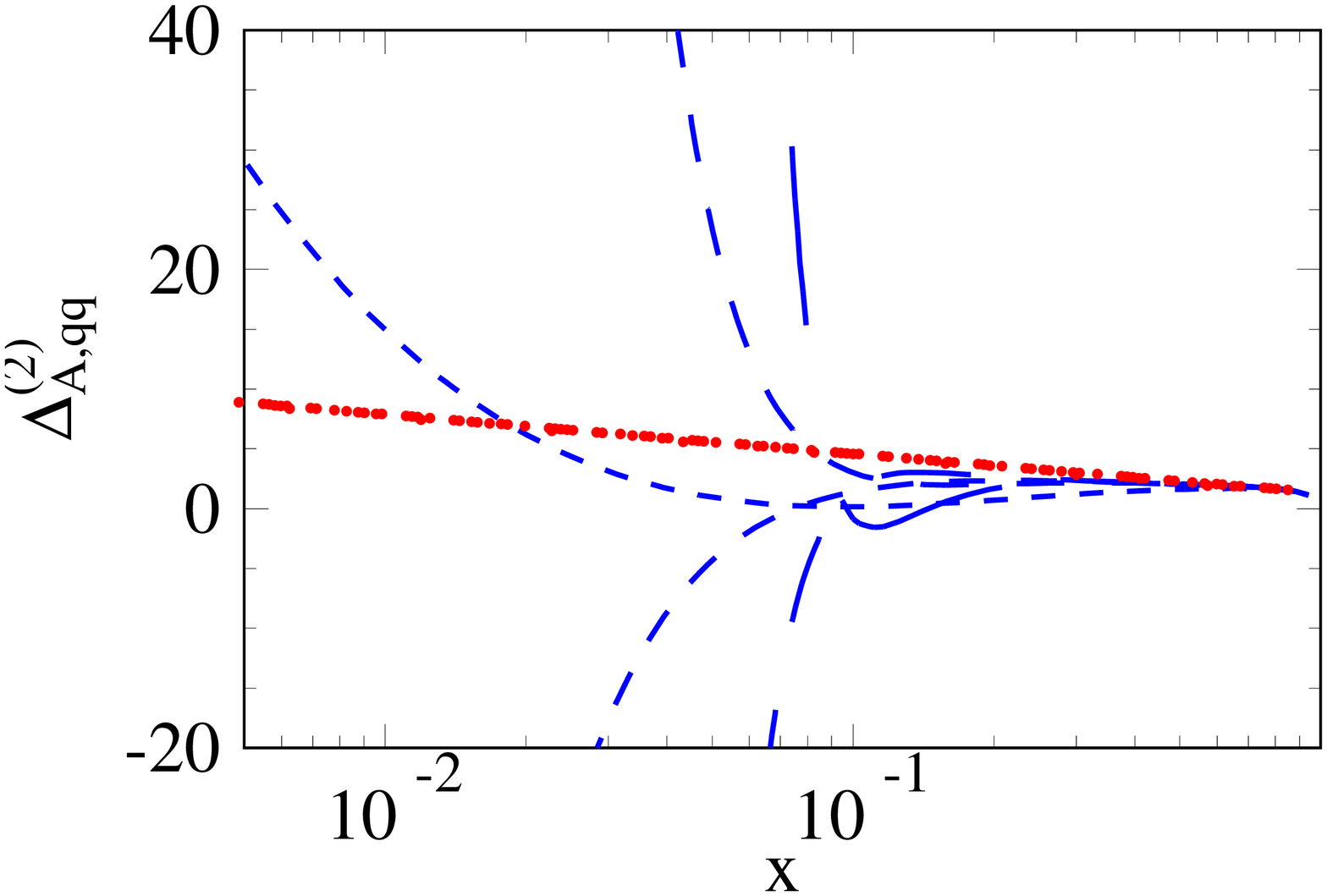}
    \\
    \includegraphics[width=0.45\linewidth]{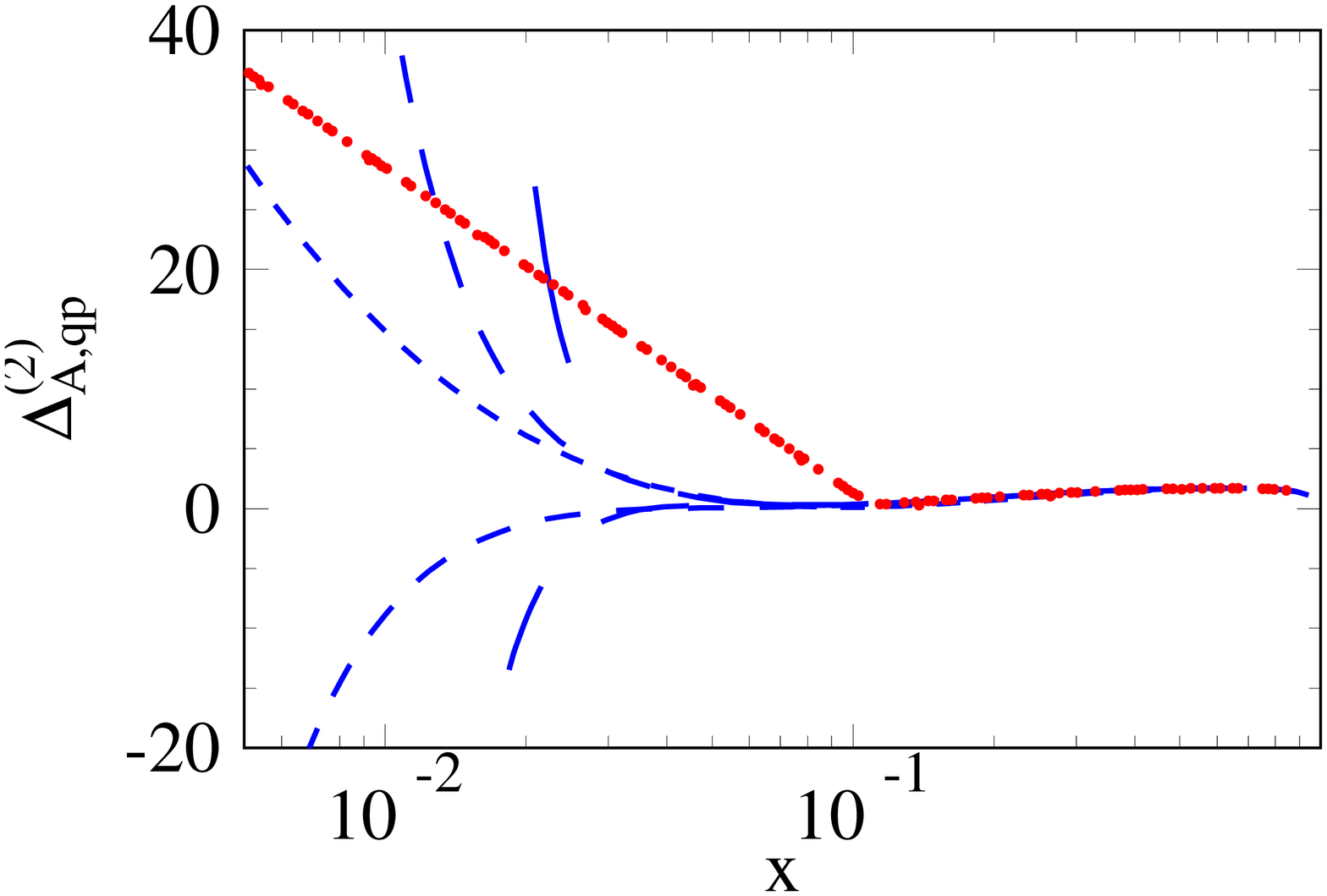}
    &
    \includegraphics[width=0.45\linewidth]{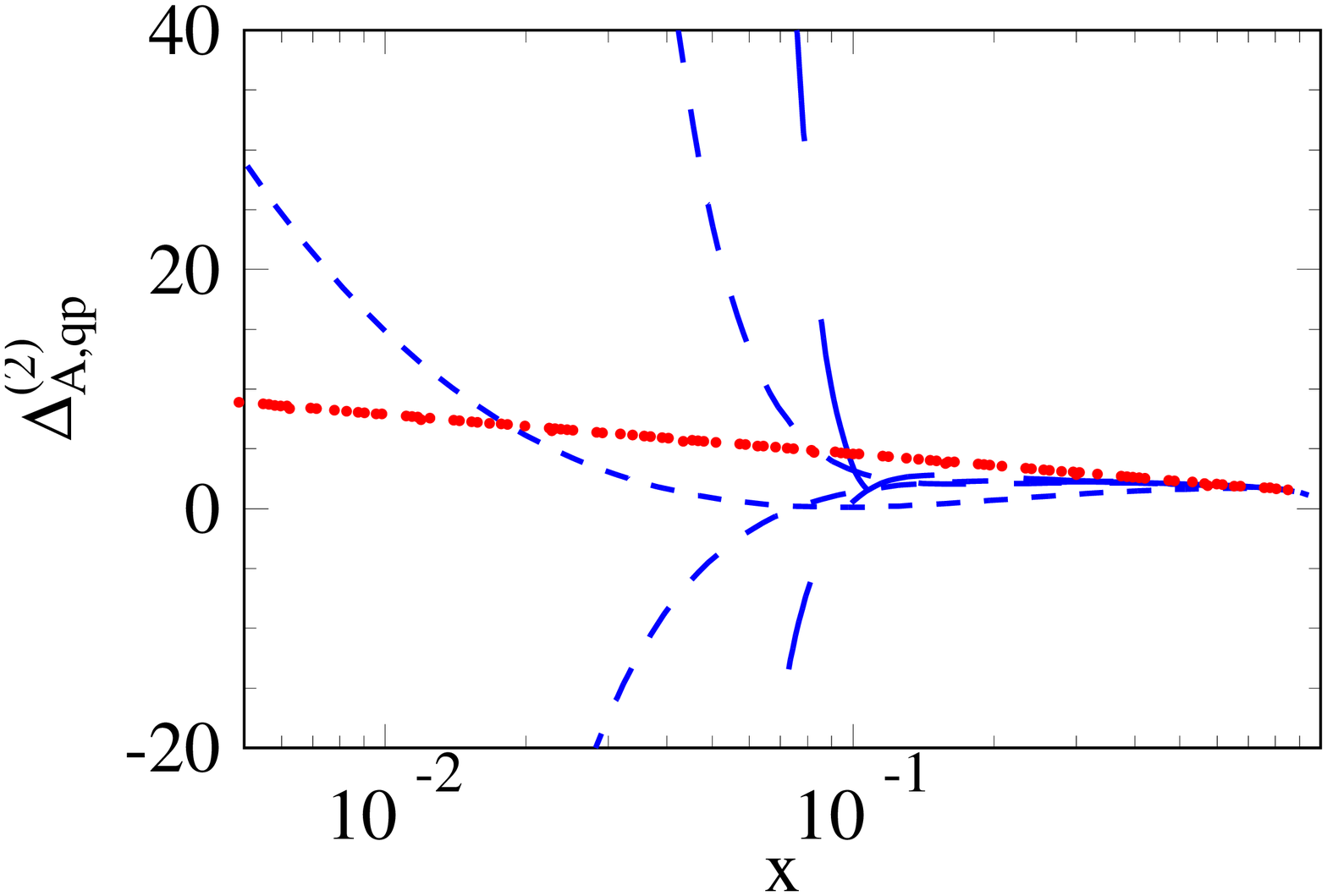}
  \end{tabular}
  \caption[]{\label{fig::nnlo_qq}
    Partonic cross section for the $q\bar{q}$ (``qb''), $qq$ and $qq^\prime$
    (``qp'') channels
    for $M_A=120$~GeV (left) and $M_A=300$~GeV (right).
    The notation is adopted from Fig.~\ref{fig::NLOpart}.}
\end{figure}

The matching procedure was also applied to the quark-initiated processes
$q\bar{q}$, $qq$ and $qq^\prime$. We show the results in
Fig.~\ref{fig::nnlo_qq} using the same notation as in
Fig.~\ref{fig::nnlo_gg}.
For the $q\bar{q}$ we expect a peculiar structure
similar to the one at NLO (cf. Fig.~\ref{fig::NLOpart}) 
which is not reproduced by our procedure.
However, since the overall contribution
from this channel is small, we may neglect the uncertainty of the matching
procedure.

The numerical contribution from the $qq$ and $qq^\prime$ channels to the total
hadronic cross section is similar to the $q\bar{q}$ one. 
For these initial states, however, we
do not expect a complicated shape of the curves and thus the matched result
through $\rho^4$ terms can be trusted both for $M_A=120$~GeV and
$M_A=300$~GeV. (The differences between the various matched results are
again small at the hadronic level.)

\begin{table}[t]
  \begin{center}
    \begin{tabular}{c|rr|rr}
      $\rho^n$ & $M_H=120$~GeV & $M_H=300$~GeV & $M_A=120$~GeV & $M_A=300$~GeV \\
      \hline
      $n=0$ &  83.5654 &   94.0264 &  95.9237   &  105.0866\\
      $n=1$ & + 3.5586 & + 24.0520 & + 6.5340   & + 44.7899\\
      $n=2$ & + 0.2190 & +  9.1031 & + 0.5330   & + 22.5153\\
      $n=3$ & + 0.0164 & +  4.2818 & + 0.0495   & + 13.0672\\
      $n=4$ & ---      & ---       & + 0.0050   & +  8.1572\\
    \end{tabular}
    \caption{\label{tab::delta_nnlo}NNLO contribution to the coefficient of $\delta(1-x)$
      in the normalization of Eq.~(\ref{eq::hatsigma}) from the various
      expansion terms $\rho^n = (M_\Phi^2/M_t^2)^n$. The renormalization scale
      has been set to $M_\Phi$.}
  \end{center}
\end{table}

By analogy to Tab.~\ref{tab::delta}, in Tab.~\ref{tab::delta_nnlo} we present
the NNLO contribution to the $\delta$-function part which is not contained in the
plots discussed before. For $M_A=120$~GeV the convergence is fast. The
$\rho$ terms still provide about 7\%, however, already the $\rho^2$ terms are
below 1\%. At $M_A=300$~GeV a good approximation to the exact result 
requires terms through $\rho^4$.
Estimating the contribution of the $\rho^n$ terms for $n\ge5$ as twice
the $\rho^4$ contribution, we expect the accuracy of about 8\%.
For $M_A=280$~GeV this number goes down to 5\%.

Since the analytic results are quite lengthy we refrain from listing them 
here explicitly. They are available on request from the authors.


\section{\label{sec::hadr}Hadronic cross sections}

In this Section we present the hadronic cross section for the individual
channels and compare in each case with the infinite-top quark mass
approximation in order to quantify the accuracy of the latter.
Although the results for the scalar Higgs boson have already been
discussed in the
literature~\cite{Harlander:2009mq,Pak:2009dg,Harlander:2009my} we present
results both for the scalar and pseudo-scalar case for comparisons.
For the numerical results we use the nominal LHC center-of-mass
energy $\sqrt{s}=14$~TeV; for $\sqrt{s}=7$~TeV the results look
qualitatively very similar.

\begin{figure}[t]
  \centering
  \begin{tabular}{cc}
    \includegraphics[width=0.45\linewidth]{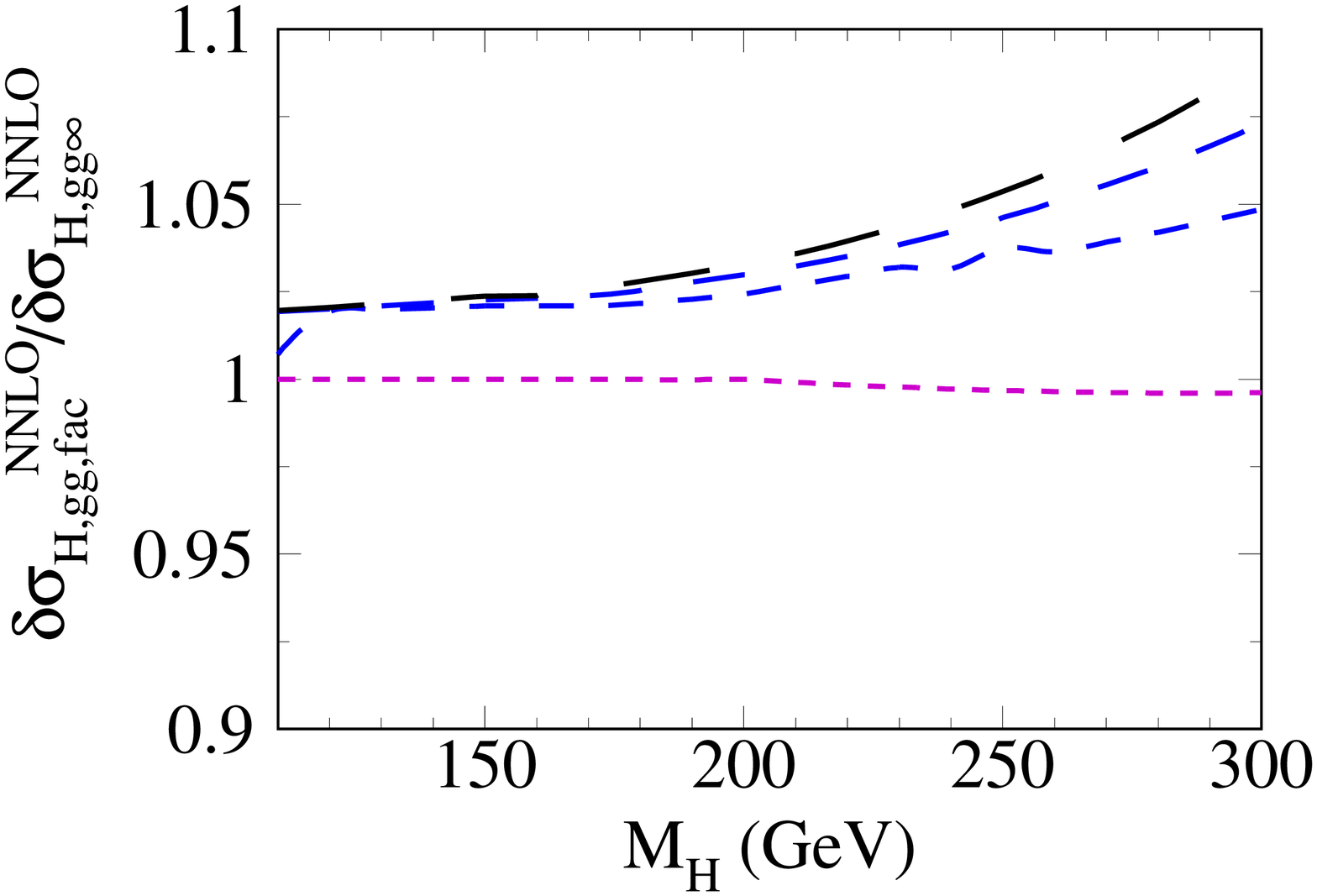}
    &
    \includegraphics[width=0.45\linewidth]{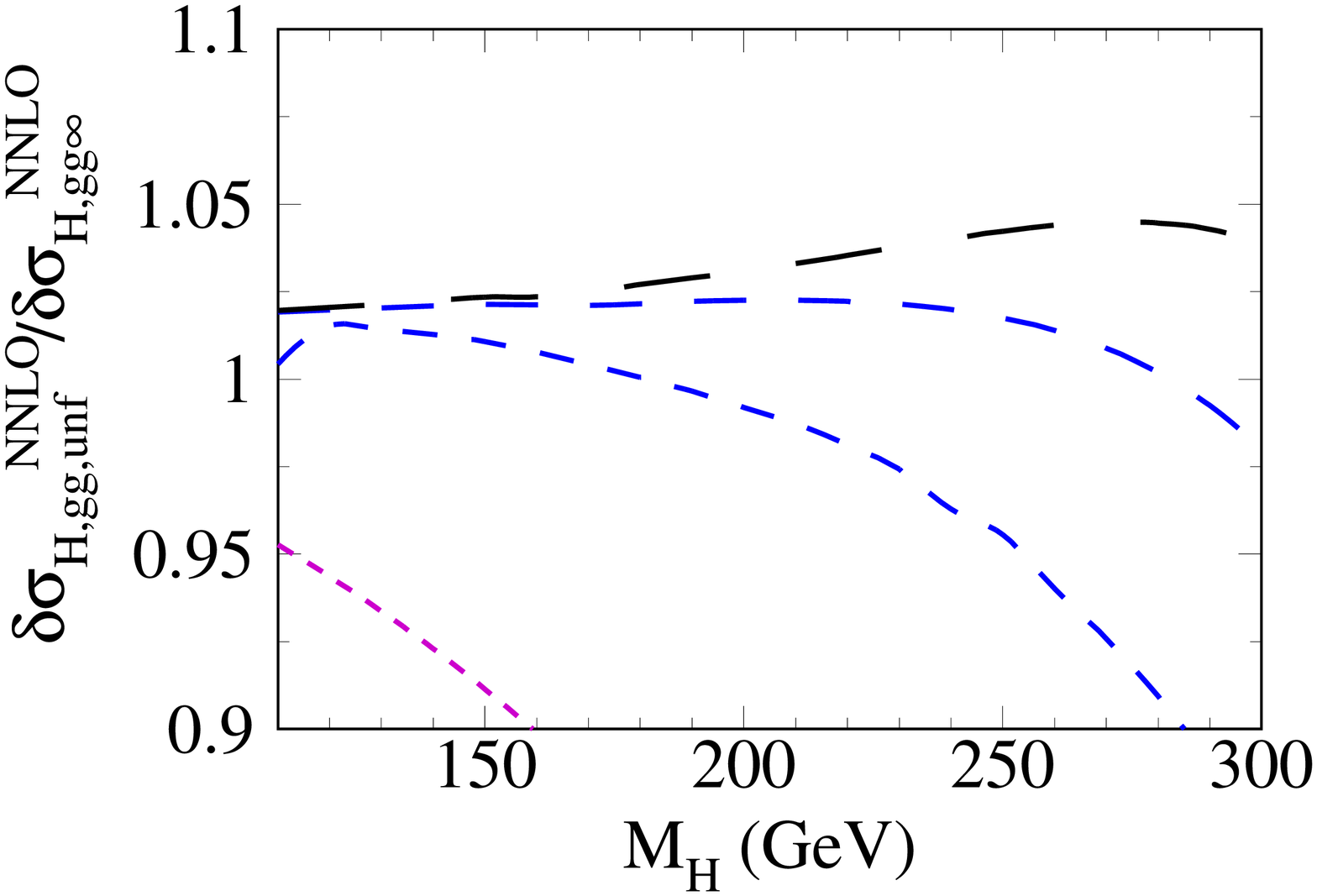}
    \\
    \includegraphics[width=0.45\linewidth]{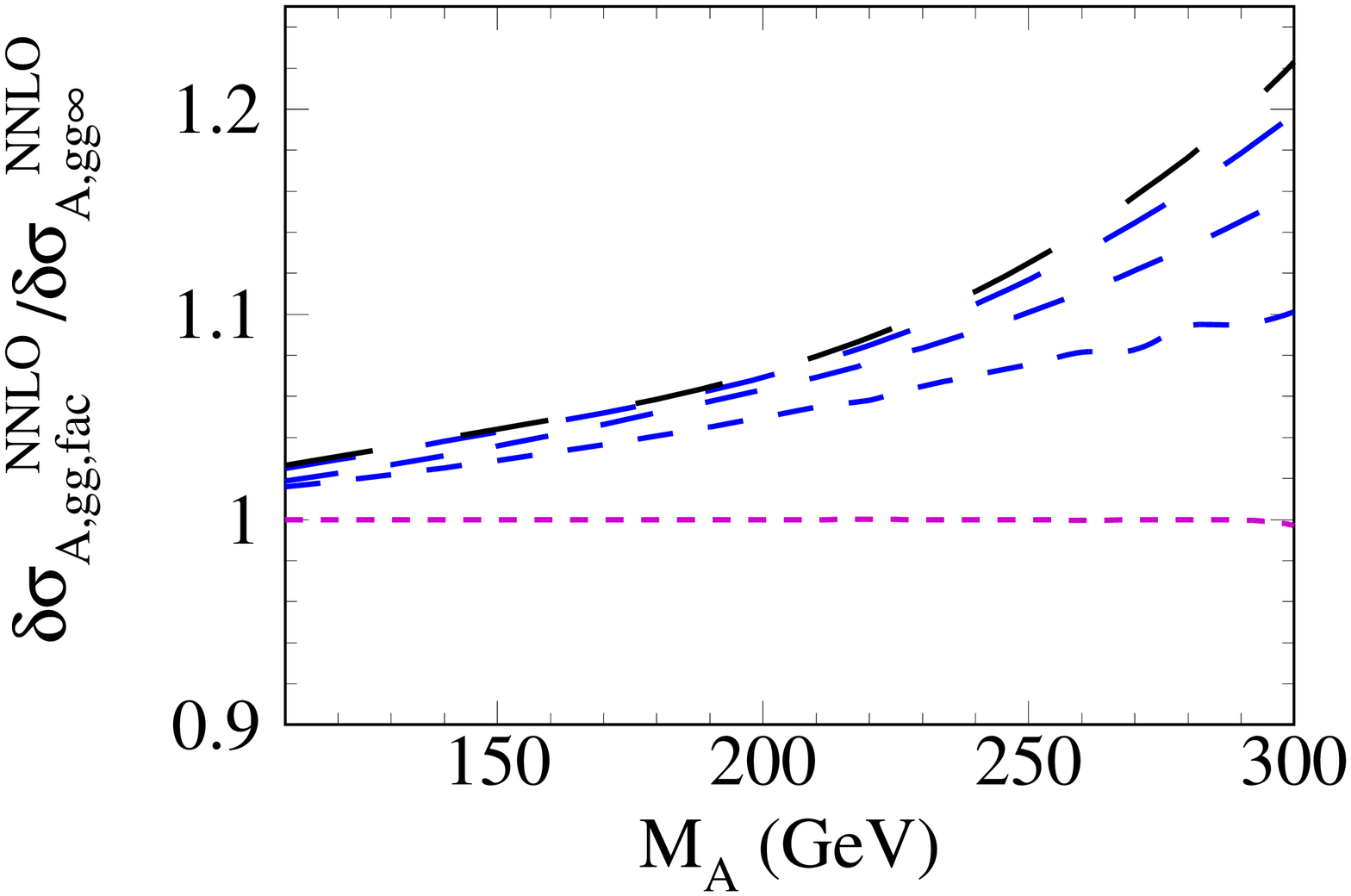}
    &
    \includegraphics[width=0.45\linewidth]{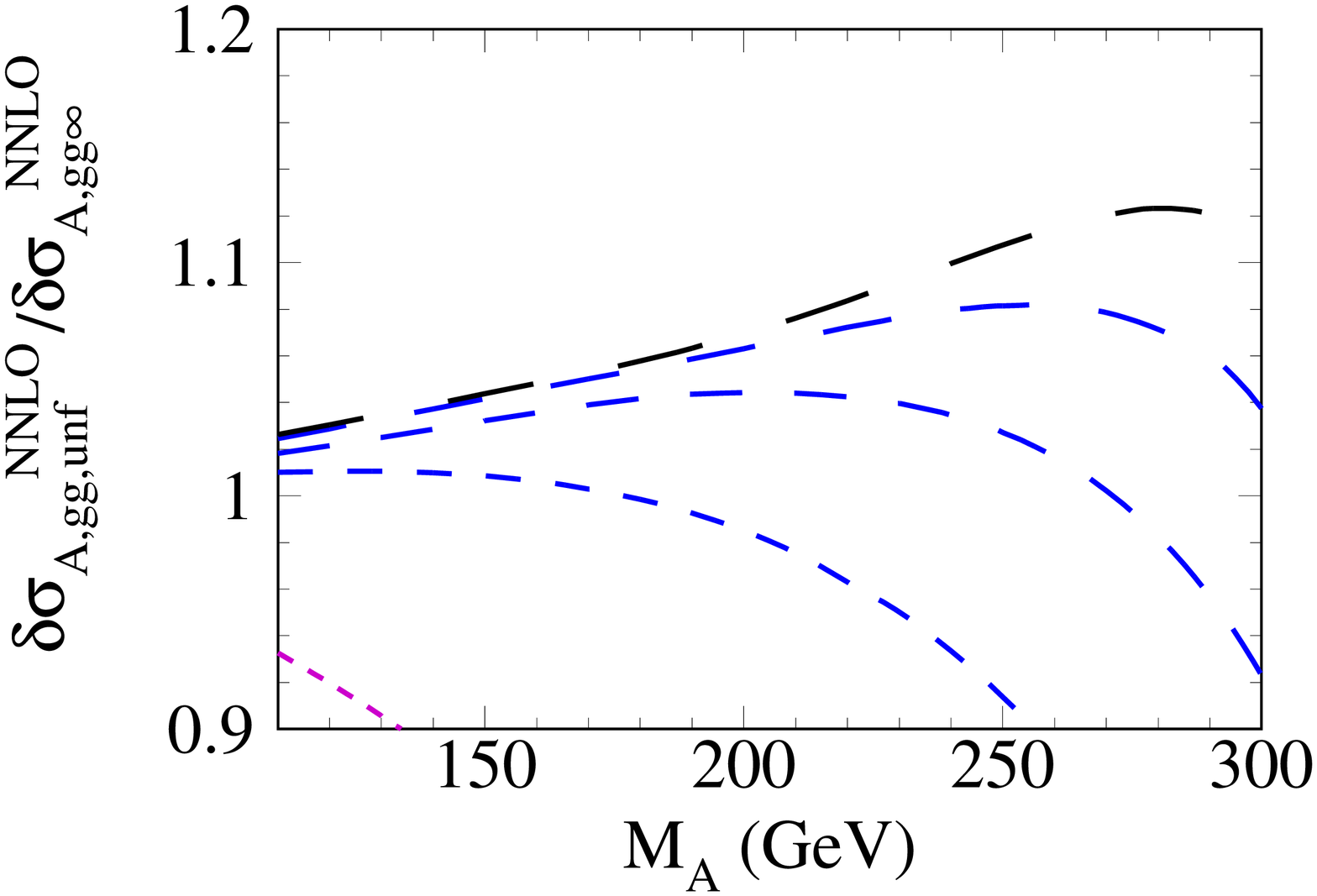}
  \end{tabular}
  \caption[]{\label{fig::hadr_gg}
    NNLO contribution to the hadronic cross section of the $gg$-channel
    normalized to the infinite-top quark mass result.
    Top: scalar Higgs boson; bottom: pseudo-scalar Higgs boson. The left plots
    the exact LO result is factorized whereas in the right plots no
    factorization is used.}
\end{figure}

Let us again start with the $gg$-induced channel. In Fig.~\ref{fig::hadr_gg}
the ratio 
\begin{eqnarray}
  \frac{\delta\sigma_{gg}^{\rm NNLO}}{\delta\sigma_{gg,\infty}^{\rm NNLO}}
  \label{eq::delsig_NNLO}
\end{eqnarray}
of the NNLO contribution to the total cross section is shown. In the numerator
the matched partonic results are used and in the denominator is the bare
infinite-top quark mass result. As before, the lines with longer dashes
include higher expansion terms in $\rho$. Note that for the scalar Higgs boson
terms through $\rho^3$ are available whereas for the pseudo-scalar case we
could compute even the $\rho^4$ terms.  Actually, in the practical calculation
it turns out that the diagrams with pseudo-scalar couplings lead to
significantly fewer terms at the intermediate steps. We believe that this is
due to elimination of many terms with anti-symmetric properties of the
$\epsilon$-tensors remaining after the $\gamma_5$-matrices.

In the left column the exact LO result is factored out both in the
numerator and the denominator\footnote{The slight deviation from unity of
  the curve including only the $\rho^0$ term is due
  to matching effects.}.
One observes that the correction for the pseudo-scalar Higgs boson is
bet\-ween 2.5\% for $M_A=120$~GeV and about 20\% for $M_A=300$~GeV which
is significantly larger than for the scalar Higgs boson, reaching at most
9\%~\cite{Harlander:2009mq,Pak:2009dg,Harlander:2009my}.
Judging from the difference of the two consecutive curves, we observe
in both cases good convergence of the $\rho$-expansion even for
$M_A=300$~GeV. This gives us some confidence that the
corrections including the highest powers in $\rho$ are approaching the
unknown exact result.

The right panels in Fig.~\ref{fig::hadr_gg} show the
ratio~(\ref{eq::delsig_NNLO}) with the numerator expanded in $\rho$
without factoring out the LO mass dependence. The plot for the scalar
Higgs boson reproduces the results in the
literature~\cite{Harlander:2009mq,Pak:2009dg,Harlander:2009my},
demonstrating that after including higher orders in $\rho$ the ratio
becomes less dependent of $M_\Phi$. The curve including $\rho^3$ terms
deviates from unity at most by about 5\%.
The situation is different for the pseudo-scalar case. For low Higgs boson
masses the corrections also converge against a few per cent, for higher mass
values, however, they amount up to about 10\%
demonstrating that the infinite-top mass result accompanied by the matching
procedure is not sufficient to approximate the exact result with 1\% accuracy
or better.

Similarly to the NLO case shown in Fig.~\ref{fig::nlo_hadr_1} we observe also
in Fig.~\ref{fig::hadr_gg} that the convergence becomes poor for Higgs boson
masses above 250~GeV if the exact LO mass dependence is not factored
out. Thus, for our final prediction we use the approach of the upper left plot in
Fig.~\ref{fig::nlo_hadr_1} where the effect of higher order terms in $\rho$
(beyond $\rho^4$) can be safely neglected.

\begin{figure}[t]
  \centering
  \begin{tabular}{cc}
    \includegraphics[width=0.45\linewidth]{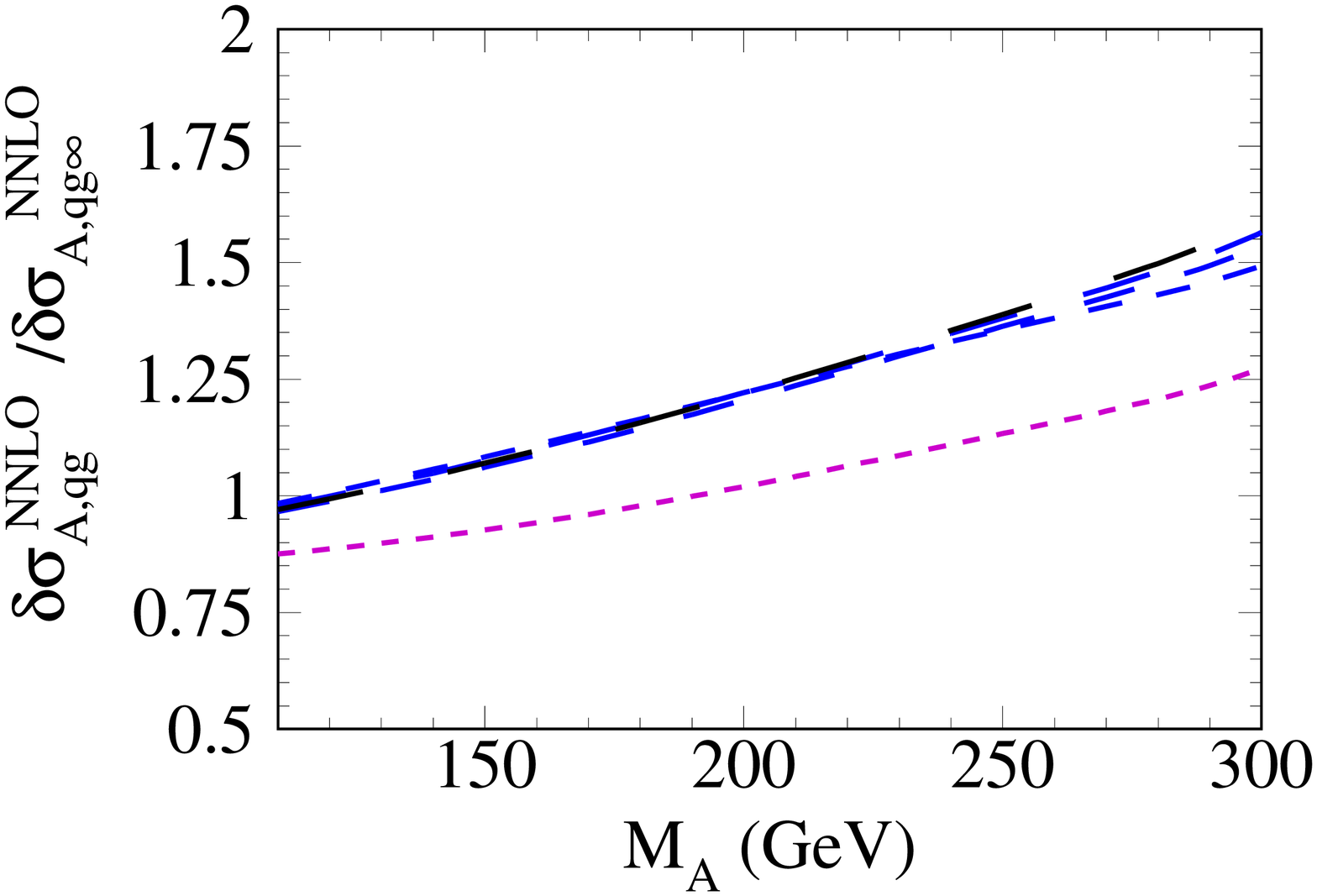}
    &
    \includegraphics[width=0.45\linewidth]{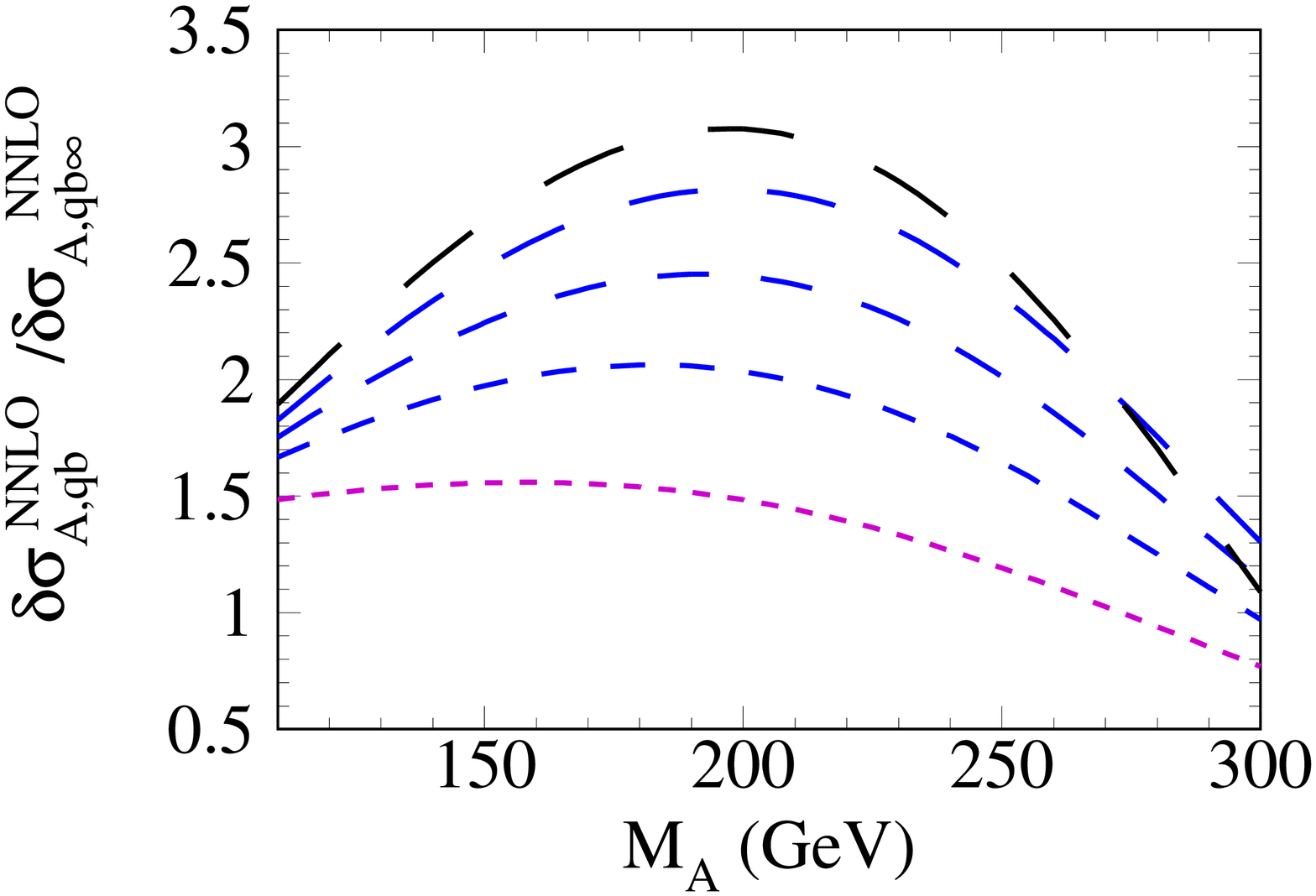}
    \\
    \includegraphics[width=0.45\linewidth]{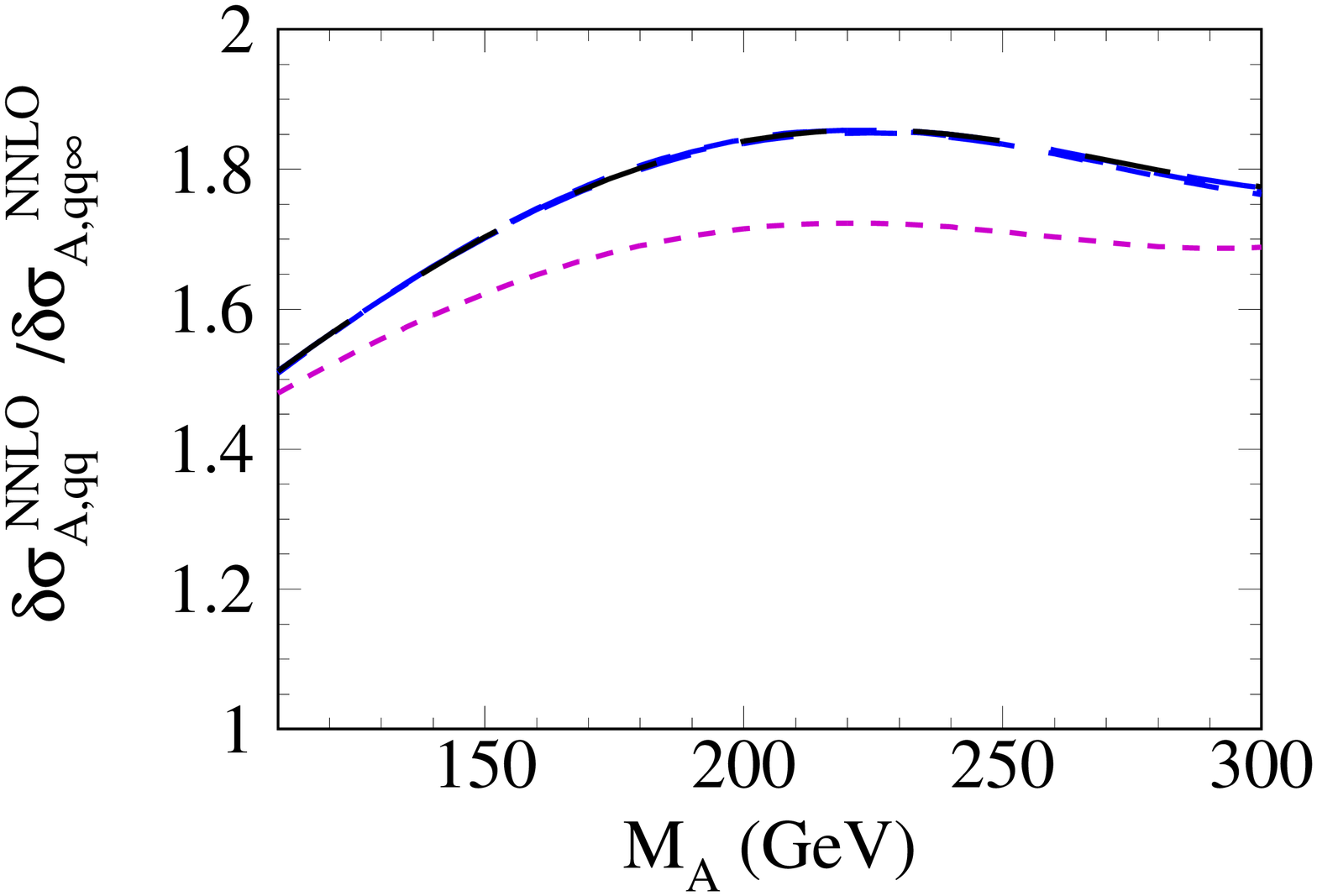}
    &
    \includegraphics[width=0.45\linewidth]{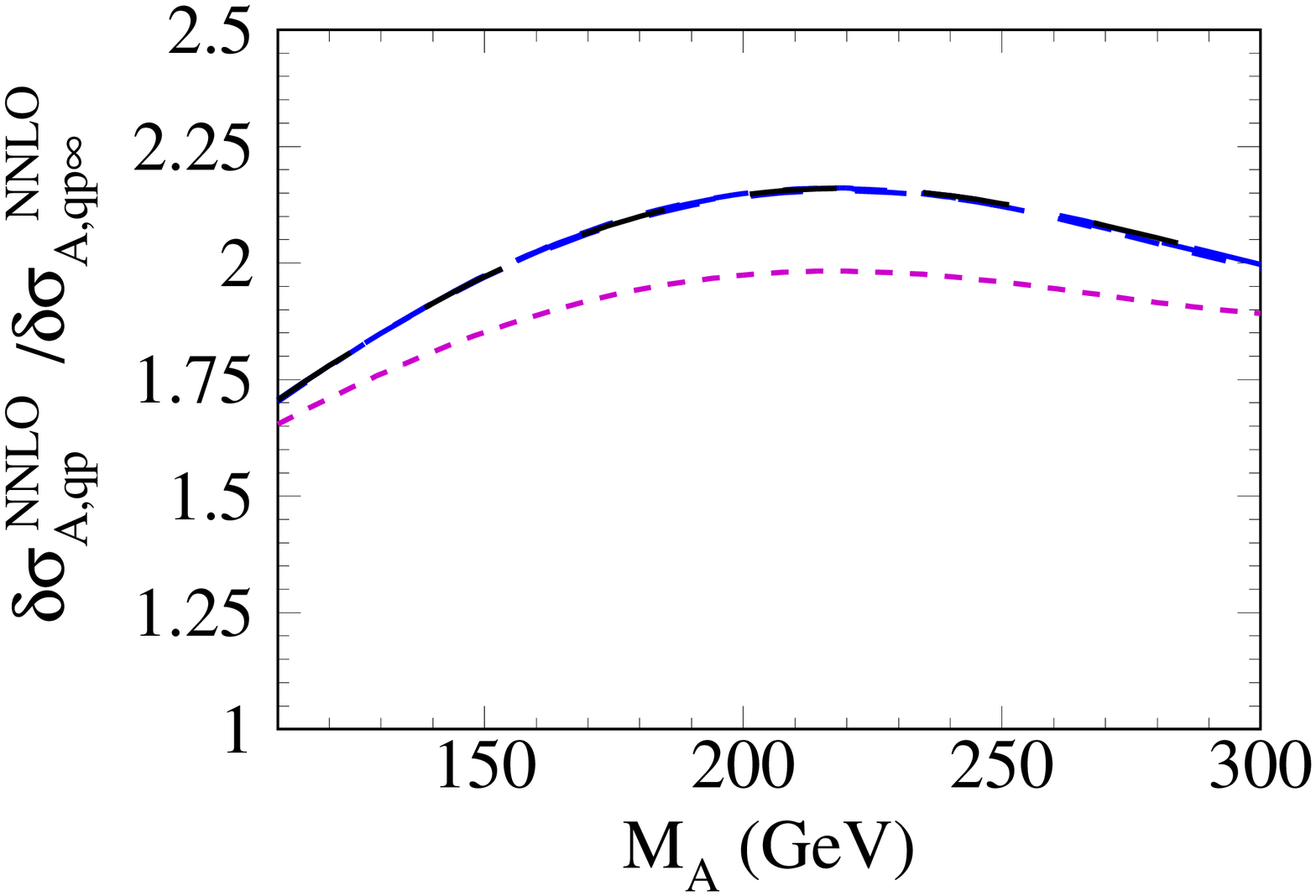}
  \end{tabular}
  \caption[]{\label{fig::hadr_quarks}
    NNLO contribution to the hadronic cross section for the inclusive
    production of a pseudo-scalar Higgs boson induced by the 
    ``$qg$'', ``$q\bar{q}$'', ``$qq$'' and ``$qq^\prime$'' channels.
    }
\end{figure}

The hadronic cross section for the quark-induced channels,
$qg$, $q\bar{q}$, $qq$ and $qq^\prime$ are shown in
Fig.~\ref{fig::hadr_quarks} with the notation similar to Fig.~\ref{fig::hadr_gg}.
In all cases we observe decent convergence with higher $\rho$ powers.
From the NLO analyses of Section~\ref{sec::nlo} we expect that the
prediction for the $qg$ channel agrees with the exact result to within about 15\%.
Note that there is a noteable correction from including the first
$\rho$-suppressed term. This behaviour can be explained by
the pattern in corresponding partonic cross section 
discussed in the previous Section (see Fig.~\ref{fig::nnlo_qg}). 

As far as the $q\bar{q}$ channel is concerned, the matched results coincide
with the effective-theory prediction within roughly a factor of two to three
for $120~\mbox{GeV}<M_A<250$~GeV which decreases to about 1.2 for
$M_A=300$~GeV.
The results for the $qq$ and $qq^\prime$ channels are shown for completeness
in the bottom row of Fig.~\ref{fig::hadr_quarks}. The deviation between the
matched and the effective-theory result in the considered Higgs boson mass
range is about 1.5 to 2.
For $\sqrt{s}=14$~TeV the overall contribution from the $qg$, $q\bar{q}$, $qq$
and $qq^\prime$ channels to the NNLO corrections
amounts to $(-8)\%$ to $(-17)\%$, 0.1\% to 0.2\%, 0.08\% and 0.3\%,
respectively.
Thus, at the current
level of accuracy it is certainly possible to use the infinite-top quark mass
result for the predictions originating from the $q\bar{q}$, $qq$ and
$qq^\prime$ channels. For $M_A=300$~GeV the infinite-top quark mass
approximation for the $qg$ channel may lead to an uncertainty of about 2\% in
the NNLO prediction of the hadronic cross section.

It is instructive to look at the individual contributions
of the NNLO pieces to the hadronic cross section. In
Tab.~\ref{tab::nnlo_split} we show for $M_\Phi=300$~GeV the contributions from
the $\delta$ function, the plus distributions and the remainder. The
infinite-top mass result is confronted with the approximations based on the
matching that incorporates corrections of order $\rho^n$ ($n=0,1,\ldots$).
One observes that both for the scalar and the pseudo-scalar Higgs boson the
virtual corrections grow by almost a factor of two compared to the
infinite-top mass values. In both cases the contribution from the plus
distributions are quite small and the contribution from the $\delta$ function
amounts to about a quarter of the remainder. The latter dominate the
corrections and amount to about 85\% in the scalar and about 80\%
in the pseudo-scalar case.
The difference between the best prediction and the infinite-top corrections
amounts to 9\% and 22\%, respectively.

For the $M_\Phi$ values below $300$~GeV the convergence of all
individual contributions significantly improves and our error
estimate (given by the size of the last known term) is below
2\% and thus is negligible.

\begin{table}[t]
  \centering
\scalefont{1.0}
  \begin{tabular}{cc}
    \begin{tabular}{c|ccc}
      $H$ & $\delta(1-x)$ & $\left[{\ldots}\right]_+$ & rest \\
      \hline
      $M_t\to\infty$ &  $0.363$& $-0.066$& $2.555$ \\
      \hline
      ``0'' &
      $0.363$ & $ -0.066$ & $ 2.544$\\
      ``1-0'' &
      $0.093$ & $ 0.002$ & $ 0.054$\\
      ``2-1'' &
      $0.035$ & $ 0.001$ & $ 0.034$\\
      ``3-2'' &
      $0.017$ & $ 0.000$ & $ 0.034$\\
      \hline
      ``3'' &
      $0.508$ & $ -0.063$ & $ 2.666$
    \end{tabular}
    &
    \begin{tabular}{c|ccc}
      $A$ & $\delta(1-x)$ & $\left[{\ldots}\right]_+$ & rest \\
      \hline
      $M_t\to\infty$ &
      $1.213$ & $ -0.194$ & $ 7.728$\\
      \hline
      ``0'' &
      $1.213$ & $ -0.194$ & $ 7.703$\\
      ``1-0'' &
      $0.517$ & $ 0.013$ & $ 0.380$\\
      ``2-1'' &
      $0.260$ & $ 0.006$ & $ 0.236$\\
      ``3-2'' &
      $0.151$ & $ 0.003$ & $ 0.185$\\
      ``4-3'' &
      $0.094$ & $ 0.002$ & $ 0.128$\\
      \hline
      ``4'' &
      $2.235$ & $ -0.170$ & $ 8.633$\\
    \end{tabular}
  \end{tabular}
  \caption{\label{tab::nnlo_split}
    Individual contributions to the quantity $\delta\sigma^{\rm
      NNLO}_{\Phi,gg}$ (i.e. the $gg$ induced part) in
    Eq.~(\ref{eq::deltasigma}) for $M_\Phi=300$~GeV 
    for a scalar (left) and pseudo-scalar Higgs
    boson (right). The first line corresponds to the infinite-top quark mass
    result and ``$i$'' represents the matched result including terms of order
    $\rho^i$. The bottom line contains the best available prediction.}
\end{table}

\begin{figure}[t]
  \centering
  \begin{tabular}{cc}
    \includegraphics[width=0.5\linewidth]{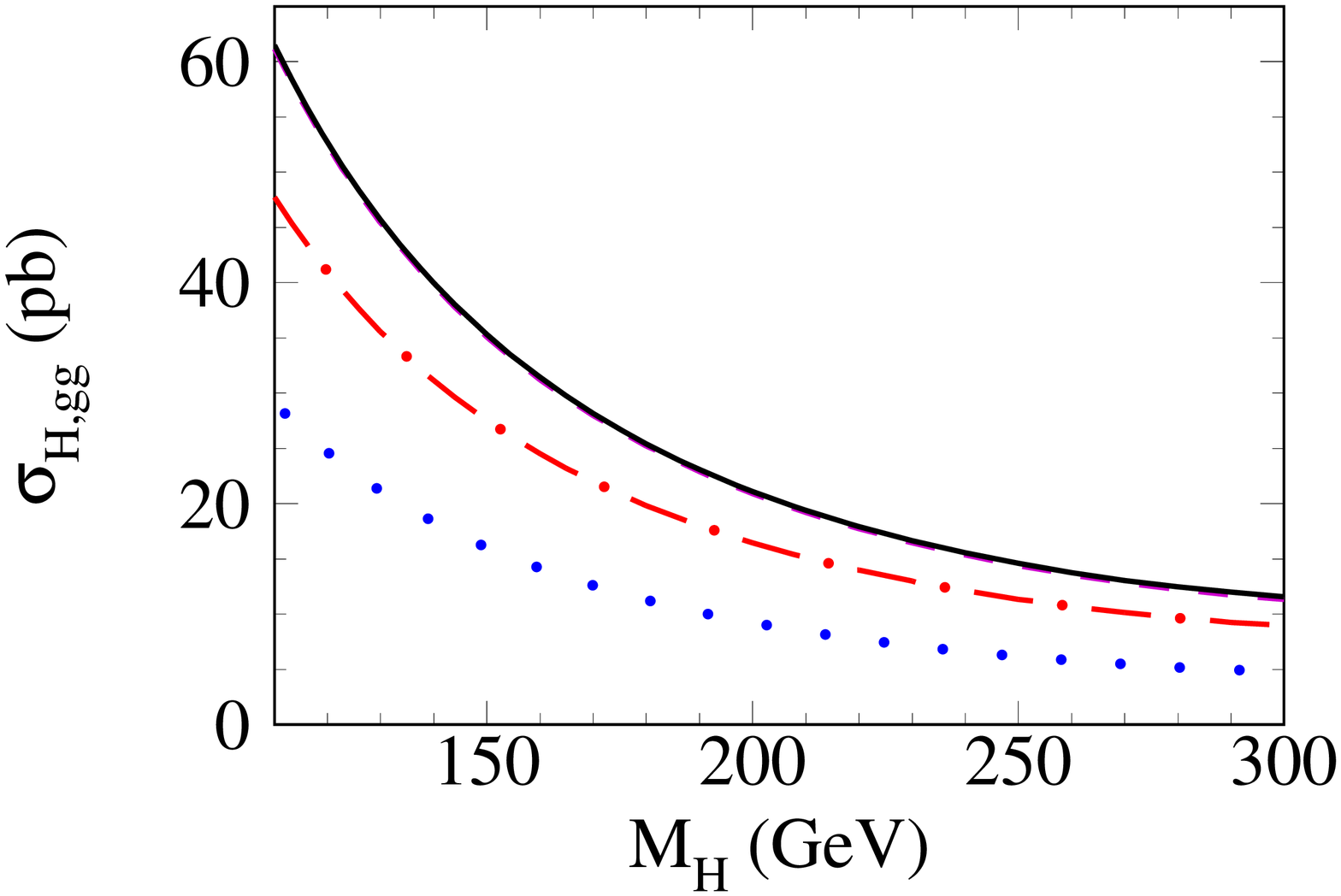}
    &
    \includegraphics[width=0.5\linewidth]{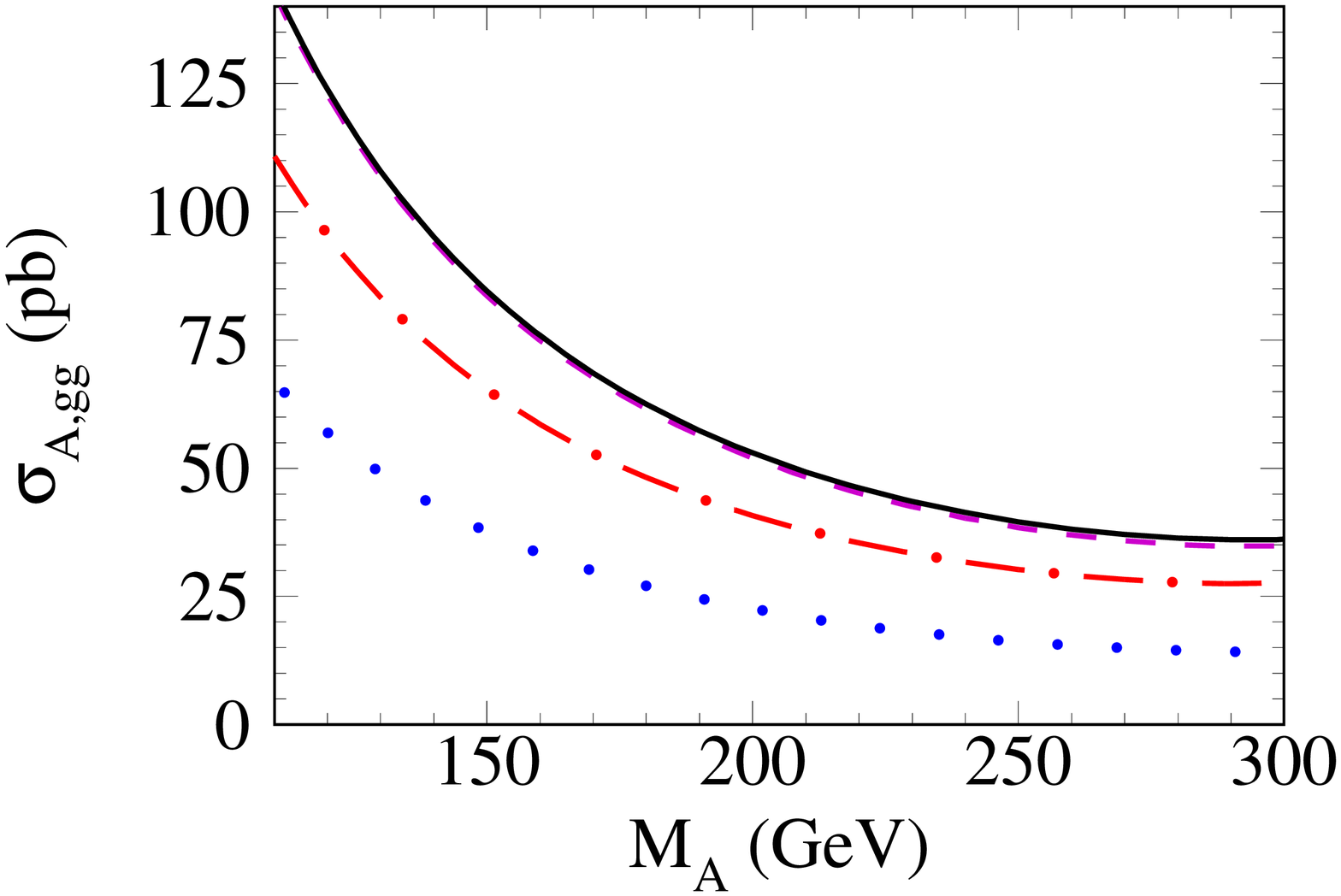}
  \end{tabular}
  \caption[]{\label{fig::sigMh}
    Total cross section for the production of a scalar (left) and pseudo-scalar
    (right) Higgs boson to LO (dotted), NLO (dash-dotted) and NNLO (solid) precision.
    The dashed line corresponds to the prediction where for the NNLO
    corrections the heavy-top approximation has been used.}
\end{figure}

Let us finally show results for the cross section of the numerically dominant
$gg$-induced contribution.  In Fig.~\ref{fig::sigMh} the LO, NLO and NNLO
predictions for the total production cross section both for a scalar (left)
and a pseudo-scalar Higgs boson (right) is plotted. The dotted and the
dash-dotted lines belong to the exact LO and NLO prediction and the solid
lines represent our best NNLO expression, i.e. the matched cross section
containing the $\rho^3$ and $\rho^4$ corrections for the scalar and
pseudo-scalar Higgs boson, respectively. For comparison we also show as the
dashed line the result which is obtained if the infinite-top quark mass
approximation is used for the NNLO term. In the scalar case one barely sees a
difference with the solid line whereas for the pseudo-scalar Higgs boson there
is a deviation of about 6\% for masses around 300~GeV.  Thus it is important
to replace the infinite-top mass results in this Higgs boson mass region by
the matched results presented in this paper in future precision studies.

\section{\label{sec::concl}Conclusions}

In this paper we consider the inclusive production of the scalar and
the pseudo-scalar Higgs boson at hadron colliders. We compute the NNLO
production cross section for Higgs bosons with masses below
approximately 300~GeV, including top quark mass effects.
The technique that we use extends the method that was recently
applied to scalar Higgs boson production to the pseudo-scalar case.
It relies on an asymptotic expansion in the limit of a large top quark mass and
a subsequent matching to the zero-mass limit. This paper contains a detailed
description of the method and several intermediate results which could be
useful in the context of other calculations. In particular we provide all one-
and two-loop four-point master integrals with an $\epsilon$ expansion
sufficient for a N$^3$LO calculation.

The main conclusions of our NNLO analysis are as follows.
\begin{itemize}
\item At NLO our approach reproduces the exact result with high precision
  (see discussion in Section~\ref{sec::nlo}).
\item Since at NLO the agreement between our approximation and the exact
  result is below 1\% for the $gg$ channel
  once expansion terms up to order $1/M_t^8$ are included we are quite
  confident that the same is true at NNLO. Thus, from the practical point of
  view our calculation is equivalent to the exact one (at least for Higgs boson
  masses below threshold).
\item Our NNLO corrections deviate from the 
  infinite-top quark mass approximation (with exact LO mass
  dependence factored out) by about 2\% for low Higgs boson masses.
  For higher masses the deviation is larger. For $M_\Phi=300$~GeV it
  amounts to about 9\% for the scalar Higgs boson and about 22\% 
  for the pseudo-scalar case.
\item This leads to the conclusion that up to an uncertainty of about 2\% for
  the scalar case and about 6\% for the pseudo-scalar case it
  is save to use the infinite-top quark mass
  approximation~\cite{Harlander:2002wh,Anastasiou:2002yz,Harlander:2002vv,Anastasiou:2002wq,Ravindran:2003um}
  for the prediction of the total cross section.
  This is a non-trivial result which so far has no fundamental explanation.

  In case one aims for a better precision the infinite-top quark mass
  results have to be replaced by the results presented in this paper.
\item
  The accuracy of the NNLO part of the 
  infinite-top quark mass result for $M_\Phi=300$~GeV 
  amounts to 6\% (15\%) for the scalar (pseudo-scalar) case at 
  Tevatron.
  For the LHC with $\sqrt{s}=7$~TeV the numbers are 8\% and 20\%.
\end{itemize}

For Higgs boson masses above approximately two times the top quark mass the
approximation used for the asymptotic expansion is not justified.
In this limit one has to rely on the good agreement between the exact and
the infinite-top mass result at NLO and assume that it extends to NNLO.


\vspace*{2em}
{\large\bf Acknowledgements}

This work was supported by the DFG through the SFB/TR~9 ``Computational
Particle Physics''.
We would like to thank Fabrizio Caola and Simone Marzani for providing us with
detailed numerical results for the NNLO production cross section in the
infinite-energy limit.


\begin{appendix}


\section{\label{app::2lMIs}Two-loop four-point master integrals}

The one- and two-loop four-point integrals have been studied for the first
time in Ref.~\cite{Anastasiou:2002yz} (see Appendix~B). Unfortunately, that
reference contains a number of misprints. We independently
evaluated these integrals by a combination of soft expansion\footnote{We
  acknowledge help with cross checks of the soft expansion by Robert
  Harlander and Kemal Ozeren.} and differential
equation methods. We furthermore
extend the results of Ref.~\cite{Anastasiou:2002yz} by adding more terms in
$\epsilon$ which are required for a third-order calculation of the Higgs boson
production cross section.
The results can be downloaded in {\tt Mathematica} format from the
webpage~\cite{ttpdata}.

The integrals published in Ref.~\cite{Anastasiou:2002yz}
suffer from multiple typographical errors.
We have also found a number of less trivial
inaccuracies. In particular, the integral of Eq.~(B.3)
is off by a factor of $2$, the right-hand side of
Eq.~(B.27) must read
\begin{eqnarray}
  = \mathcal{P}^2 z^{-2\ep}\left[
    \frac{\log(z)^2}{2\ep} - 4\zeta_3 - 4\zeta_2\log(z)
    + 4{\rm Li}_3(z) + \frac{\log^3(z)}{2} - 2\log^2(z)
  + \mathcal{O}(\ep)\right],
\end{eqnarray}
the right-hand side of Eq.~(B.22) should begin with
\begin{eqnarray}
 = \mathcal{P}^2 z^{-\ep}(1-z)^{-2\ep}\left\{
   {\rm Li}_2(1-z) + \frac{\log^2(z)}{2} + ... \right\},
\end{eqnarray}
the two terms $-4\log(1-z)^2 - 8$ of the $\mathcal{O}(\ep^0)$
contribution in Eq.~(B.21) must be replaced with
$+4\log(1-z)^2 + 8$, etc.


\section{\label{app::conv}Convolution of partonic cross section and splitting functions}

The singularities associated with the collinear radiation of quarks
and gluons from the incoming partons are described by convolutions
of the partonic cross sections $\sigma_{ij}(x)$ with the
splitting functions $P_{jk}(x)$ where a convolution of the
two functions $f(x)$ and $g(x)$ is defined as
\begin{eqnarray}
  \left[f \otimes g \right](x)
    = \int_0^1 {\rm d}x_1 {\rm d}x_2 \delta(x - x_1 x_2) f(x_1) g(x_2).
\end{eqnarray}

The functions that appear in the LO, NLO, and NNLO cross sections
and LO and NLO splitting functions include combinations of HPLs
of weight one, two, and three with the factors $x$, $1-x$, and $1+x$,
and the generalized functions $\delta(1-x)$ and
$\left[\frac{\ln^k(1-x)}{1-x}\right]_+$. It is convenient to
transform those functions to the Mellin space, where
the convolutions turn into products of Mellin images:
\begin{eqnarray}
  M_n\left[f(x)\right] &=& \int_0^1 x^{n-1} f(x) dx, \label{eqn:mellinn} \\
  M_n\left[\left[f \otimes g\right](x)\right] &=& M_n\left[f(x)\right] M_n\left[g(x)\right].
\end{eqnarray}
The Mellin transforms of various functions present in the NNLO Higgs
boson production were studied in the literature~\cite{Blumlein:1998if}.
We, however, decided to relate all required results to a limited set
of Mellin transforms of HPLs with a certain maximum weight.

Let us consider the Mellin transforms of HPLs of weight zero and one:
\begin{eqnarray}
  M_n[1] &=& \frac{1}{n}\,,\nonumber\\\nonumber
  M_n[\HPL{0}{x}] &=& - \frac{1}{n^2}\,,\\ \nonumber
  M_n[\HPL{1}{x}] &=& \frac{\HS{1}{n}}{n}\,,\\ 
  M_n[\HPL{-1}{x}]&=& - \frac{(-1)^n}{n} \left( \HS{-1}{n} + \ln{2}
  \right) + \frac{\ln{2}}{n}\,,
  \label{eqn:mttab}
\end{eqnarray}
where $H_i$ are the HPLs of weight 1~\cite{Maitre:2005uu}.
The harmonic sums $\HS{...}{n}$ are defined as follows:
\begin{eqnarray}
  \HS{}{n} = 1,~~
  \HS{a,\vec{b}}{n} = \sum_{i=1}^n f_a(i)~ \HS{\vec{b}}{i},~~
  f_a(i) = \left\{
    \begin{array}{ll}
      i^{-a}, &  a \ge 0\,, \\
      (-1)^i ~i^a, & a < 0\,.
    \end{array}\right.
  \label{eqn:hsums}
\end{eqnarray}
One may define the weight of those sums as the
sum of the absolute values of their indices.

Similarly to Eqs.~(\ref{eqn:mttab}), Mellin images of HPLs of higher weights contain
harmonic sums of higher weights and various transcendental numbers originating
from various HPLs evaluated at $x = 1$. We choose the representations
where each Mellin image contains only $n$ as the argument of sums and integer
powers of $n$ in the denominators, and no
products of sums (the corresponding algebra is discussed later).

Some related Mellin transforms can be found by index
shifting and integration by parts:
\begin{eqnarray}
  M_n\left[ x^k f(x)\right] &=& M_{n+k}\left[ f(x) \right], 
  \label{eqn:shift}  \\
  M_n\left[ \frac{d}{dx} f(x) \right] &=& x^n f(x) |_0^1
    - (n-1) M_{n-1}\left[ f(x) \right].
  \label{eqn:deriv}
\end{eqnarray}
In the latter relation, the boundary term has to be properly defined.
In the limit $x\to 0$ it vanishes since we always consider $n$
higher than the order of the highest pole that $f(x)$ may have at $x=0$.
If $f(x)$ is some HPL, then its limit at $x\to 1$ may be either
zero, a non-zero constant, or singular as
$\ln^k(1 - x)$. Such a singularity is present e.g. in $\HPL{1}{x} = - \ln(1-x)$,
so that Mellin transform of $\frac{d}{dx} \HPL{1}{x} = \frac{1}{1-x}$
does not exist.
On the other hand, such functions can be regularized by turning them into
a plus-distribution, and $\left[\frac{1}{1-x}\right]_+$
has a well-defined Mellin transform,
$-\HS{1}{n-1}$.
We notice that this result can be obtained if we artificially drop the
logarithmically divergent contributions from the boundary term
in Eq.~(\ref{eqn:deriv}). In general, we may define the regularized
derivative $\hat{\partial}_x$ acting as follows:
\begin{eqnarray}
  && M_n\left[ \hat{\partial}_x f(x) \right] = R\left[f(x)\right]
    - (n-1) M_{n-1}\left[ f(x) \right]\,
\end{eqnarray}
where
\begin{eqnarray}
  && R[g_a(x)\ln^a(1-x) + g_b(x)\ln^b(1-x) + ... + g_0(x)] = g_0(1)\,, 
  \\ \nonumber
  && ~ a,b,... > 0,~~ g_k(1) \ne 0~~ \forall k > 0\,.
  \label{eqn:rderiv}
\end{eqnarray}
The extraction of divergent logarithms from HPLs is implemented e.g.
in the function \verb+HPLLogExtract+ of the HPL.m
package~\cite{Maitre:2005uu}. 

Using this definition and the inverse Mellin transform,
we may establish relations between the derivatives of HPLs and
the common generalized functions:
\begin{eqnarray}
  \hat{\partial}_x~ 1 &=& \delta(1 - x), \nonumber\\
  \hat{\partial}_x \HPL{1}{x} &=& \left[\frac{1}{1-x}\right]_+, \nonumber\\
  \hat{\partial}_x \HPL{11}{x} &=& -\left[\frac{\ln(1-x)}{1-x}\right]_+, \nonumber\\
  \hat{\partial}_x \HPL{111}{x} &=& \frac{1}{2}
  \left[\frac{\ln^2(1-x)}{1-x}\right]_+, 
  \nonumber\\
  \hat{\partial}_x \HPL{101}{x} &=& \frac{\pi^2}{6} \left[\frac{1}{1-x}\right]_+
    + \frac{\HPL{01}{x} - \zeta_2}{1-x}, ~~\mbox{etc.}
\end{eqnarray}
Thus, it is not necessary to separately consider Mellin images of
functions related to the derivatives of HPLs, such as
$\frac{\HPL{...}{x}}{1+x}$, $\left[\frac{\HPL{...}{x}}{1-x}\right]_+$, etc.

At a given weight of sums, one may interpret the
relations obtained by extending Eq.~(\ref{eqn:mttab}) as a system of
equations, 
and complete it with the generalized derivatives of every
line (excluding trivial relations).
It is possible to solve this system for the monomials
$1/n^k$, $\HS{...}{n}/n^k$, and $(-1)^n \HS{...}{n}/n^k$ in
order to determine their inverse Mellin transforms.
While writing this paper, we have learned about the
work \cite{Ablinger:2011te} that in particular mentions
some very general algorithms to compute Mellin images
of HPLs of arbitrary weights. Unfortunately, at this
moment the program implementing those algorithms
is not yet available to the public.

If a Mellin image that we deal with is a linear combination
of such known terms (with possibly shifted indices),
its inverse transform is trivial to find with the help of the
linearity of $M_n[f(x)]$ and Eq.~(\ref{eqn:shift}).
However, to find the convolution of two functions we need
to multiply the two Mellin images and the resulting terms
may have a more complicated structure. Nevertheless, it is
possible to transform the expressions to the required shape.

First, the factorized denominators $(n+a)^{-i} (n+b)^{-j} ...$
must be decomposed by partial fractioning.
Second, the arguments of harmonic sums must be brought to agreement
with the denominators using the definition Eq.~(\ref{eqn:hsums}):
$\HS{a,\vec{b}}{n+1} = \HS{a,\vec{b}}{n} + f_a(n+1) \HS{\vec{b}}{n+1}$.
Finally, the products of sums must be transformed into
linear combinations of sums of higher orders. The corresponding
algebra originates from the obvious identity
\begin{equation}
  \left(\sum_{i=1}^n a_i\right) \left(\sum_{j=1}^n b_j\right)
  = \sum_{i=1}^n \left(a_i \sum_{j=1}^i b_j\right)
  + \sum_{i=1}^n \left(b_i \sum_{j=1}^i a_j\right)
  - \sum_{i=1}^n \left(a_i b_i\right),
\end{equation}
which is applied recursively. For every product, it is then possible
to arrive at the decomposition such as
$\HS{-1,2}{n} \HS{-1}{n}
  = 2 \HS{-1,-1,2}{n}
  + \HS{-1,2,-1}{n}
  - \HS{-1,-3}{n}
  - \HS{2,2}{n}$.

The above algorithm has been implemented as a Mathematica program.
Since the HPL package~\cite{Maitre:2005uu} is not capable of computing
integrals Eq.~(\ref{eqn:mellinn}) with arbitrary $n$, the table of
Mellin transforms (similar to Eq.~(\ref{eqn:mttab}))
has to be pre-computed. The regular form of the
relations Eq.~(\ref{eqn:mttab}) and the further results greatly
simplifies the solution. We have been able to compile the
table for the transforms of HPLs up to weight five, which should
be sufficient to evaluate convolutions relevant to the NNNLO
correction to the Higgs boson production.
For that purpose one has to consider among other contributions the
convolution of the NNLO partonic $gg$-induced cross section with the
LO splitting function. The former contains contributions like
$H_{010}(x)/(1-x)$ which can be written in terms of generalized derivatives of
HPLs of weight 4. The LO splitting function has contributions from HPLs of
weight~1 thus resulting in quantities of weight 5.
Similary, the convolution of the NLO partonic cross section with the
NLO splitting function involves HPLs of weight~2 and~3, respectively,
again leading to HPLs of weight~5.


\end{appendix}



\end{document}